\documentclass[twocolumn]{aastex631}

\usepackage{siunitx}
\usepackage{natbib}
\usepackage{bm}
\usepackage{mathtools}
\usepackage{multirow}
\usepackage{CJK}         
\usepackage{hyperref}

\begin{document}
\begin{CJK*}{UTF8}{gbsn}

\title{Disk Structure May Determine AGN Variability and Explain the Accretion Disk Size Problem: No Broad Line Region Required}





\author[0000-0002-1174-2873]{Amy Secunda}\thanks{E-mail: asecunda@flatironinstitute.org}
\affil{Center for Computational Astrophysics, Flatiron Institute, New York, NY 10010, USA}

\author[0000-0002-2624-3399]{Yan-Fei Jiang (姜燕飞)}
\affil{Center for Computational Astrophysics, Flatiron Institute, New York, NY 10010, USA}

\author[0000-0002-5612-3427]{Jenny E. Greene}
\affil{Department of Astrophysical Sciences, Princeton University, Peyton Hall, Princeton, NJ 08544, USA}

\begin{abstract}
Variability is a crucial probe of active galactic nuclei (AGN) disks, but the impact of the structure and internal physics of the disk on this variability is poorly understood. We perform 3D multi-frequency radiation magnetohydrodynamic simulations of the $1300-5940~\rm{\AA}$ emitting region of an AGN disk. This turbulent disk simulation happens to stochastically have different disk thickness above and below the midplane. Light curves emitted from the thin top of the disk are broadly consistent with traditional variability models, where the X-ray irradiation emitted from the corona is the main driver of variability in UV-optical light curves. Conversely, the UV-optical light curves emitted from the thicker bottom of the disk are not strongly correlated with the X-ray light curve and have disk continuum reverberation mapping lags that are $3-5\times$ longer than the light travel time. While these features are inconsistent with traditional variability models, they are in good agreement with numerous recent observations. Our simulations allow us to interpret these recent observations with a new model for AGN disks, where the thicker disk leads to greater X-ray absorption, less X-ray reprocessing at larger disk radii, and a stronger signal from intrinsic variability driven by fluctuations in the AGN disk. In our simulations, we can measure the inflow timescale of these fluctuations as a long lag between light curves emitted from different disk radii. Understanding how disk structure impacts AGN variability and reverberation mapping can allow us to predict the thickness of AGN disks using UV-optical AGN light curves.
\end{abstract}

\section{Introduction}
\label{sec:intro}

Active galactic nuclei (AGN) accretion disks are important environments for investigating accretion physics and drive powerful radiation and outflows that are believed to play a significant role in shaping how galaxies form and evolve \citep{Fabian2012,Kormendy2013}. Because most AGN are too distant to resolve spatially, studying how AGN corona, accretion disks, and broad line regions (BLRs) vary over timescales of hours to years is the main way we can probe the structure and internal physics of AGN. In particular, there are many different methods for using AGN variability to study the AGN accretion disk. For example, many studies fit different models to AGN variability and relate these models to different AGN properties \citep[e.g.,][]{MacLeod:2010,Kelly:2009,Kozlowski2010,MacLeod2012,Yu:2022,Burke2021,Burke:2023,Beard:2025,Petrecca:2024,Petrecca:2026}. Another popular method is disk continuum reverberation mapping (RM) which measures time lags between the variability in different wavelength AGN light curves on the light-crossing time to measure the radial temperature profile and radial extent of AGN disks \citep[e.g.,][]{Krolik:1991,Collier:1999,Sergeev:2005,Cackett:2007,Cackett:2018,Cackett:2021,Edelson:2015,Edelson2017,Edelson:2019,Fausnaugh:2016,Jiang:2017,Starkey2017,Homayouni:2022}.

In traditional models for AGN variability and RM, time-varying X-rays are emitted by the corona and move outwards along the disk where they are absorbed, reprocessed, and re-emitted based on the local blackbody temperature of that radial region of the disk in the UV or optical. In this model X-ray variability is responsible for driving the variability observed in UV and optical light curves emitted by the AGN disk. As a result, UV-optical variability should lag the X-ray variability by the light-crossing timescale to each different wavelength-emitting region, and all wavelength light curves should be highly correlated.

However, there is growing evidence that this traditional X-ray-driven RM model does not capture the complete picture of AGN variability, casting doubt on the \cite{Shakura1973} standard thin disk model that has been used for decades to model AGN disks. For example, there are a number of AGN for which the X-ray and UV-optical light curves are only moderately \citep[e.g.][]{Edelson:2019,Kara:2023,Edelson:2024} or weakly \citep{Cackett:2023,Buisson:2018,Morales:2019} correlated. There are several proposed explanations for this lack of correlation, such as a rapidly varying corona temperature and height \citep{Panagiotou:2022B} or the absorption of X-rays by intervening material \citep{Kara:2021}. Alternatively, intrinsic variability in UV-optical light curves driven by disk fluctuations instead of X-ray variability would lower the correlation between X-ray and UV-optical light curves. There is observational evidence for this intrinsic variability in many AGN light curves \citep[e.g.,][]{Arevalo:2009,F92020,Neustadt:2022,Beard:2025,Hagen:2025,Petrecca:2026}, and radiation magnetohydrodynamic (MHD) simulations show that intrinsic variability may naturally arise from fluctuations in the UV-optical emitting region of the disk, due to magnetorotational instability (MRI) driven turbulence \citep{BalbusHawley1991,Secunda:2024,Secunda:2025} or convection caused by increased opacity in the UV-optical region of the disk \citep{JiangBlaes2020,Secunda:2024,Secunda:2025}. 


There are also a growing number of RM campaigns that observe continuum lags that are $\approx3\times$ longer (or more) than predicted by the standard thin disk model \citep[e.g.,][]{Steyn:2026,Guo:2022,Wang:2025,Lewin:2025}. Sometimes splitting longer baseline light curves from an individual AGN into different $\sim 100$~day epochs reveals changes between epochs from well-correlated lags on timescales consistent with the thin disk model to poorly-correlated lags on timescales longer than the thin disk model \citep[e.g.][]{Lewin:2024,Hagen:2025,Zhou:2025}. Contamination from diffuse continuum emitted by the more distant BLR is often invoked to explain these longer than anticipated lags \citep[e.g.,][]{Cackett2021,Steyn:2026,Lewin:2023,Wang:2025,Best:2024}. However, non-instantaneous reprocessing timescales \citep{Salvesen:2022}, changes in the radial temperature profile and disk albedo due to X-ray heating \citep{Secunda:2024,Secunda:2025,Fausnaugh:2016,Kammoun:2021,Langis:2024}, the impact of magnetic fields \citep{Sun:2020,Ren:2024}, or biases in lag measurements from intrinsic variability \citep{Secunda:2023} could also produce longer than anticipated lags and should be investigated.

Finally, in addition to evidence for intrinsic variability in UV-optical AGN light cures, there is also evidence that we are observing the propagation of the fluctuations driving this intrinsic variability in multi-waveband AGN light curves \citep{Neustadt:2022,Neustadt:2023,Stone:2023,F92020,Yao:2022,Secunda:2023,Hagen:2025}. Observing these propagating fluctuations as long lags between intrinsic variability in different wavelength AGN light curves offers the exciting prospect of measuring the inflow timescale and testing the standard thin disk model. The standard thin disk model predicts an inflow timescale,
\begin{equation}
\label{eq:inf}
    \tau_{\rm inflow} =260~\rm{year} \frac{M}{10^8~M_\odot} \frac{0.1}{\alpha} \left( \frac{H/R}{0.01} \right)^{-2}
    \left( \frac{R}{30~r_g} \right)^{3/2},
\end{equation}
where $M$ is the supermasssive black hole (SMBH) mass, $\alpha$ is the \cite{Shakura1973} viscosity parameter, $H$ is the height of the disk, $R$ is the cylindrical disk radius, and $r_g$ is a gravitational radius. In this thin disk scenario, the long lag would be too long to measure in human lifetimes. However, these recent observations of propagating fluctuations suggest that AGN disks may, at least in some cases, be thicker than usually assumed, $H/R>0.01$. Understanding the propagation of fluctuations better using simulations is crucial to improving our ability to detect these long lags.

The recent development of accurate, high-speed numerical methods to calculate radiative transfer coupled to MHD evolution allows us to study how both radiation and magnetic fields impact AGN disk structure and turbulence \citep{Jiang:2013,JiangBlaes2020,Jiang:2025,Hopkins:2023,Secunda:2024,Secunda:2025,Kaul:2025}. These simulations show that AGN disks are not static and that fluctuations in the disk can impact the light curves they emit, even when time-varying X-rays are irradiating the disk \citep{Secunda:2024,Secunda:2025}. Here, we expand on these simulations by performing 3D multi-frequency radiation MHD disk wedge simulations of an AGN disk over a resolved radial range of $50-250~r_g$. We irradiate our simulated AGN disk with time-varying X-rays emitted by a mock corona located within the inner boundary of our simulation. These simulations allow us to test traditional models for RM and understand the underlying physics behind poorly correlated X-ray and UV-optical light curves, longer than predicted light-crossing lags, and long lags on the inflow timescale.

We describe our four multi-frequency radiation MHD simulations in Section \ref{sec:methods:sims} and the methods we use throughout the paper for light curve analysis in Section \ref{sec:methods:lcs}. We outline the properties of our simulated AGN disks in Section \ref{sec:results:disk}. In Section \ref{sec:results:lcs} we present the light curves from our simulations, starting in Section \ref{sec:gray} with the light curves from our simulation without X-ray irradiation. We then discuss light curves from our simulations that are consistent with traditional RM models (Section \ref{sec:top}), light curves that are inconsistent with traditional RM models (Section \ref{sec:departures}), and the lags that we observe in our light curves on the inflow timescale (Section \ref{sec:results:long}). In Section \ref{sec:discuss:compare} we compare our simulated light curves with observations to provide physical explanations for departures from traditional RM models. Finally, in Section \ref{sec:discuss:caveats} we provide a thorough overview of the caveats of our simulations, and in Section \ref{sec:conclude} we provide the conclusions of the paper.





\section{Methods}
\label{sec:methods}
\subsection{Simulation Set-up}
\label{sec:methods:sims}
We use the MHD code {\sc athena++} \citep{Stone:2020} with the radiation transport module developed by \cite{Jiang2021} and \cite{Jiang2022} to perform multi-frequency radiation MHD simulations of an AGN disk on a spherical polar coordinate ($r,\theta,\phi)$ grid centered on a $10^8~M_\odot$ SMBH. The full set of ideal MHD equations is solved, and coupled to the radiation transport equations by evolving specific intensities over discrete angles. To avoid initializing a torus and evolving it to a steady state, we restart the simulations in this paper from the gray radiation MHD simulation, AGNUVB0.03, from \cite{Jiang:2025}, where the radiation was integrated over all frequencies and there was no X-ray irradiation. The mass accretion rate in this simulation was a few percent Eddington (see Figure \ref{fig:mdot_t} in Appendix \ref{app:disk}).

Unlike the multi-frequency radiation MHD shearing box simulations of the UV-emitting region of an AGN disk in \cite{Secunda:2024} and \cite{Secunda:2025}, the simulations in this paper are configured as a 3D wedge, with a broad radial range. The wedge dimensions  are $(r,\theta,\phi) \in [50~r_g,10^3~r_g]\times[0,\pi]\times[0,0.39]$. The simulations have (32, 32, 4) root level cells and five levels of static mesh refinement with maximum refinement of $dr/r=0.3\%$ at $(r,\theta,\phi) \in [50~r_g,250~r_g]\times[87^{\circ},93^{\circ}]\times[0,0.39]$ in order to resolve the inner disk and the disk mid-plane. The extent of this simulation was chosen for computational feasibility and in order to simulate the UV-emitting region. In the analysis in this paper, we only use simulation output from within the cylindrical disk radius, $R<250~r_g$, where the mid-plane is sufficiently resolved. The simulation is initialized with a single magnetic-field loop and becomes magnetically dominated during the early evolution. Additional details of the simulation set-up and early evolution are available in \cite{Jiang:2025}.

We restart the simulation from \cite{Jiang:2025} with the multi-frequency radiation solver from \cite{Jiang2022} to divide the previously gray radiation integration into two radiation frequency bins, a high-frequency bin for the X-ray irradiation ($h\nu>1$~keV, where $h$ is Planck's constant in keV~s) and a low-frequency bin for the local blackbody radiation ($h\nu<1$~keV). The effective temperatures at the radii studied in these simulations produce blackbody radiation that peaks from $1300-5940$~\AA, which includes UV and optical wavelengths. We use the same tabulated opacities from {\sc opal} \citep{IglesiasRogers1996} and {\sc tops} \citep{tops}, for the UV-optical and X-ray group, respectively, as in \cite{Secunda:2025}. When restarting, the radiation solver will split the pre-existing gray radiation in each simulation grid cell into each frequency bin based on the local blackbody temperature. Nearly all radiation is placed in the low-frequency bin.

Because the simulation does not extend to the innermost stable circular orbit or to the outer edge of the AGN disk, boundary conditions for the inner and outer limit of the simulation need to be defined. We implement these boundary conditions by adding two ghost cells that are not part of the active simulation at $r<50~r_g$ and $r>1000~r_g$. For the gas, magnetic fields, and UV-optical flux we use outflow boundary conditions in these ghost cells, as in \cite{Jiang:2025}\footnote{For outflow boundary conditions the gas, magnetic flux, and UV-optical flux are copied from the first active cell in the simulation if they are moving out of the simulation and are set to zero if they are moving into the simulation}. For the X-ray flux we tweak the inner boundary condition in order to inject X-ray irradiation as if it comes from a corona located in two axisymmetric rings with cylindrical radius $R=10~r_g$ and cylindrical heights $z=\pm30~r_g$. In practice, this means that we set the X-ray specific intensity in each ghost cell to the time dependent X-ray specific intensity along the discrete angle corresponding to the cell location relative to the corona location. 

We use the method in \cite{Timmer1995}, which assumes Gaussian-distributed light curves, to simulate an X-ray light curve with the variability power spectral density (PSD), PSD$\propto 1/f^{\alpha_{\rm in}}$, where $f$ is the variability frequency and $\alpha_{\rm in}=2$. We chose this power-law based on empirical power-law fits to X-ray light curves \citep[e.g.][]{Markowitz:2003} and to match the X-ray light curve from the Power-Law Run in \cite{Secunda:2025}. Our X-ray light curve has a timestep of $dt=2.3\times10^{-4}$~days, which is slightly smaller than the timestep of the simulation ($dt=2.5\times10^{-4}$~days), in order to probe variability down to the shortest possible timescales. We run our simulations for $T\approx 4500$~days, printing the emitted luminosity every $dt=0.23$~days.

We test the impact of the injected X-ray luminosity on the disk and the light curves emitted by the disk by running three simulations where the only difference is the total luminosity of the injected X-ray light curve. We normalize the luminosity of the injected X-ray light curve to different factors of the total disk luminosity from before we inject X-rays. The total injected X-ray luminosity is equal to the total pre-injection disk luminosity in our Medium X-ray run, 5 times the pre-injection disk luminosity in our High X-ray run, and 0.5 times the pre-injection disk luminosity in our Low X-ray run. For comparison, we also perform a fourth Gray run without injected X-rays. Figure \ref{fig:mdot_t} in Appendix \ref{app:disk} shows the accretion rate $\dot{M}$, calculated as the mass flux at the inner radial boundary, as a function of time and radius for our simulations with and without injected X-rays. The accretion rate in our simulations rapidly re-equilibrates after we begin injecting X-rays at $t=0$. The accretion rate remains similar to the Gray run accretion rate for the High and Low X-ray runs, but slightly increases compared to our Gray run at $r<130$ for the Medium X-ray run.


\subsection{Light Curve Analysis}
\label{sec:methods:lcs}

The main goal of this paper is to study the properties of the light curves emitted by our simulated AGN disks. In order to generate these light curves, we integrate the flux emitted in the positive radial-direction and the outward $\theta$-direction (positive $\theta$ below the disk and negative $\theta$ above the disk) from simulation cells at a cylindrical height between $230<z< 250~r_g$ for all possible cylindrical radii, $R$. We end up with six distinct light curves from six different cylindrical radii from $R=34-219~r_g$. The effective temperatures at these radii correspond to blackbody emission with peak wavelengths in the range $1300-5940$~\AA. We tested the impact on our results of changing the height at which we measure the outgoing flux and found that at lower heights the flux at $R>100~r_g$ was not converged. We treat the light curves in our simulation that are emitted from the top and bottom of the disk completely separately, as if we are analyzing light curves from nearly face-on disks, where each side would be viewed independently.

We note that because our light curves are drawn from simulations, they are noiseless and error-free. These light curves also have a constant cadence, eliminating the need to interpolate or model the data to perform the fast Fourier transforms required for our analysis. As a result, we can get an unbiased view of our light curve properties that is only possible with simulations.

In Section \ref{sec:results:lcs} we use several common techniques to analyze our emitted light curves.  First, we fit a bending power-law model to the variability frequency-binned PSDs of these light curves, 
\begin{equation}
\label{eq:bend}
    P(f) = A \, f^{-\alpha_l} \left[1 + \left(\frac{f}{f_b}\right)^{\alpha_h - \alpha_l}\right]^{-1},
\end{equation}
where $A$ is the normalization, $\alpha_l$ is the low-frequency slope, $\alpha_h$ is the high-frequency slope, and $f_b$ is the bending frequency. We also calculate the coherence as a function of variability frequency between the injected X-ray light curve and the emitted light curves using the frequency-binned mean cross-spectrum, $\langle C(f) \rangle$, and the frequency-binned mean PSDs of the injected and emitted light curves, $\langle P_i(f) \rangle$ and $\langle P_e(f) \rangle$, respectively. The coherence is then \citep{Vaughan:1997},
\begin{equation}
\label{eq:gamma}
    \Gamma^2(f) = \frac{|\langle C(f) \rangle|^2}{\langle P_i(f) \rangle\langle P_e(f) \rangle},
\end{equation}
and the statistical uncertainty in each bin is,
\begin{equation}
\label{eq:sig_gamma}
    \sigma (\Gamma^2) =\sqrt{\frac{2^3}{m(f)}} (1-\Gamma^2(f)),
\end{equation}
where $m(f)$ is the number of samples in each bin \citep{Bendat:1986}. This coherence measures the fraction of variability in the emitted light curve at a given frequency that can be predicted from a linear transform of the X-ray light curve \citep{Bendat:1986}.

Next, we recover lags between the variability in the injected and emitted light curves and lags between the variability in different radii emitted light curves using two methods. For the first method, we calculate the cross-correlation coefficient as a function of time delay, $\tau_{\rm lag}$ \citep{Gaskell1987,GaskellSparke1986}. The highest coefficient indicates the most probable lag. In our light curves, the traditional light-crossing time lag tends to be the most probable lag. Additional lags, such as the long lag (see Section \ref{sec:results:long}), can also be found by identifying peaks in the distribution of correlation coefficients for the cross-correlated light curves that are not present in the distribution of correlation coefficients for the auto-correlated light curve. In practice, we measure the light-crossing lag as the lag with the highest coefficient between $\pm5$~days, and consider a long lag to be present if there is an additional peak in the distribution of correlation coefficients compared to the auto-correlated light curve distribution on timescales longer than 20~days.

Often, these additional peaks can be subtle, so we use a second Fourier method described by \citep{Uttley:2014} that is better able to capture multiple lag signals. In Equation \ref{eq:gamma}, we use the magnitude of the complex cross-spectrum to measure the coherence between two different light curves. This Fourier method uses the phase of this complex cross-spectrum, $\Phi$, to measure the lag as a function of frequency through,
\begin{equation}
    \label{eq:phase}
    \tau_{\rm lag} (f) = \Phi/2 \pi f.
\end{equation}
Because this method measures the lag as a function of frequency, it may show multiple lags that are present in the light curves. To search for long lag signals in Section \ref{sec:results:long}, we focus primarily on the lags present in the lowest frequency bins, because previous real and mock observations suggest the long lag signal is strongest in those bins \citep{Yao:2022,Secunda:2023}. 

We will mostly analyze the full simulated light curves, with baselines of $T \approx 4500$~days and a constant cadence of $dt=0.23$~days. However, some light curves have $\sim50$~day flares that occur roughly once per simulation run over $\sim10$~years of simulation time (see Section \ref{sec:departures:other}). These flares make it very difficult to measure accurate cross-correlation lags on the light-crossing timescale, so we cut off these light curves, before the flares occur to measure these lags in Section \ref{sec:top}. In Section \ref{sec:departures:other}, we also briefly examine how our cross-correlation lag measurements change if we measure them over a shorter baseline light curve to compare our results with shorter timescale RM studies. To perform this comparison, we divide our light curves into intervals of as many segments of 100, 200, 300, 400, 500, 600, 1000, 2000, and 3000~days as possible. 

Finally, we note that RM actually measures the lags between the variability in different wavelength light curves corresponding to different radial temperature regions of an AGN disk. In our simulation, we know the radius of the light curve directly. Therefore, true RM is actually more complicated than what we show here. 


\begin{figure*}
    \centering

    \includegraphics[width=\columnwidth]{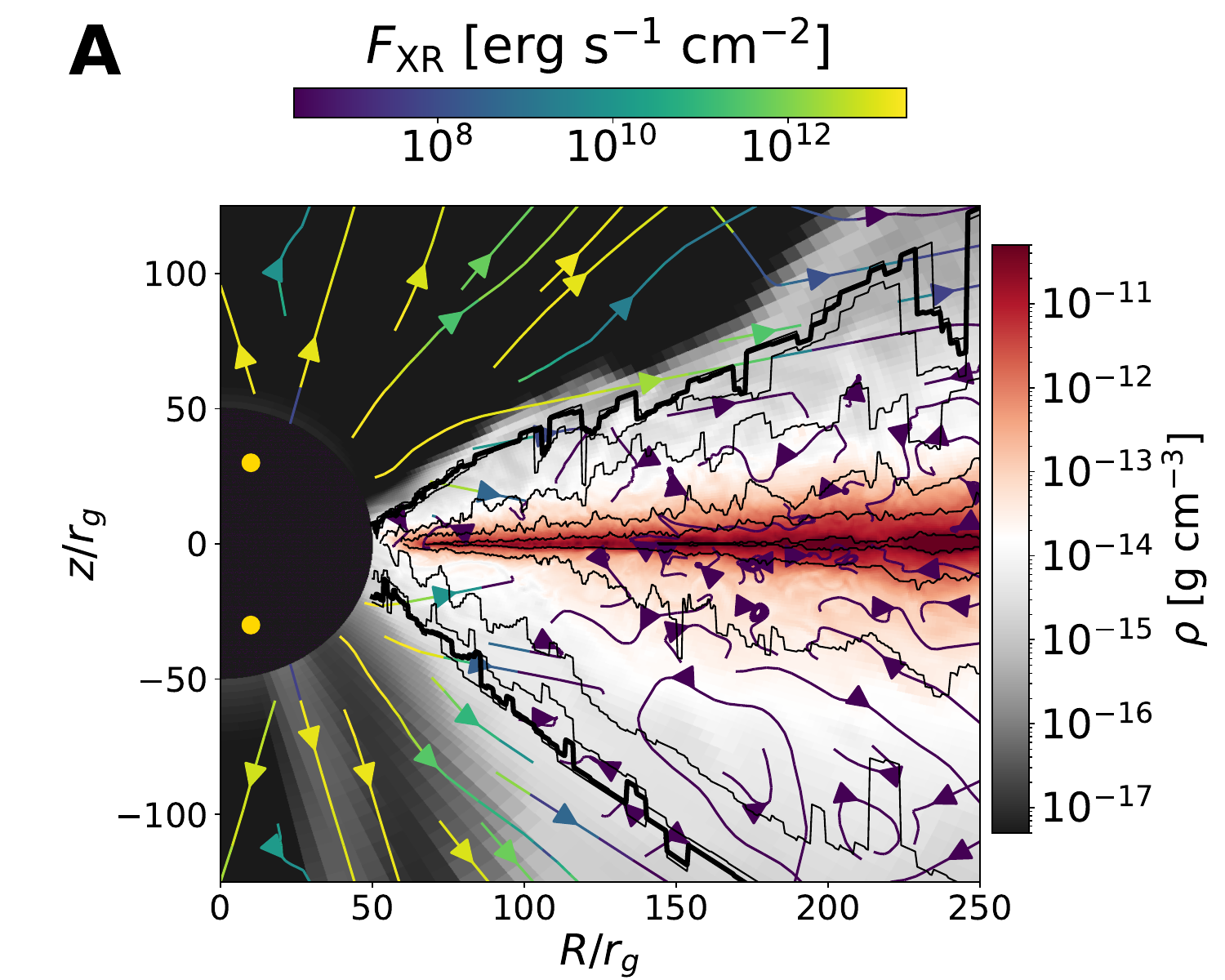}
    \includegraphics[width=\columnwidth]{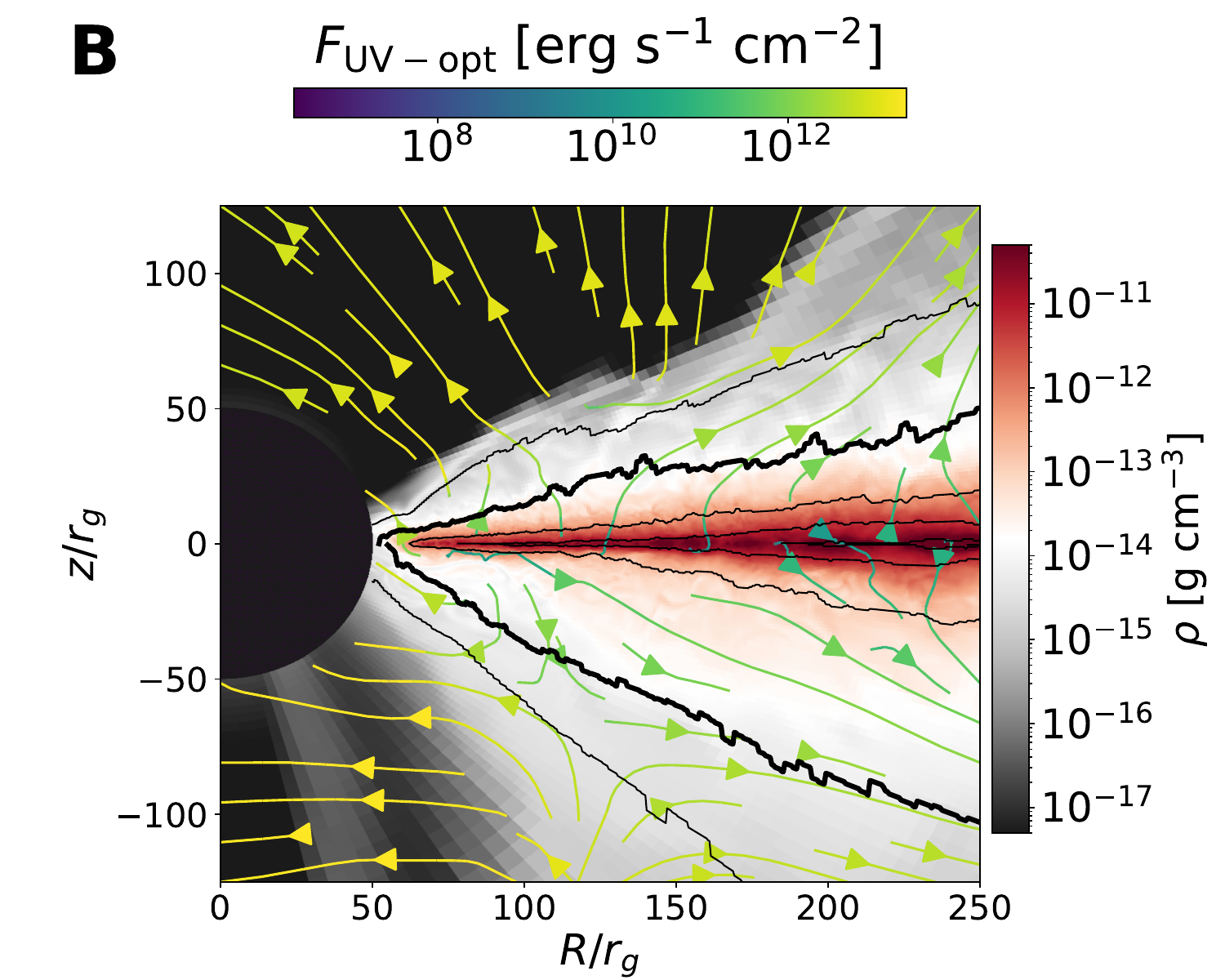} \\
    \includegraphics[width=\columnwidth]{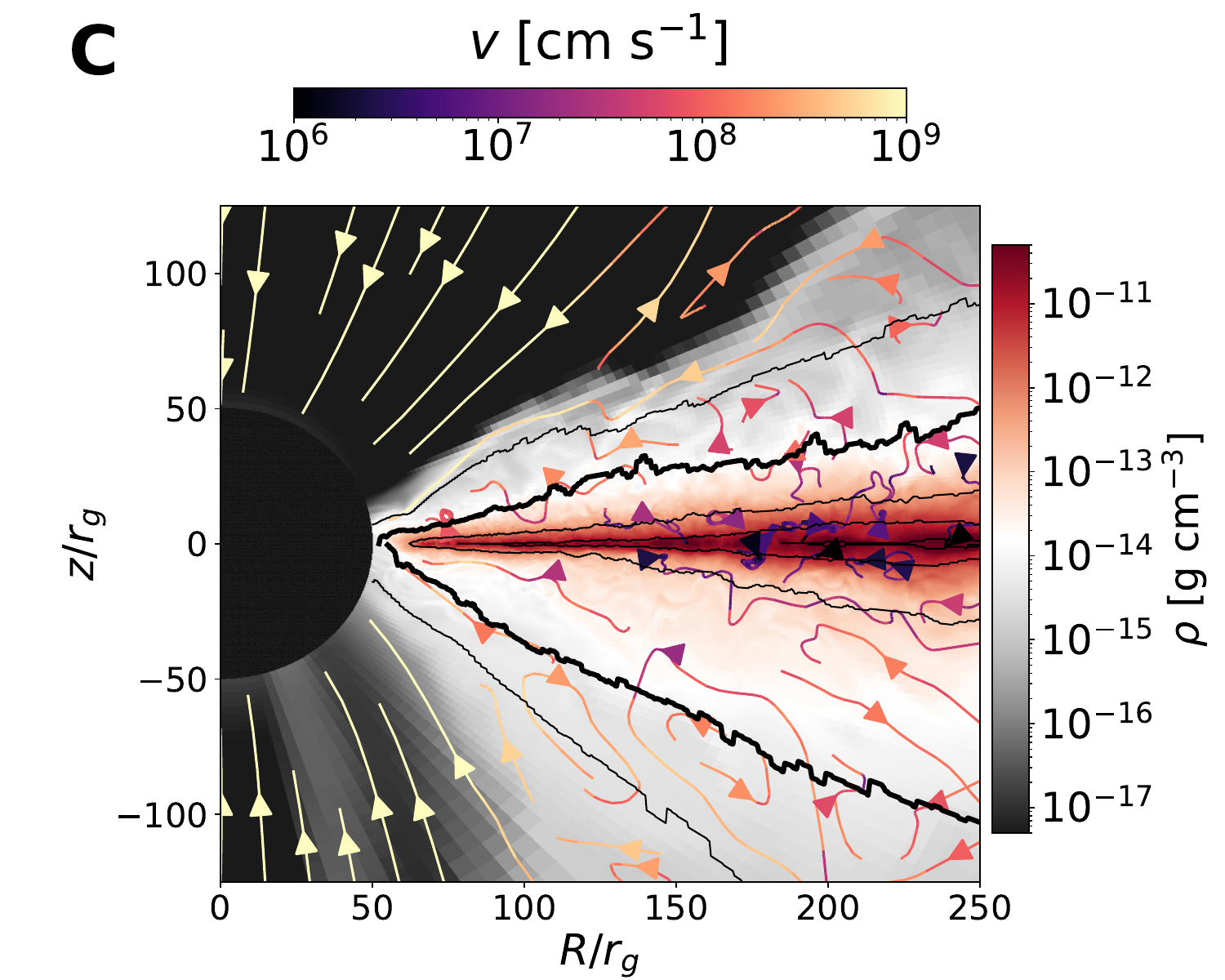}
    \includegraphics[width=\columnwidth]{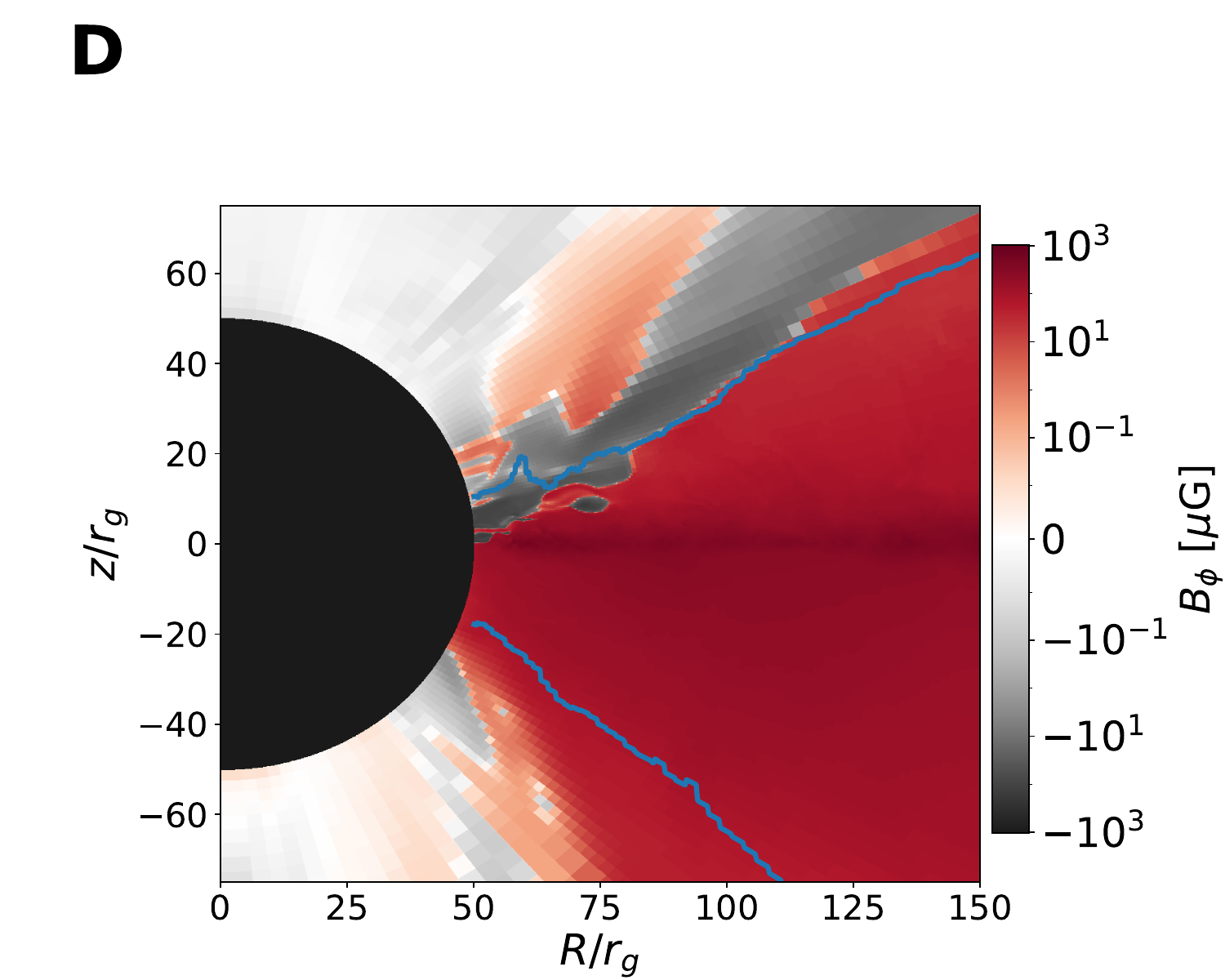}
    \caption{\textbf{Panels A-C:} Azimuthally-averaged gas density for a snapshot at time $t=600$~days for the High X-ray run. The streamlines show the azimuthally-averaged $r$- and $\theta$-components of the X-ray flux (panel A), the UV-optical flux (panel B), and the gas velocity (panel C). The thick black lines show the location of the azimuthally-averaged X-ray (UV-optical) photosphere in panel A (panels B and C), while the thin black lines show contours of optical depth increasing logarithmically by factors of ten. We show the simulated location ($(R,z)=(10,\pm30)~r_g$) of the axisymmetric X-ray emitter as yellow circles in panel A. The disk density decreases more slowly below the midplane leading to more X-ray absorption and the UV-optical flux getting advected along with the flow near the photosphere. \textbf{Panel D:} Snapshot of the azimuthal component of the magnetic field for a single $\phi$-index at $2950$~days for the Medium X-ray run. The blue line shows the location of the X-ray photosphere. There is strong reconnection occurring near the upper disk photosphere. This reconnection leads to a flare in the light curves emitted from the top of the disk at this time (see left panel of Figure \ref{fig:mid_lcs}).}
    \label{fig:rho}
\end{figure*}

\begin{figure*}
    \centering
    \includegraphics[width=0.49\linewidth]{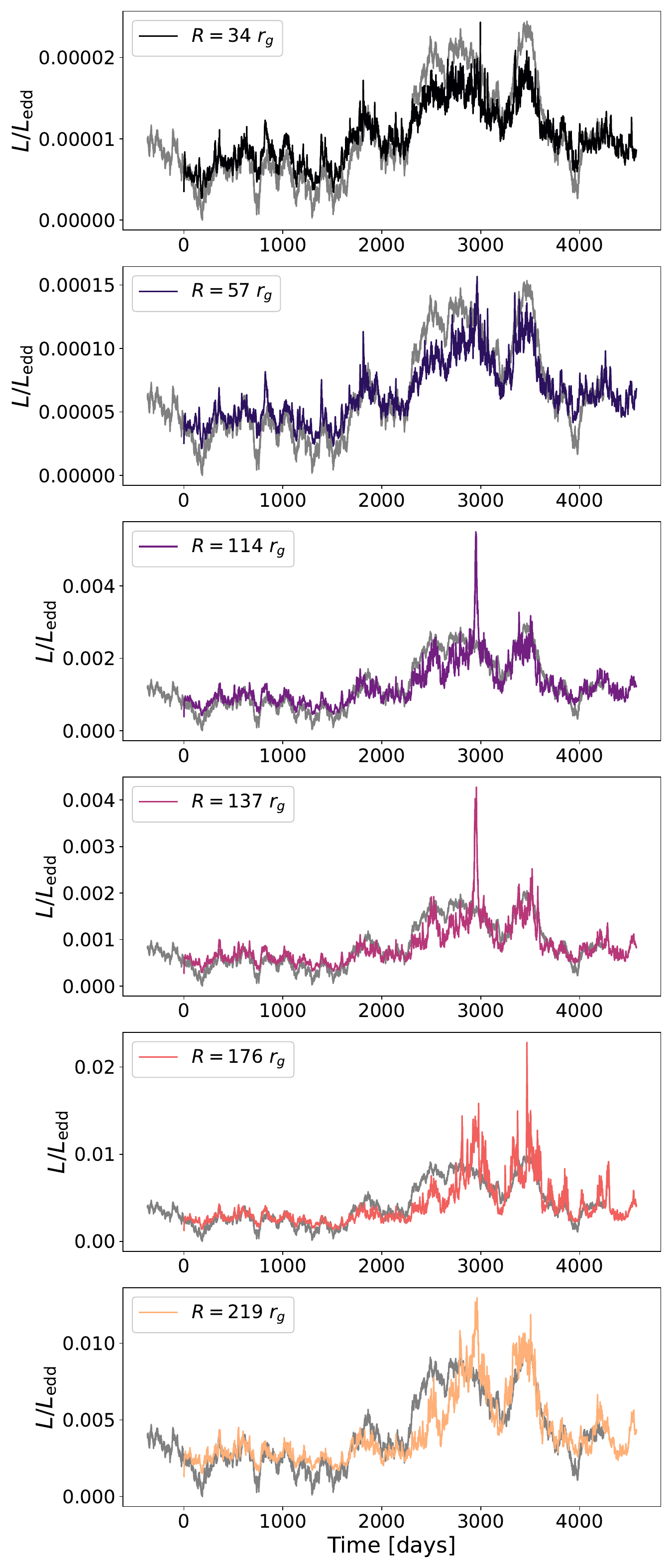}
    \includegraphics[width=0.49\linewidth]{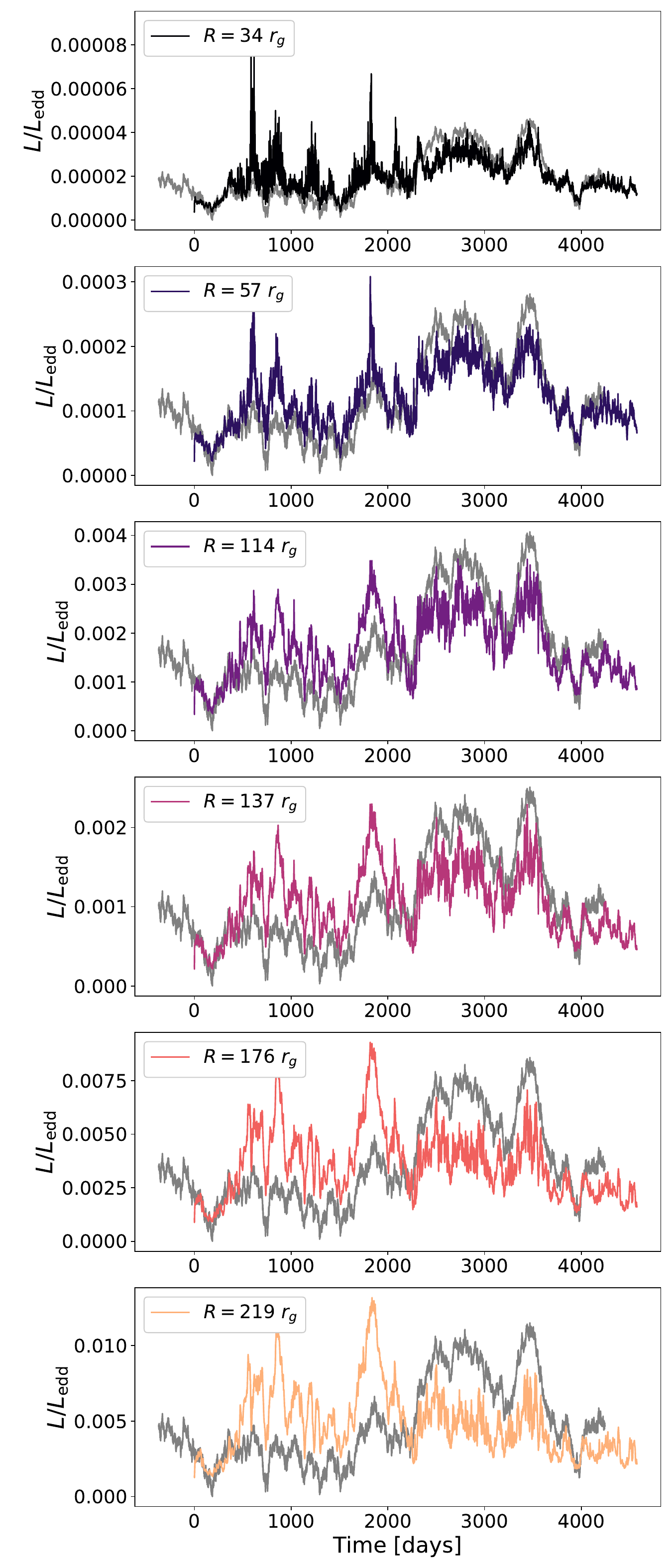}
    \caption{Light curves from the Medium X-ray run emitted from the top and bottom of the disk in the left and right panel, respectively, for six different radial regions. The effective temperatures at these radii would produce blackbody radiation that peaks in the UV and optical between $1300-5940$~\AA. We show the injected X-ray light curve (rescaled for comparison) in gray. Light curves emitted from the top of the disk are more strongly correlated with the X-ray light curve.}
    \label{fig:mid_lcs}
\end{figure*}

\section{Disk Properties}
\label{sec:results:disk}

Before presenting the light curves from our simulations, we will first outline the simulated disk structure that arises self-consistently from evolving the MHD equations coupled to the radiative transfer equations. We show a 2D snapshot of the azimuthally-averaged gas density, $\rho$, at $t=600$~days for the High X-ray run in panels A-C of Figure \ref{fig:rho}. The streamlines show the X-ray flux (panel A), UV-optical flux (panel B), and gas velocity (panel C). The contours show the optical depth from $\tau=0.1$ upward, increasing logarithmically by factors of ten. The thickest contour shows the location of the photosphere ($\tau=1$). The top and bottom of the disk exhibit very different disk structures that impact the flux absorbed and emitted by the disk.


The top of the disk is thin, with a UV-optical photosphere at $z\lesssim 25~r_g$ at all radii. The velocities within the top photosphere appear mainly turbulent (panel C), and the X-ray irradiation directly reaches the X-ray photosphere (panel A). The emitted radiation mostly moves in the $+z$-direction, directly away from the disk (panel B). In Section \ref{sec:top} we will see how this classic thin disk structure produces emitted light curves that are consistent with traditional X-ray-driven RM models. 

The structure of the bottom of the disk is different from that of the top of the disk, which we will see in Section \ref{sec:departures:bottom} has a large impact on the light curves emitted from the bottom of the disk. The UV-optical photosphere for the bottom of the disk is significantly farther out than the upper disk photosphere because the gas density decreases more gradually, leading to a higher scattering optical depth. This puffed up structure is driven by convection arising from entropy inversions within the UV-optical photosphere below the disk midplane. We see these entropy inversions in Figure \ref{fig:entropy} in Appendix \ref{app:disk}. The circular convective motion of the gas is also captured by the velocity vectors below the midplane in Panel C of Figure \ref{fig:rho}.

Because the density decreases more gradually below the disk, the X-ray photons pass through a significantly larger region at moderate densities ($\rho\sim 10^{-14}$~g~cm$^{-3}$) relative to the X-ray photons approaching the disk from the top. As a result, the X-ray irradiation is often absorbed before reaching the photosphere, especially at larger radii (panel A). The UV-optical flux also has a much larger radial component at the photosphere for the bottom of the disk compared to the top of the disk (panel B). The radial component is larger because the optical depth is high enough for the radiation to be transported along with the gas flow, which tends to be more coherent (see panel C), and because of the increased scattering opacity. 

The bottom of the disk remains significantly thicker than the top of the disk over the 4500~day timescale of our simulation runs. However, both sides of the disk are MRI turbulent, and the top of the disk occasionally becomes convectively unstable, increasing the height of the upper disk photosphere over shorter timescales. This change in disk structure does impact the light curves emitted over some epochs (see Section \ref{sec:departures:other}).

Panel D of Figure \ref{fig:rho} shows a strong magnetic reconnection event that occurs at $t=2950$~days in our Medium X-ray run, marked by the sharp change from a positive to a negative azimuthal magnetic field near the upper X-ray photosphere (blue line). A strong enough magnetic reconnection event can drive a flare in the light curves emitted from these regions. While the azimuthal magnetic field changes sign at the bottom of the disk as well, the change is more mild and farther out from the photosphere of the disk, especially at larger radii. Therefore we mainly see flares in light curves emitted from the top of the disk. We discuss this departure from traditional AGN variability models in Section \ref{sec:departures:other}.

The asymmetric disk structure in Figure \ref{fig:rho} is present in all four of our simulations because it is a feature of the initial gray simulation from \cite{Jiang:2025}. This structure occurs because turbulence in the gray simulation leads to a stochastic fluctuation in gas density that grows over time to produce a more puffed out disk for the bottom of the disk. Because the turbulent fluctuations leading to this difference between the top and bottom of the disk are stochastic, there is no guarantee that this precise disk structure would arise, and we cannot predict how common a thin or thick disk or an asymmetric disk would be based on this one initial simulation. In fact, while the asymmetry remains over the course of our $\sim 10$ simulation year runs, at other times during the $\sim100$ simulation year run of the Gray simulation in \cite{Jiang:2025}, there are many significant changes in the AGN disk structure. 

Nonetheless, our simulation shows that these AGN parameters can stochastically produce either a thin disk with a photosphere at $|z|\lesssim 25~r_g$ at most radii or a puffed out disk with a photosphere at $50~r_g<|z|<100~r_g$ at most radii. Here, we treat the emission from the top and bottom of the disk separately, because an observer performing variability analysis on a nearly face-on disk would only observe emission from one side. In the following sections, we show how this top and bottom disk structure impacts the variability in the light curves emitted from these different sides. 


Future work will explore in more depth the impact of X-ray irradiation on the disk structure, the disk temperature, and the disk spectrum. Here, we note that the X-ray irradiation does not have a large impact on the average density structure of the disk. In addition, while the X-ray irradiation clearly drives temperature fluctuations in the region near the photosphere, where the disk is not thermalized, on average the irradiation only slightly increases the temperature in the thermalized region (see Figure \ref{fig:t_of_tau} in Appendix \ref{app:disk}). This result is consistent with the results of the shearing box simulations in \cite{Secunda:2024} and \cite{Secunda:2025}.

\section{Light Curves}
\label{sec:results:lcs}

\begin{figure*}
    \centering
    \includegraphics[width=0.49\linewidth]{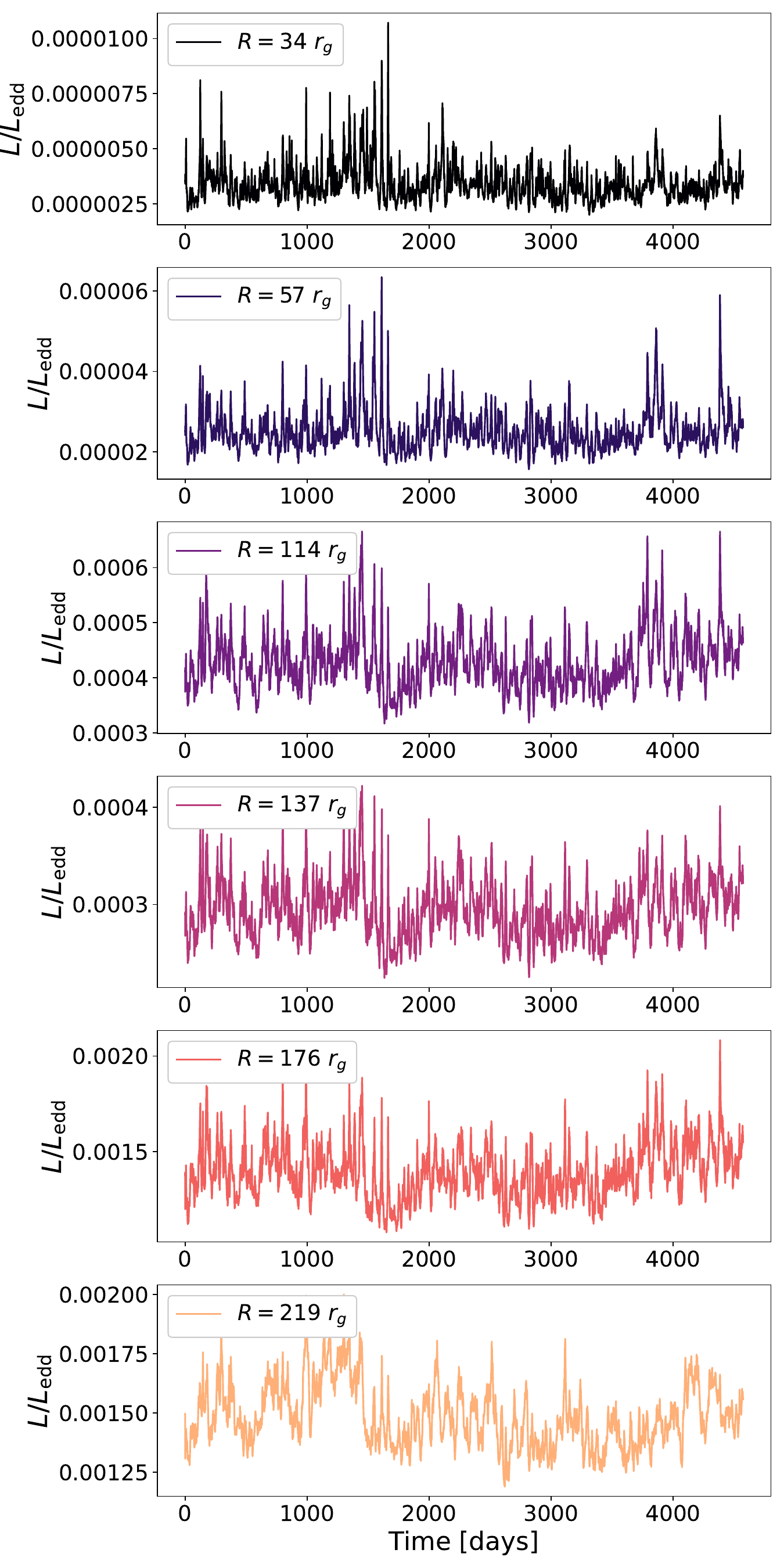}
    \includegraphics[width=0.49\linewidth]{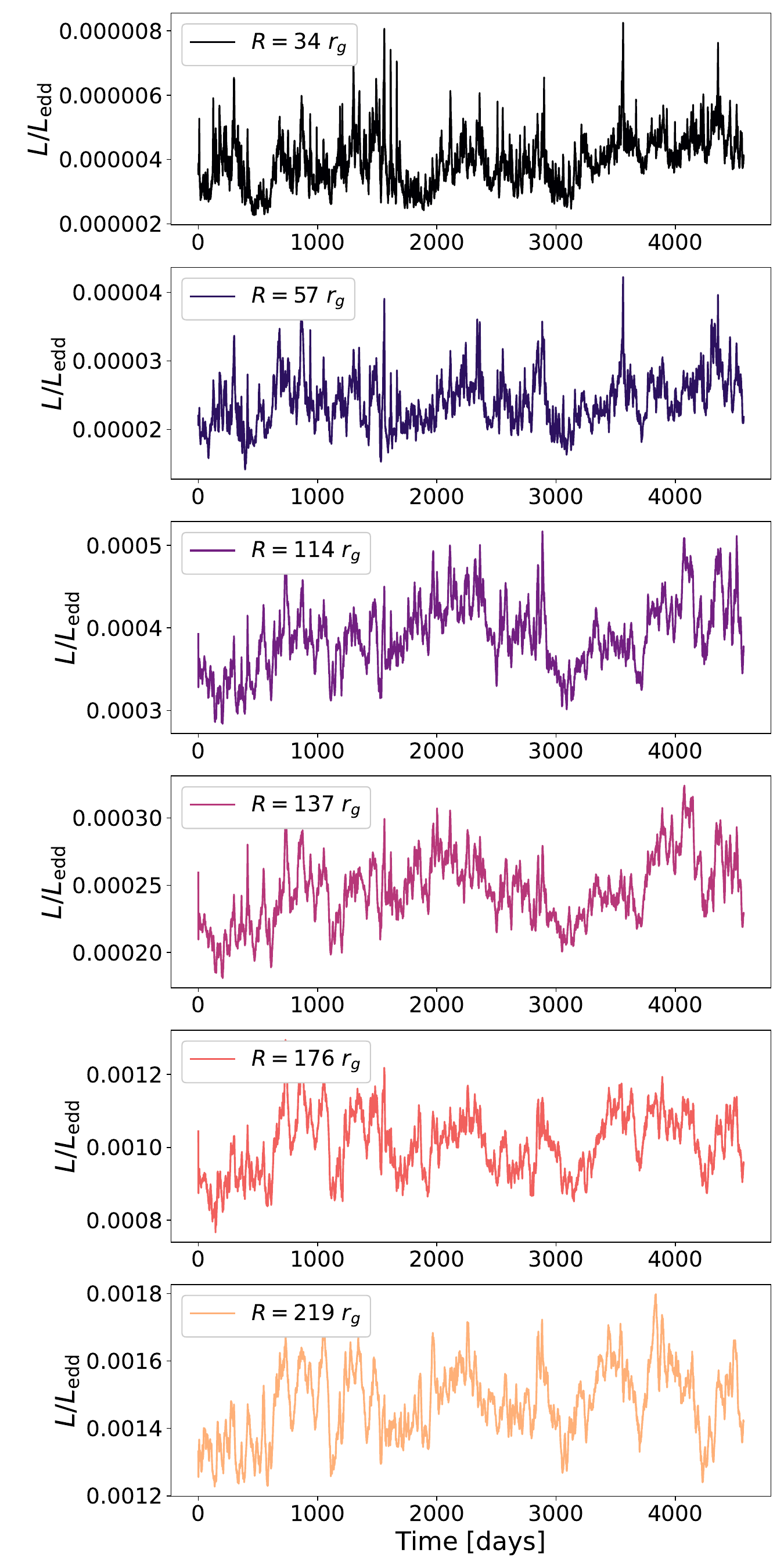}
    \caption{Light curves from the Gray run emitted from the top and bottom of the disk in the left and right panel, respectively, for six different radial regions. Even though this run has no X-ray irradiation, the light curves still show variability over a broad range of timescales.}
    \label{fig:gray_lcs}
\end{figure*}

In this section, we analyze the light curves emitted from different disk radii for our four simulations. These four simulations have different amounts of X-ray luminosity ranging from no X-ray irradiation, in our Gray run, to five times the disk luminosity before injecting X-rays, in our High X-ray run. As an example, we show the light curves emitted from six different radii from $R=34~r_g$ to $R=219~r_g$ from our Medium X-ray run in Figure \ref{fig:mid_lcs}, along with the injected X-ray light curve rescaled for comparison in each panel in gray. The effective temperatures at these radii would produce blackbody radiation that peaks in the UV and optical between $1300-5940$~\AA. 

We start in Section \ref{sec:gray} by analyzing the light curves from our Gray run to understand the intrinsic variability generated in the turbulent gas disk, which will also appear in the light curves from our runs with X-ray irradiation. Next, in Section \ref{sec:top}, we present the light curves emitted from the tops of our simulated disks, which most closely resemble the predictions of traditional X-ray-driven RM models due to the thin disk structure of the upper disk. Then, in Section \ref{sec:departures:bottom}, we analyze the light curves emitted from the bottom of the disk. Due to the thicker disk structure for the bottom of our disk, we find that the variability in these light curves is inconsistent with traditional RM models. In Section \ref{sec:departures:other}, we highlight a few ways in which the light curves emitted from the top of the disk are also inconsistent with traditional RM models. Finally, we study how the propagation of density fluctuations in our simulated AGN disks produces long lags in the emitted light curves in Section \ref{sec:results:long}.

\subsection{Light Curve Variability Without X-ray Irradiation}
\label{sec:gray}

\begin{figure*}
    \centering
    \includegraphics[width=\columnwidth]{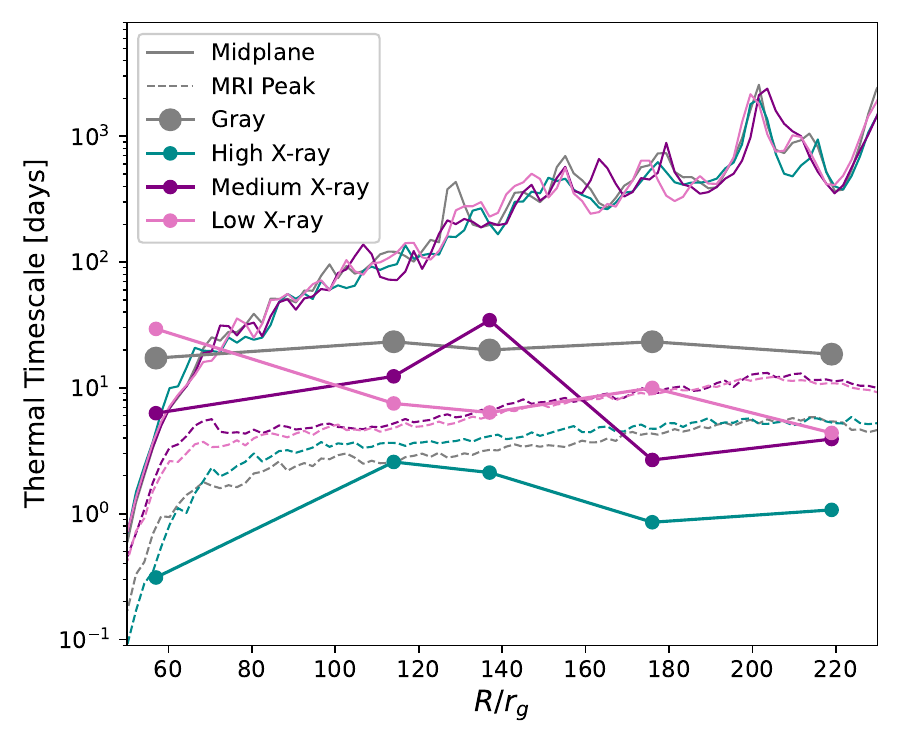} 
    \includegraphics[width=\columnwidth]{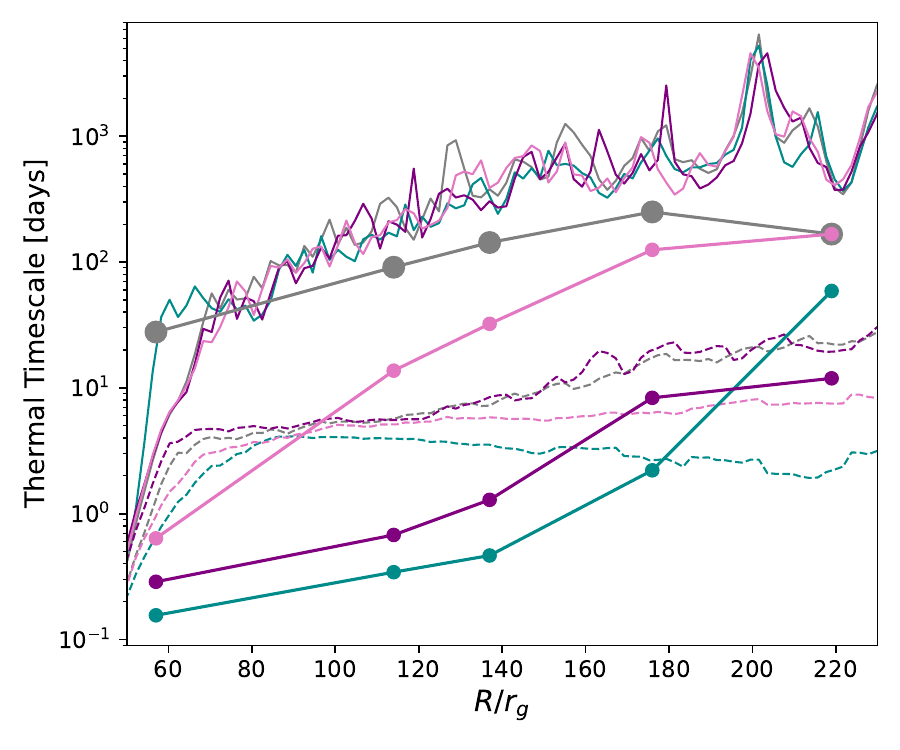} 
    \caption{The time- and azimuthally-averaged thermal timescale from the midplane (solid lines) and from where MRI turbulence peaks (dashed lines) as a function of radius, $R$, for our four runs. For comparison, we show the bending timescales we fit to the light curve PSDs at each radius from our four different simulations. Comparing the bending and thermal timescales can probe the height in the disk where the variability originates.}
    \label{fig:thermal}
\end{figure*}

\begin{figure}
    \centering
    \includegraphics[width=\linewidth]{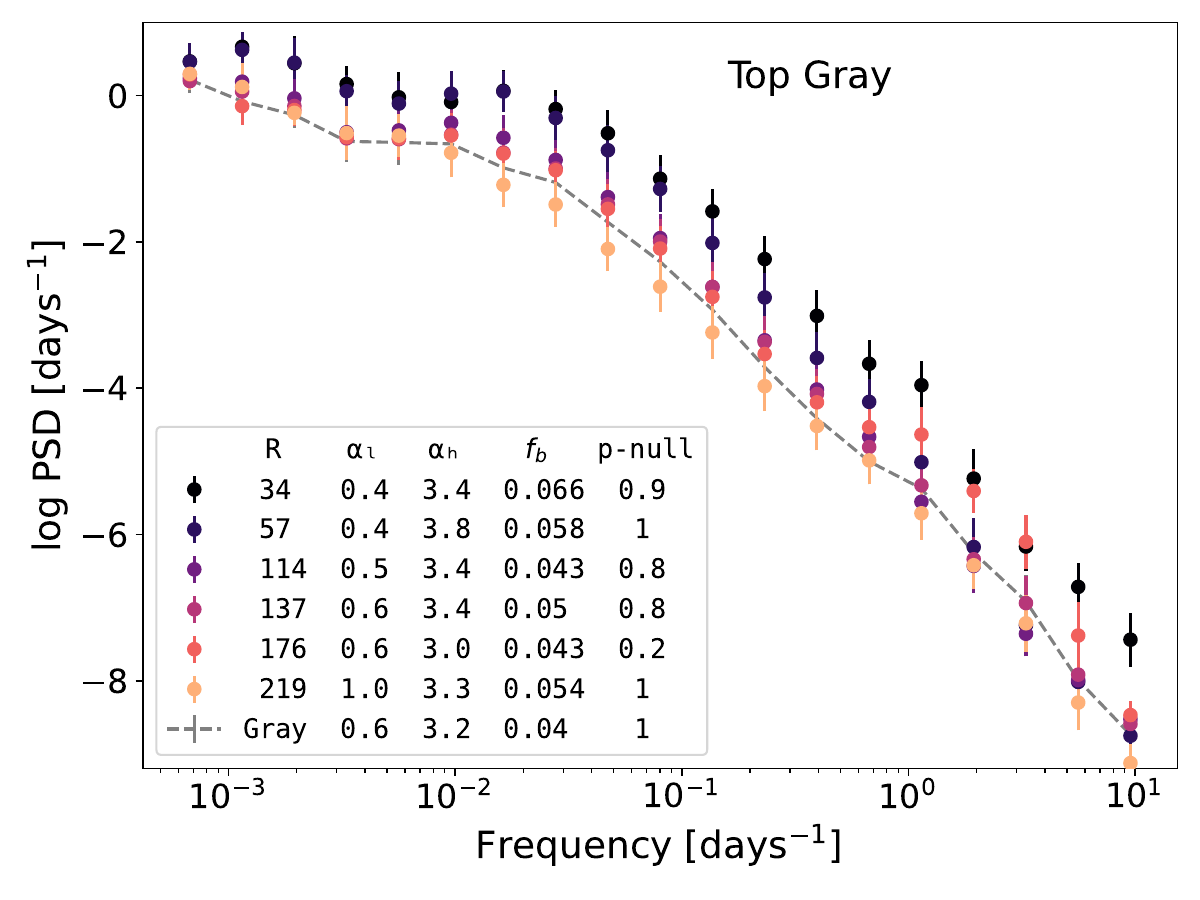} \\
    \includegraphics[width=\linewidth]{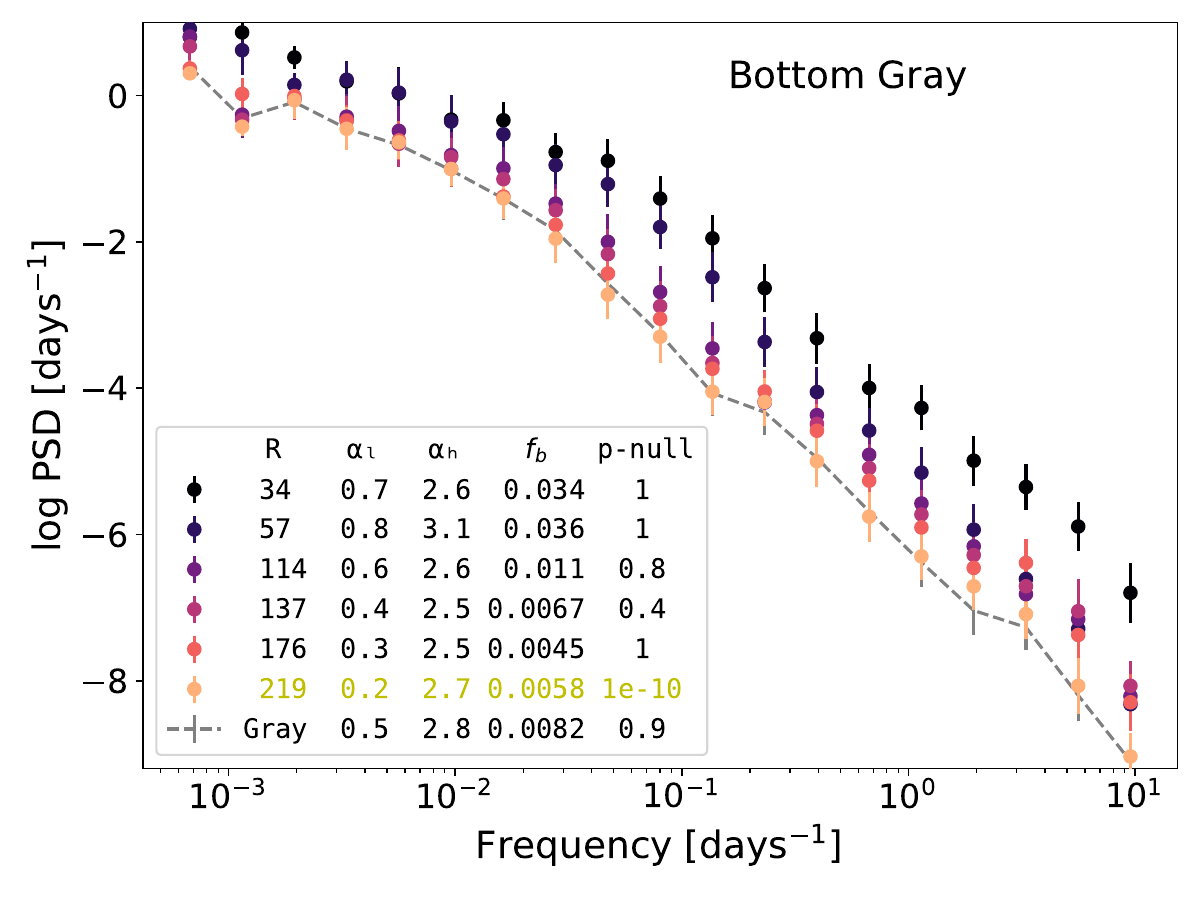} \\
    \includegraphics[width=\linewidth]{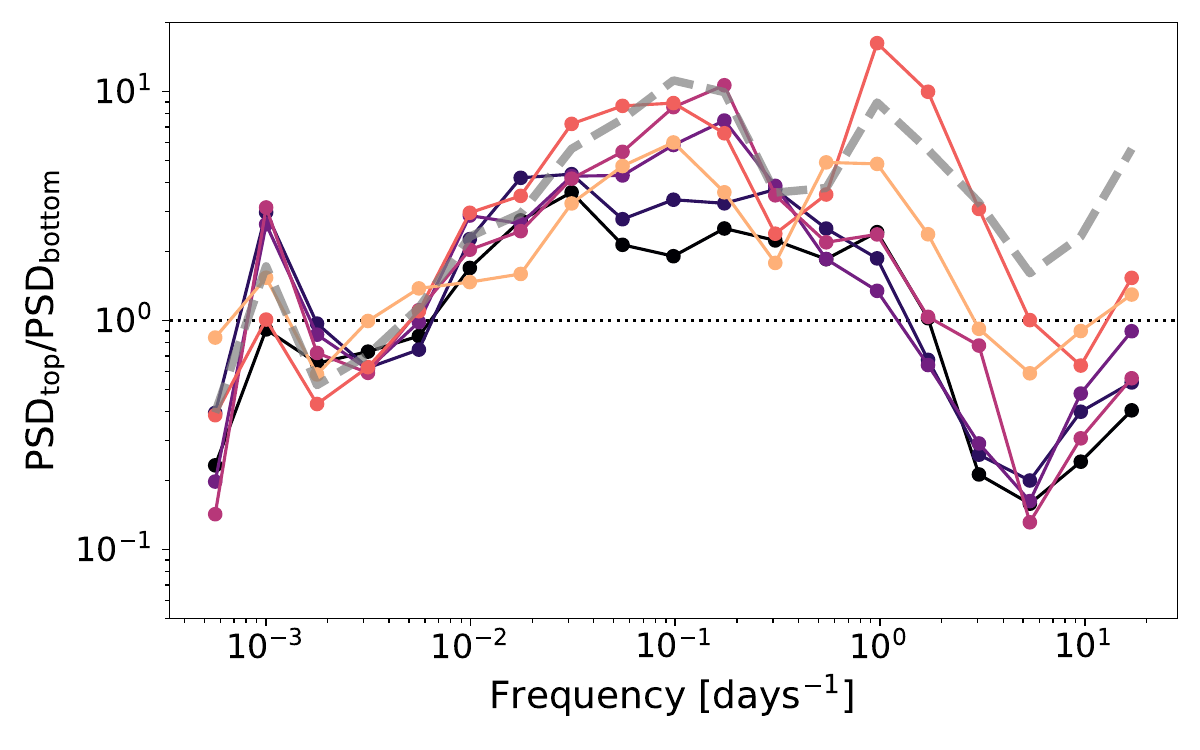}
    \caption{Frequency-binned PSDs of light curves from the Gray run without injected X-rays emitted from from six different radii, $R$. The top (middle) panel shows the PSDs for light curves emitted from the top (bottom) of the disk and bottom panel shows the ratio between the two PSDs (bottom panel). The points show the mean of each bin and the error-bars show the standard deviation of that bin. The dashed line shows the PSD for the light curve summed over all radii. The labels give the parameters and p-null value for the best-fit bending power-law at each radius. Inner radii light curves tend to have greater variability and the light curves emitted from top of the disk have more variability at intermediate frequencies than the light curves emitted from the bottom.}
    \label{fig:gray_psd}
\end{figure}

\begin{figure*}
    \centering
    \includegraphics[width=0.49\linewidth]{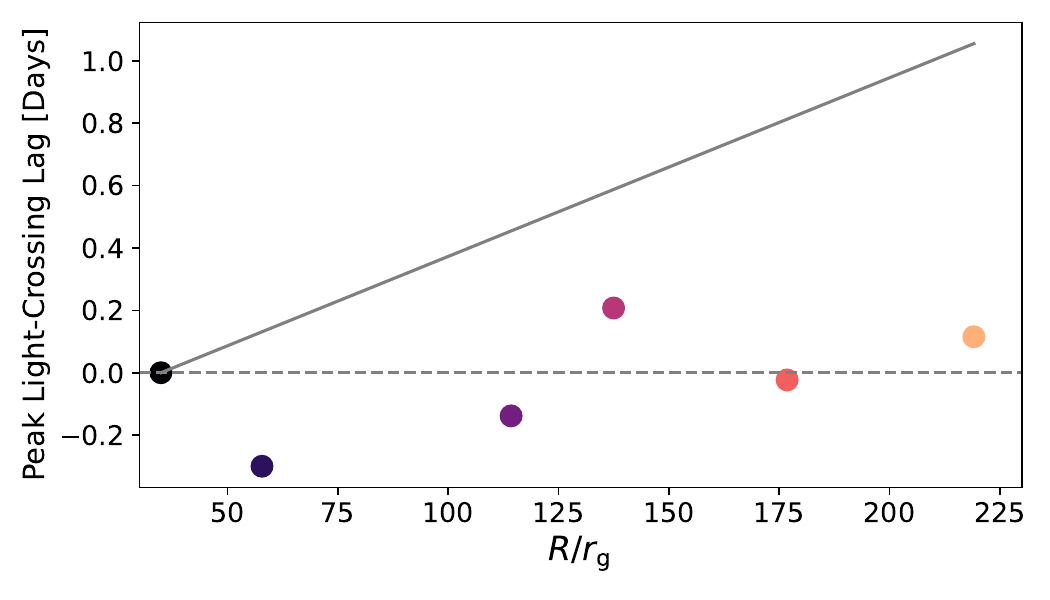}
    \includegraphics[width=0.49\linewidth]{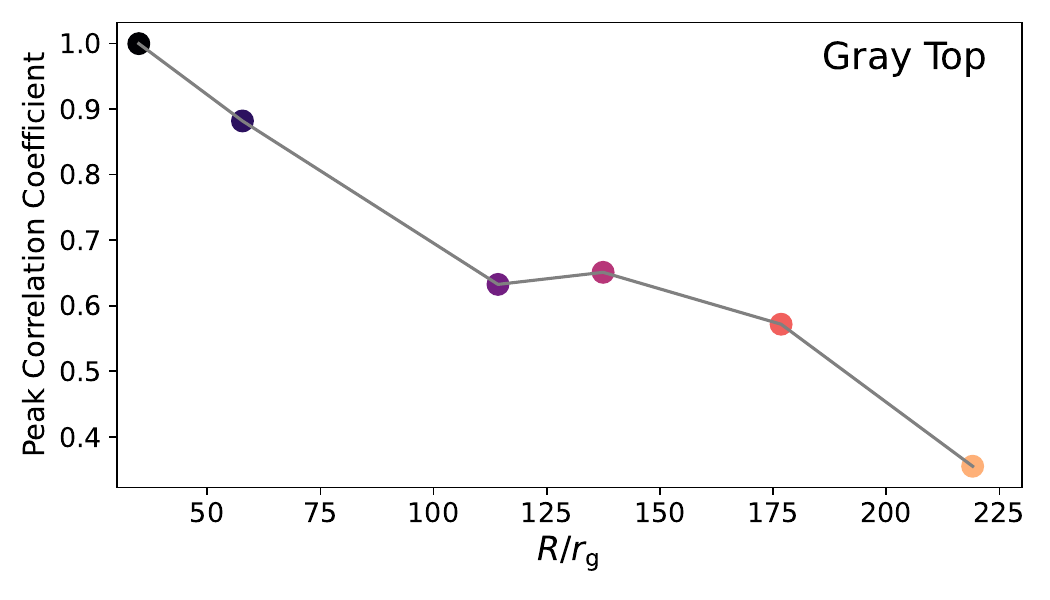} \\
    \includegraphics[width=0.49\linewidth]{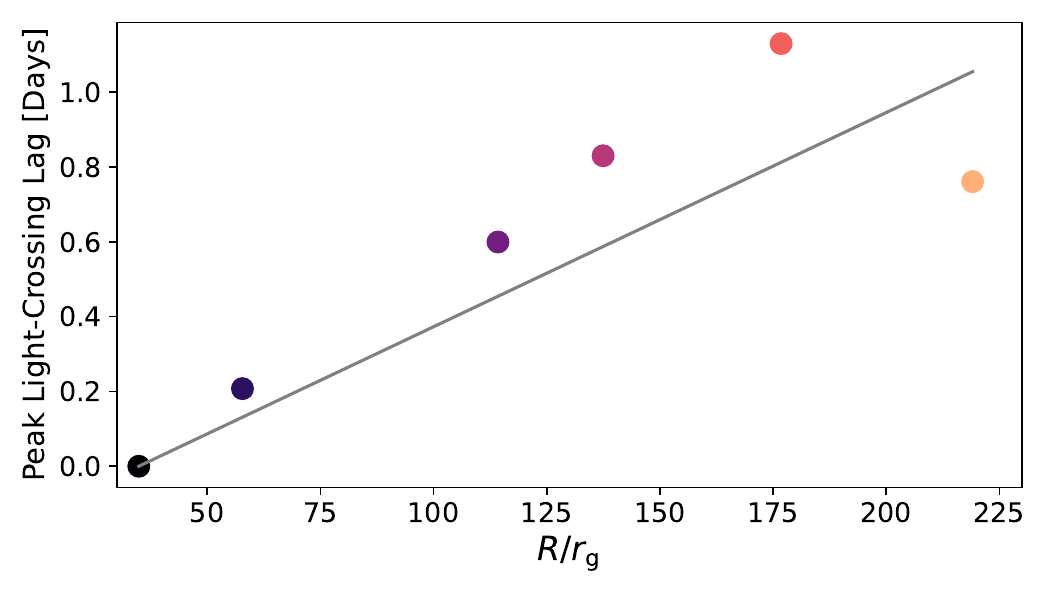}
    \includegraphics[width=0.49\linewidth]{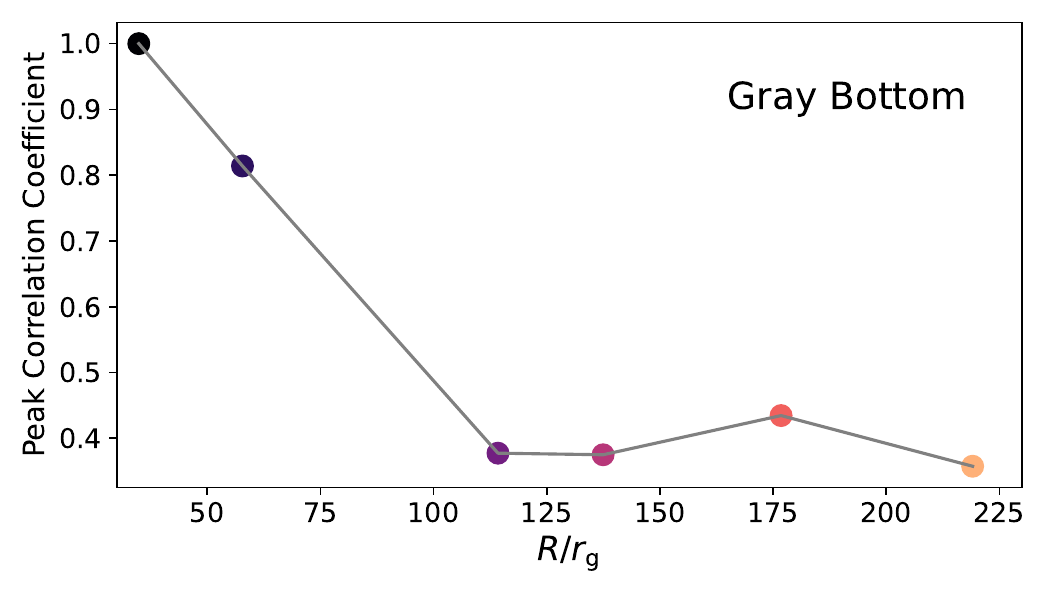} 
    \caption{The top and bottom panels are for light curves emitted from the top and bottom of the disk, respectively, for the Gray run without injected X-rays. \textbf{Left panel:} The light-crossing lag, measured as the peak of the cross-correlation function between $\pm5$~days, as a function of radius, relative to the light curve emitted at $R=36~r_g$. The solid line shows the light-crossing time from $R=36~r_g$. The dashed line shows zero lag. \textbf{Right panel:} The correlation coefficient of the cross-correlation lags in the left panel. Without X-ray irradiation driving variability the different radii emitted light curves are not well-correlated and the lags between them are inconsistent with the light travel time.}
    \label{fig:gray_shorts}
\end{figure*}

\begin{figure*}
    \centering
    \includegraphics[width=0.49\linewidth]{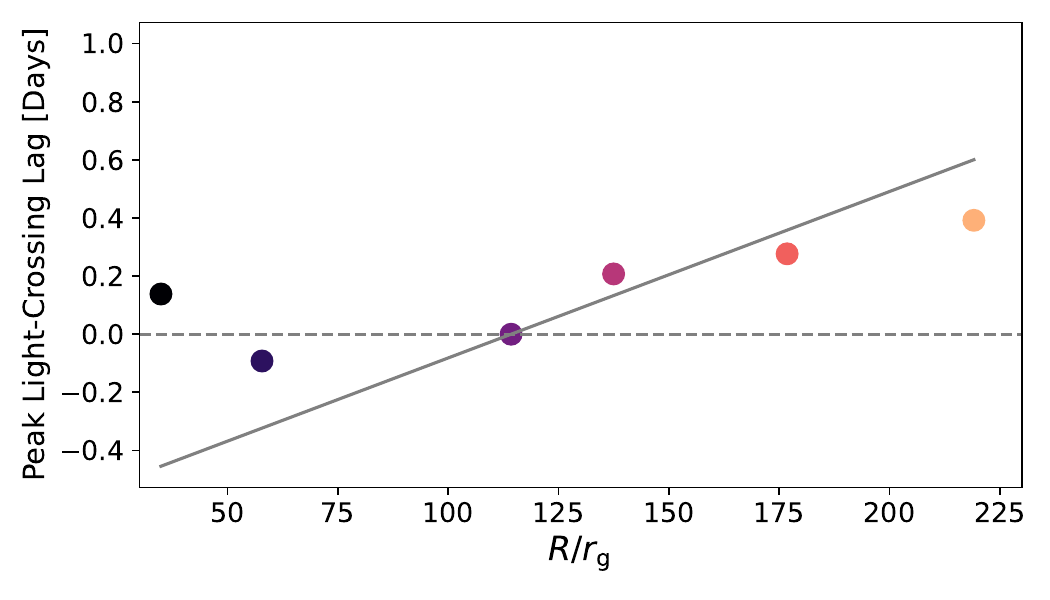}
    \includegraphics[width=0.49\linewidth]{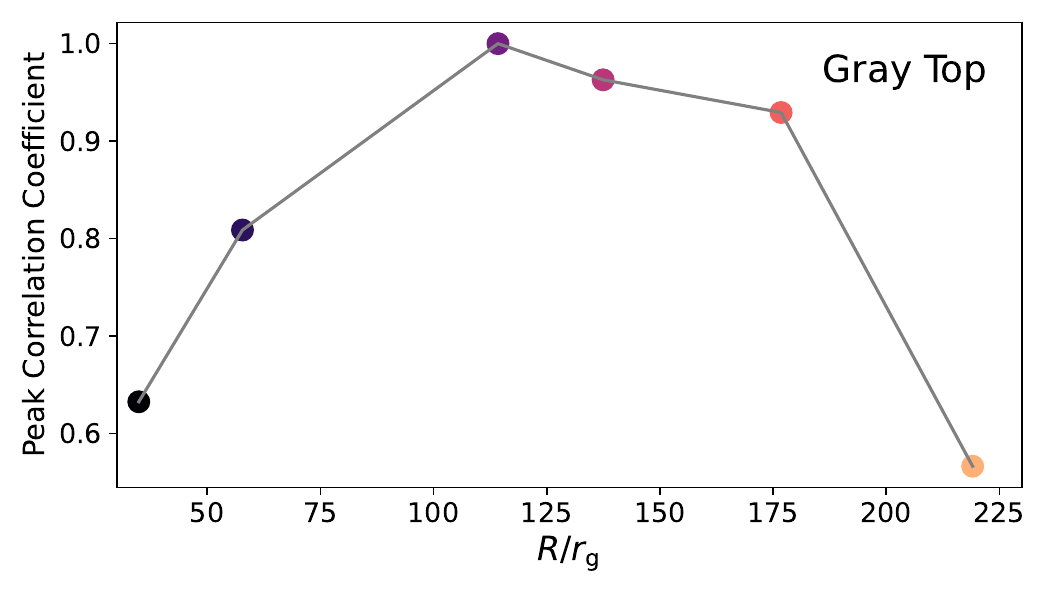} \\
    \includegraphics[width=0.49\linewidth]{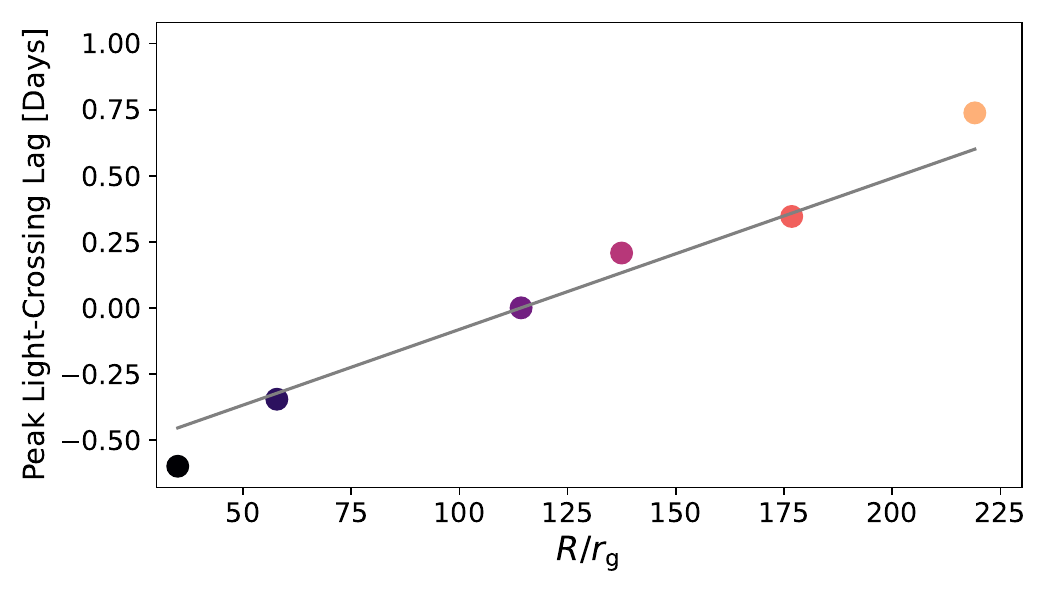}
    \includegraphics[width=0.49\linewidth]{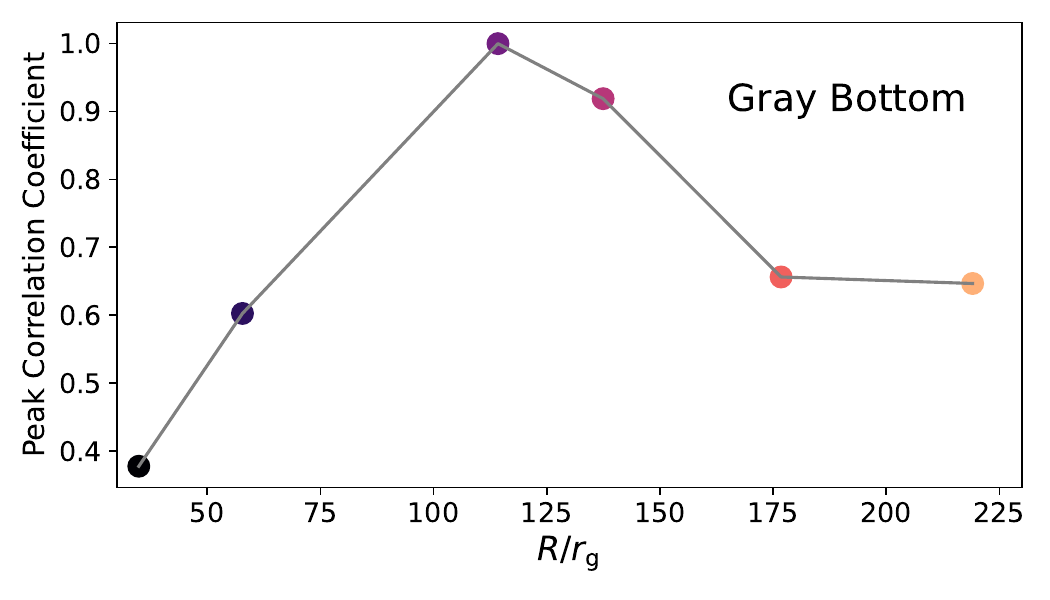} 
    \caption{Same as Figure \ref{fig:gray_shorts} but instead of using the light curve emitted at $R=36~r_g$ as a reference we show peak cross-correlation lags (left panel) and correlation coefficients (right panel) relative to the light curve emitted from $R=114~r_g$. The lag measurements are closer to the light-crossing time (especially at $R>100~r_g$) for this reference radius and the correlations between this reference light curve and other radii light curves are higher on average, although the correlations still decrease at radii farther away from the reference radius.}
    \label{fig:gray_shorts2}
\end{figure*}

We first describe the variability of light curves emitted from our Gray run without injected X-rays in order to understand the intrinsic variability produced by our simulated gas disk. We show the light curves emitted from the top and bottom of the disk in this Gray run in the left and right panels of Figure \ref{fig:gray_lcs}, respectively. The light curves in Figure \ref{fig:gray_lcs} clearly vary over a broad range of timescales, even without any X-ray irradiation. The velocity vectors in Panel C of Figure \ref{fig:rho} show that the gas disk can be very turbulent all the way up to the disk photosphere due to MRI and convection (see Panel C of Figure \ref{fig:rho} and Figure \ref{fig:entropy} in Appendix \ref{app:disk}), with turbulent velocities up to 1\% the speed of light. The light curves in Figure \ref{fig:gray_lcs} illustrate that the individual turbulence in different regions does not average or cancel out over a given radial range but instead produces intrinsic variability in the light curves emitted by the gas disk on timescales of hours to years.


We can further examine the properties of this intrinsic variability by looking at the frequency-binned PSDs of these light curves in the top and middle panel of Figure \ref{fig:gray_psd}. We provide the best-fit parameters for a bending power-law (see Equation \ref{eq:bend}) fit to these PSDs in the legends along with their p-null values. A single bending power-law is generally a good model for our simulated light curve PSDs.

UV-optical AGN light curves are often modeled as a damped random walk \citep[DRW, e.g.,][]{Kelly:2009}, which is a bending power-law with low-frequency slope $\alpha_l=0$ (white noise) and high-frequency slope $\alpha_h=2$ (red noise). For our simulated light curves, the best-fit power-laws at low-frequencies have $0.3<\alpha_l<1$, which is significantly steeper than the DRW. The high frequency slopes tend to be $50-100\%$ steeper than the DRW high frequency slope as well, because without X-ray irradiation there appears to be less variability at higher frequencies. 

The bending frequencies for the light curves in our Gray simulation mostly decrease with radius from $f_b= 0.7$~days$^{-1}$ to $f_b= 0.4$~days$^{-1}$ for the top of the disk and from $f_b= 0.4$~days$^{-1}$ to $f_b<0.01$~days$^{-1}$ for the bottom of the disk. This decrease in $f_b$ leads to distinct PSDs for the emission from each radius with greater variability in the light curves emitted from smaller radii, particularly at higher frequencies. These distinctions are most prominent for light curves emitted from the bottom of the disk. We expect these bending timescales to increase with radius because the various timescales of the disk, the dynamical, thermal, and inflow timescales, all increase with radius.

The thermal timescale, in particular, is often related to the damping/bending timescale of DRW/bending power-law models fit to AGN light curves \citep[e.g.,][]{Kelly:2009,Kozlowski2010,MacLeod:2010,Burke2021}. We compare the average thermal timescale with the bending timescales of light curves emitted from the top and bottom of the disk in the left and right panel of Figure \ref{fig:thermal}, respectively. We calculate the thermal timescale as $t_{\rm therm}=\tau(z)(250- z)/c$, where $c$ is the speed of light, and $\tau(z)$ is the optical depth integrated from $250~r_g$, where we measure the outgoing luminosity, to height $z$. In Figure \ref{fig:thermal} the solid lines show the thermal timescale measured from the midplane and the dashed lines show the thermal timescale measured from the height in the disk where the MRI turbulence peaks.

The bending timescales for both sides of the disk typically fall between the two thermal timescales. However, the bending timescales for light curves emitted from the top of the disk are more consistent with the thermal timescale measured from where the MRI peaks, suggesting the variability in these light curves is driven by the MRI. On the other hand, the bending timescales for the light curves emitted from the bottom of the disk are only slightly shorter than the thermal timescale in the midplane. These longer thermal timescales indicate that the convection that takes place in the bottom of the disk can propagate variability generated on longer timescales in the midplane up to the photosphere of the disk, allowing us to measure the midplane thermal timescales in AGN light curves. To compare the variability in light curves emitted from the top and bottom of the disk in more detail, we show the ratio of the mean PSD in each frequency bin for the top and bottom light curves in Figure \ref{fig:gray_psd}. We find that there is more variability in the light curves emitted from the top of the disk at $10^{-2}<f<1$~days$^{-1}$, consistent with the shorter thermal timescale near the disk surface where the MRI peaks.

It is traditionally assumed that any intrinsic variability in UV-optical AGN light curves comes from the propagation of fluctuations on the inflow timescale \citep{Lyubarskii1997,Hagen:2024,Hagen:2025}, which should be at least an order of magnitude longer than the thermal timescale ($\tau_{\rm inflow} \sim (H/R)^{-2}\tau_{\rm therm} $). However, we see that the bending timescales are consistent with the thermal timescale, implying that there is intrinsic variability in these light curves that is much shorter than the inflow timescale. Therefore, while propagating fluctuations play a role (see Section \ref{sec:results:long}), gas perturbations driven locally by the MRI and convection on shorter timescales are also important in driving variability.

In X-ray-driven RM models, the reprocessing of time-varying X-rays would lead to correlated variability in light curves emitted from different radii with lags between this correlated variability on the light-crossing timescale. In this Gray run, there is no X-ray irradiation to drive variability, but we still see significant variability in the light curves emitted at different radii. There are numerous examples of AGN where different UV wavebands are well-correlated despite the X-ray and UV being weakly correlated \citep[e.g.,][]{Buisson:2018,Morales:2019, Cackett:2023}. To test whether the variability in our Gray run is reprocessed from one disk radius to the next through disk self-irradiation in lieu of X-ray reprocessing, we show the peak lags and correlation coefficients in the left and right panel, respectively, of Figures \ref{fig:gray_shorts} and \ref{fig:gray_shorts2}. In Figure \ref{fig:gray_shorts} we use the light curve emitted from $R=36~r_g$ as the reference light curve and in Figure \ref{fig:gray_shorts2} we use the light curve emitted from $R=115~r_g$ as the reference light curve. The top (bottom) panels of Figures \ref{fig:gray_shorts} and \ref{fig:gray_shorts2} show the cross-correlation results for light curves emitted from the top (bottom) of the disk.

We find that without the X-ray irradiation driving variability at all radii, only emitted light curves at nearby radii will be highly correlated. When we use an inner radii light curve emitted from $R<60~r_g$ as a reference light curve, as in Figure \ref{fig:gray_shorts}, the outer radii light curves are not very well-correlated. The lags also tend to be longer than the light-crossing timescale (shown as the solid gray line in the left panels) for light curves emitted from the bottom of the disk, and roughly zero or negative for light curves emitted from the top of the disk. When we use the $R=115~r_g$ light curve as a reference in Figure \ref{fig:gray_shorts2}, the lag measurements become more consistent with the light-crossing timescale. However, the lags are less consistent with the light-crossing timescale and correlations are lower for the disk radii most distant from the reference radius. Therefore, it does seem necessary to have sufficient X-ray irradiation to drive well-correlated variability at different radii. Otherwise, we see that as you move away from the radius of the reference light curve, the other emitted light curves will become increasingly dominated by their own intrinsic variability rather than reprocessed variability from disk self-irradiation.



\subsection{Traditional X-ray-Driven RM Models and The Top of the Disk}
\label{sec:top}

\begin{figure*}
    \centering
    \includegraphics[width=0.49\linewidth]{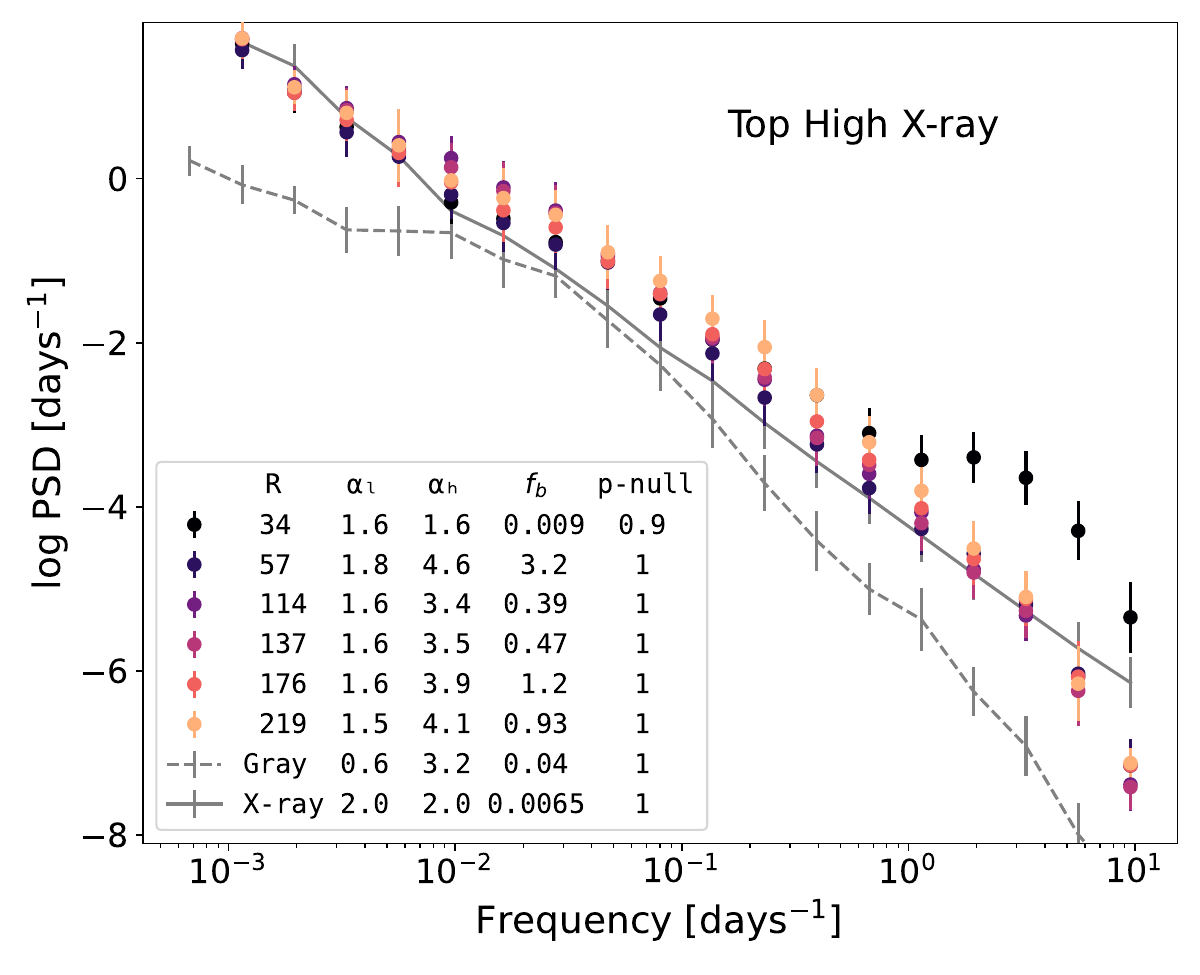}
    \includegraphics[width=0.49\linewidth]{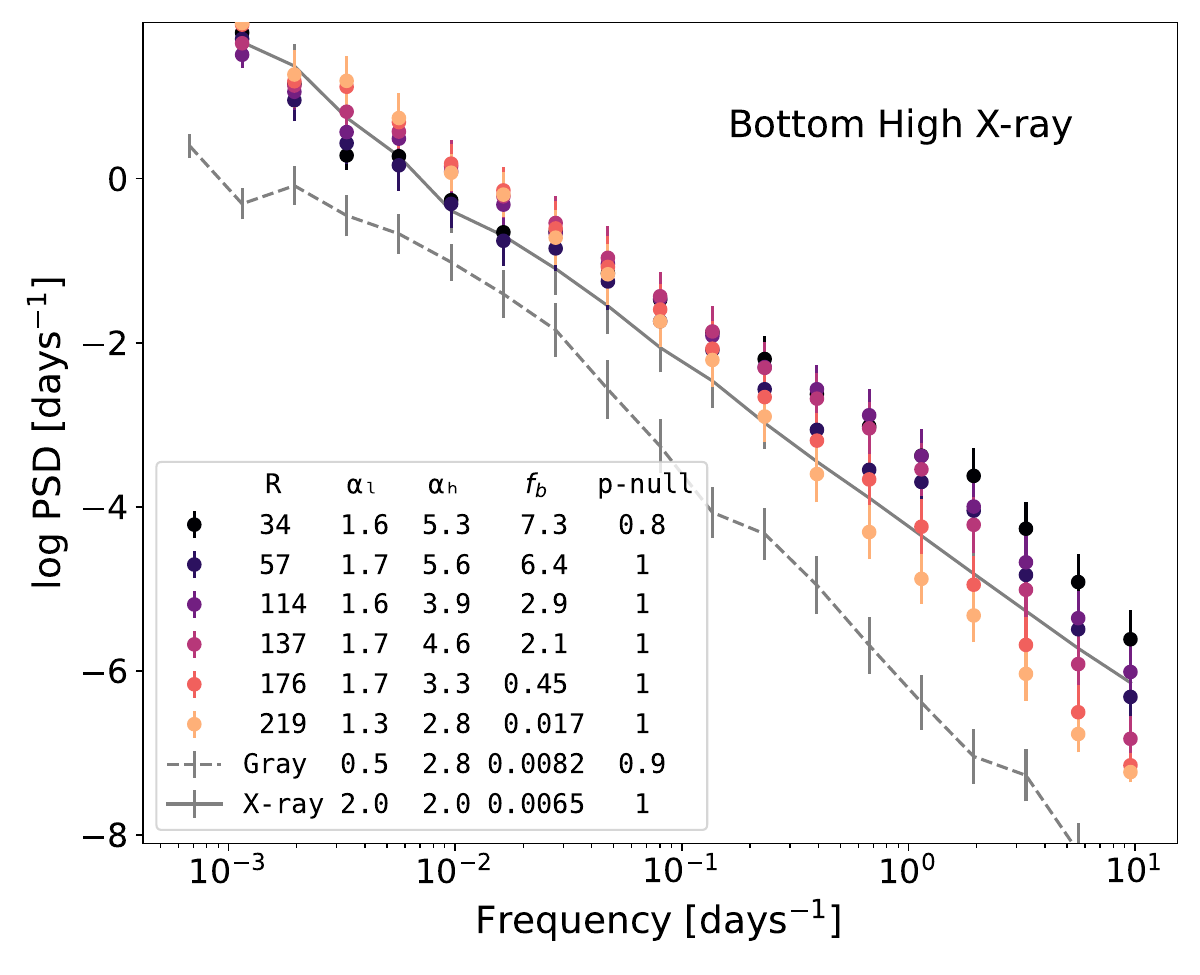} \\ 
    \includegraphics[width=0.49\linewidth]{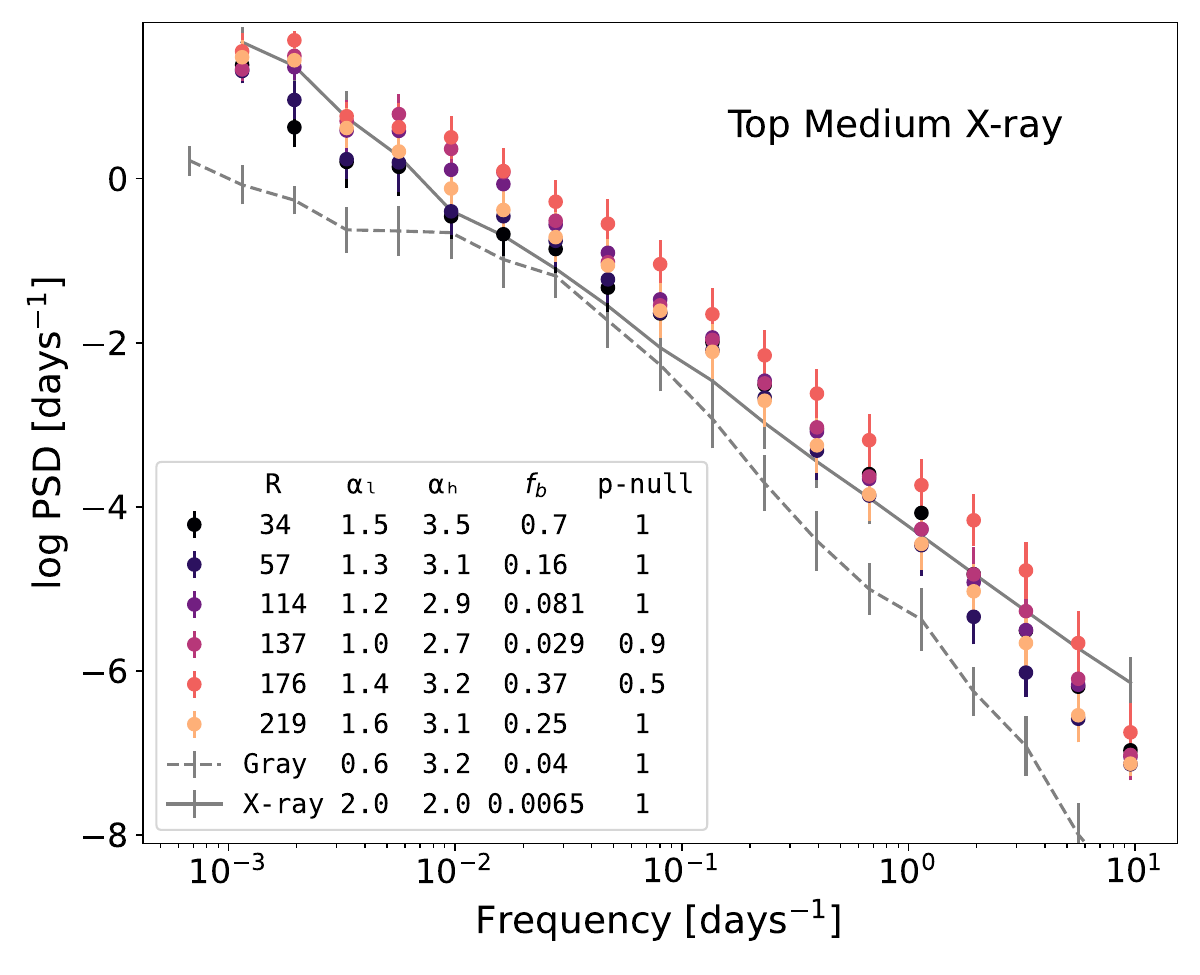}
    \includegraphics[width=0.49\linewidth]{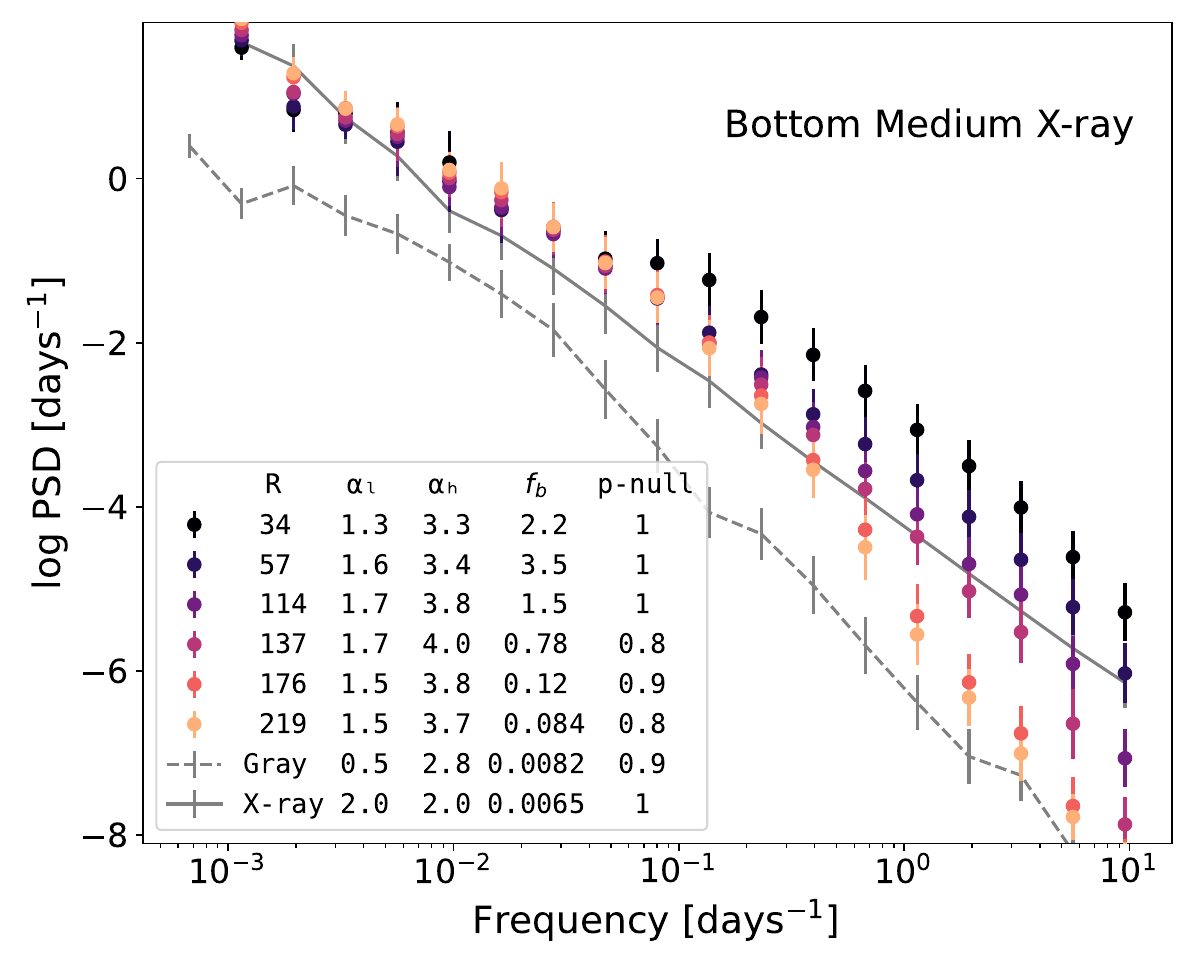} \\ 
    \includegraphics[width=0.49\linewidth]{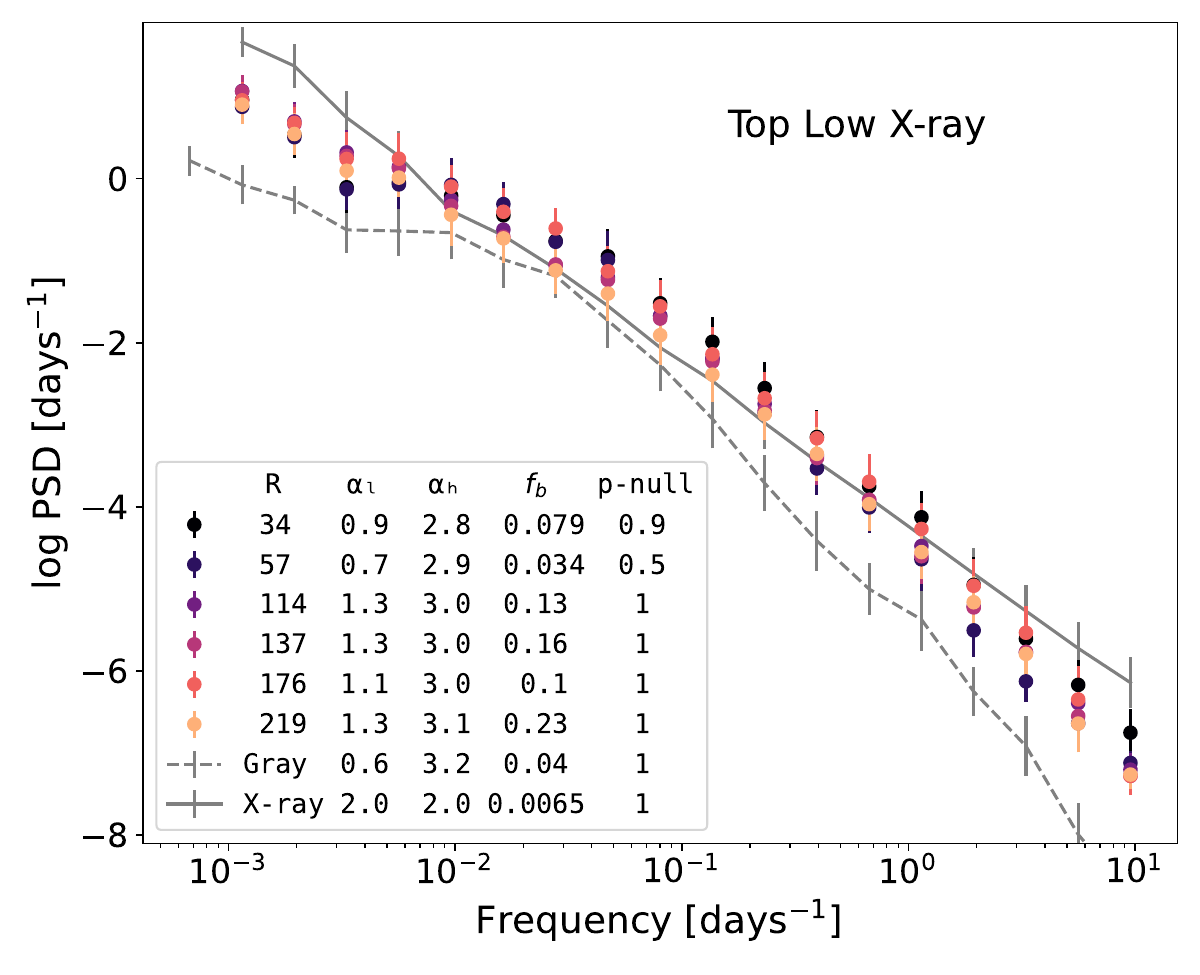}
    \includegraphics[width=0.49\linewidth]{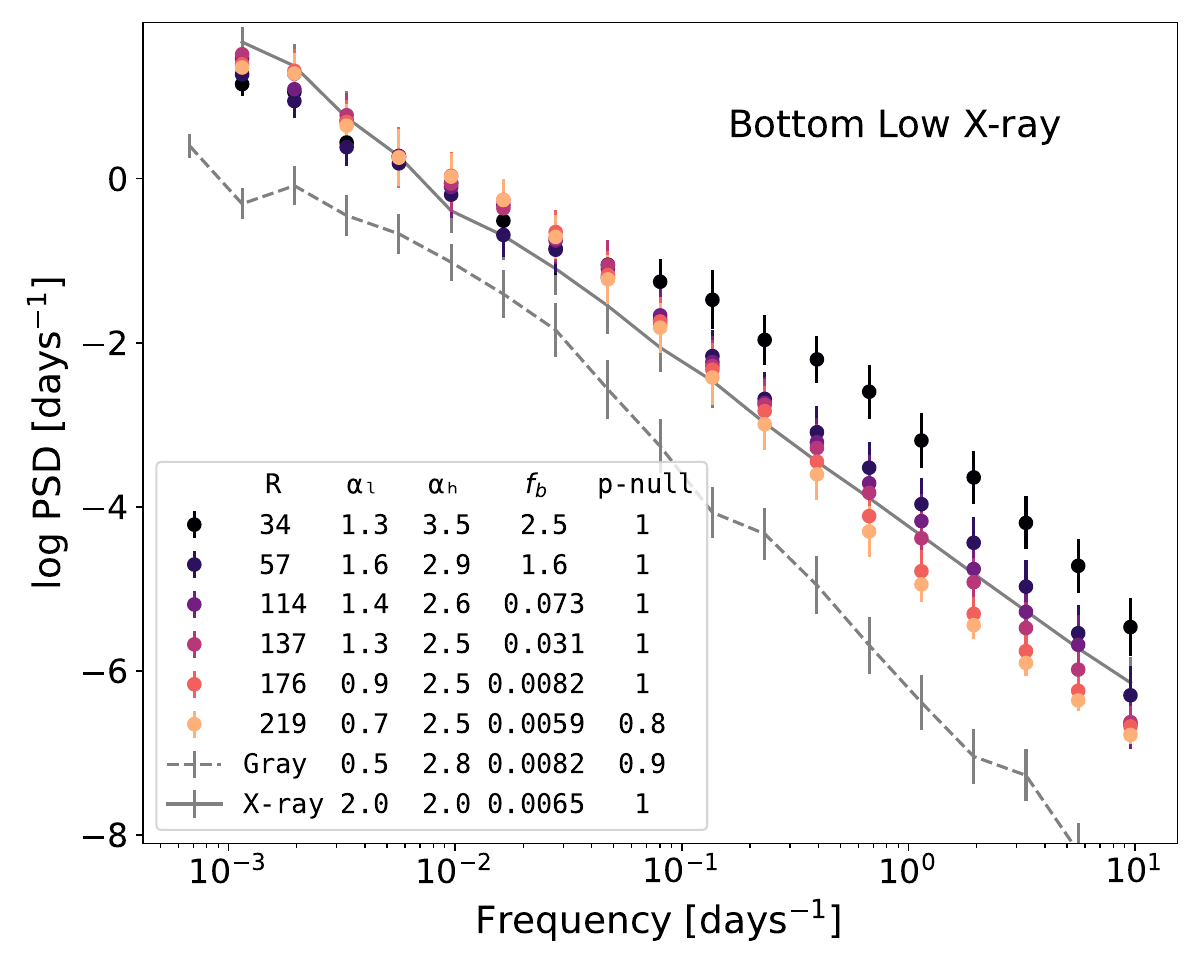} \\ 
    \caption{Frequency-binned PSDs of light curves emitted from the top and bottom of the disk in the left and right panel, respectively, for six different radii, for the High X-ray, Medium X-ray, and Low X-ray runs (from top to bottom). The solid gray line shows the injected X-ray light curve, while the dashed gray line shows the PSD for the light curve from the Gray run without X-ray irradiation summed over all radii. The points show the mean of each bin and the error-bars show the standard deviation of that bin. The PSDs of light curves emitted from the top of the disk are very similar to the X-ray PSD, but the PSDs of light curves emitted from the bottom of the disk are different from each other and the X-ray PSD. The label gives the parameters and p-null value for the best-fit bending power-law at each radius, $R$.}
    \label{fig:psd}
\end{figure*}

\begin{figure*}
    \centering
    \includegraphics[width=0.49\linewidth]{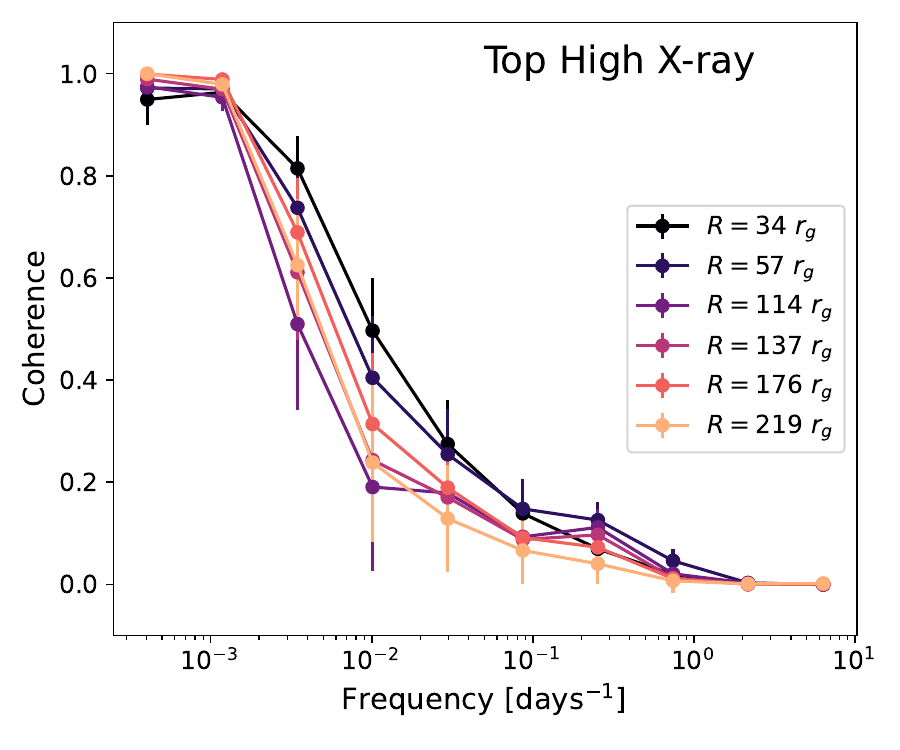}
    \includegraphics[width=0.49\linewidth]{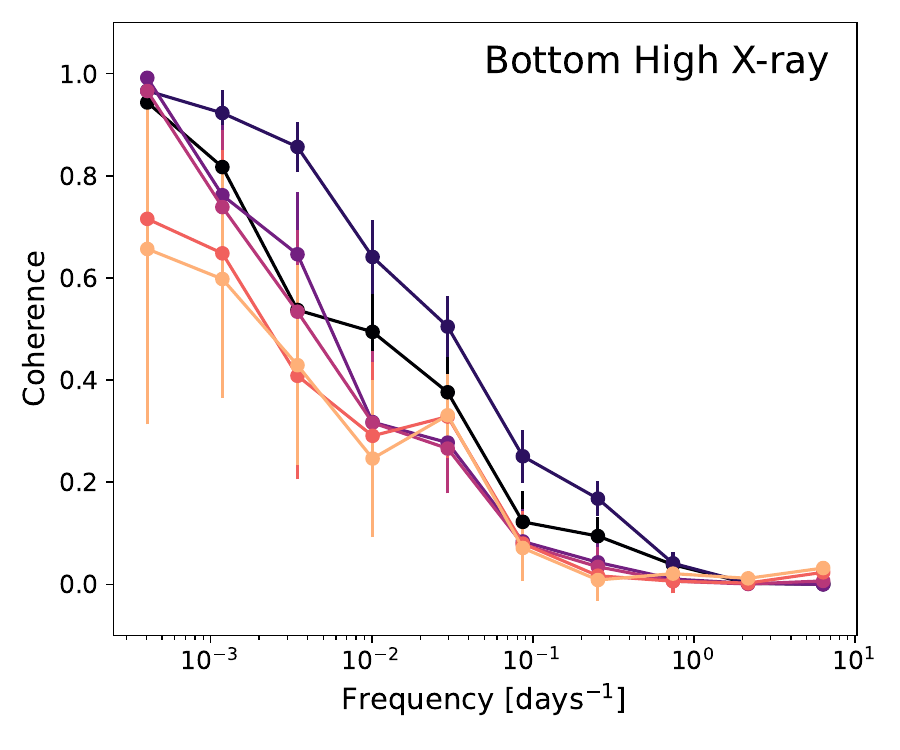} \\
    \includegraphics[width=0.49\linewidth]{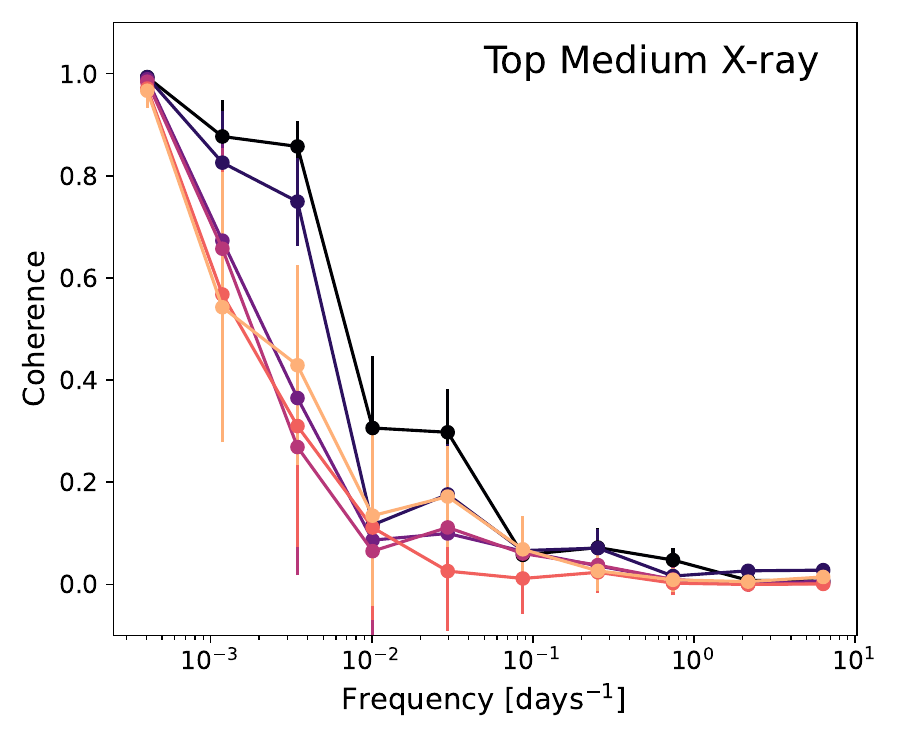}
    \includegraphics[width=0.49\linewidth]{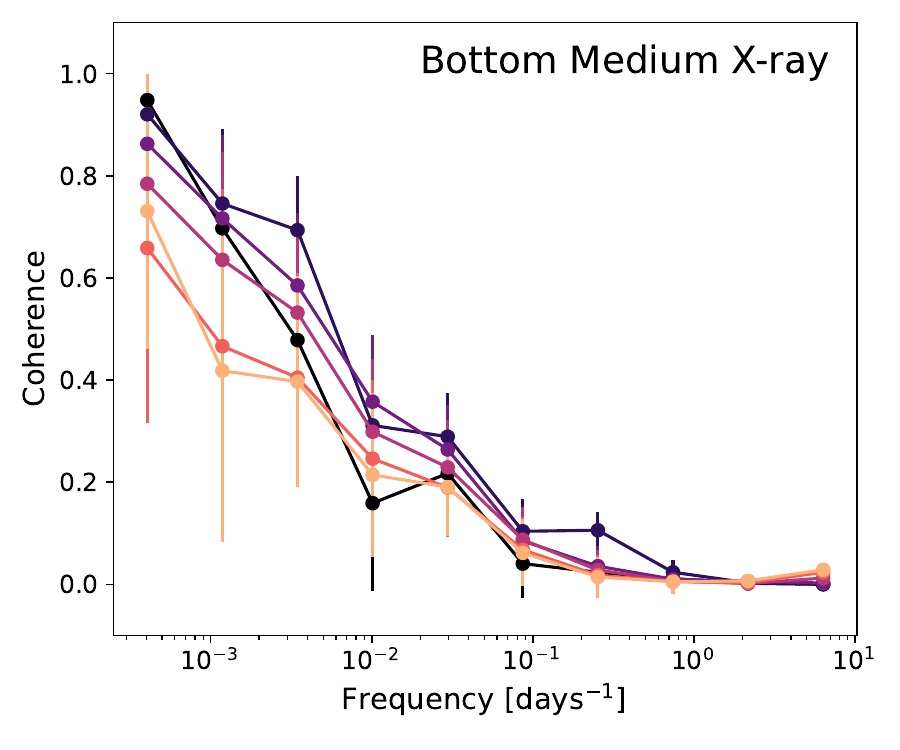} \\
    \includegraphics[width=0.49\linewidth]{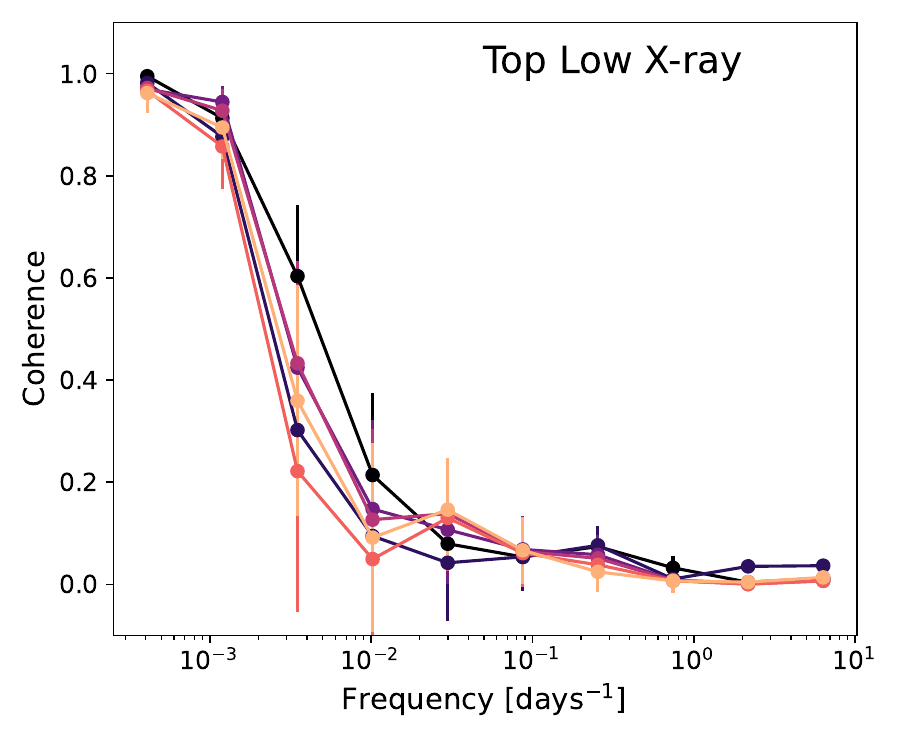}
    \includegraphics[width=0.49\linewidth]{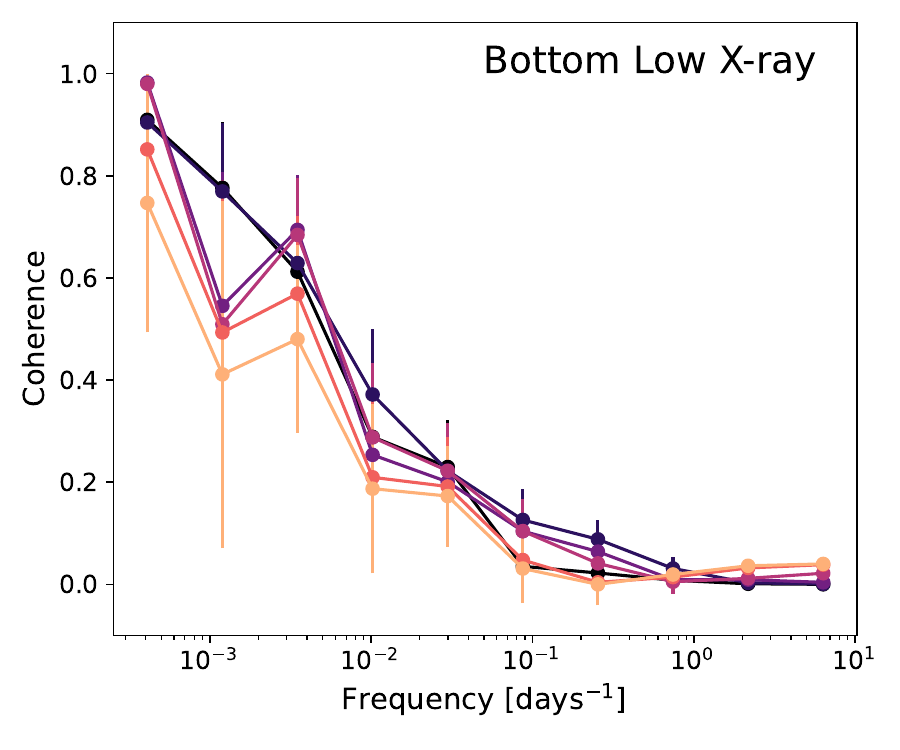}
    \caption{The coherence between the injected X-ray light curve and the light curves emitted from six different radii for the High, Medium and Low X-ray runs (from top to bottom). The error bars give the statistical uncertainty in each bin (see Equation \ref{eq:sig_gamma}). For light curves emitted from the top of the disk (left panel) the coherence is $\Gamma^2 \approx 1$ on the longest timescales, but the maximum coherence is lower for light curves emitted from the bottom of the disk (right panel).}
    \label{fig:coherence}
\end{figure*}

\begin{figure*}
    \centering
    \includegraphics[width=0.49\linewidth]{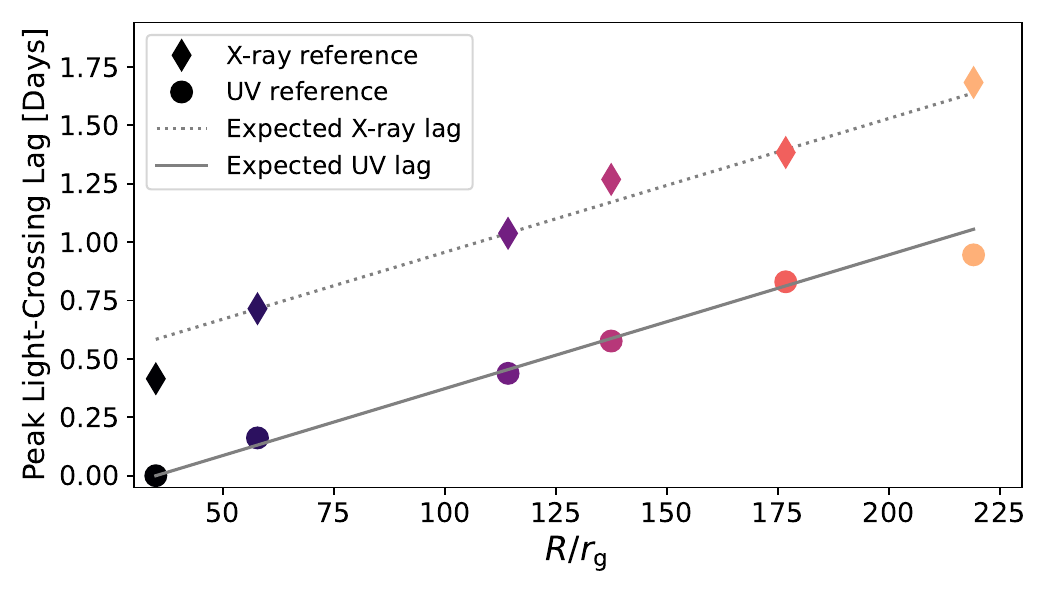}
    \includegraphics[width=0.49\linewidth]{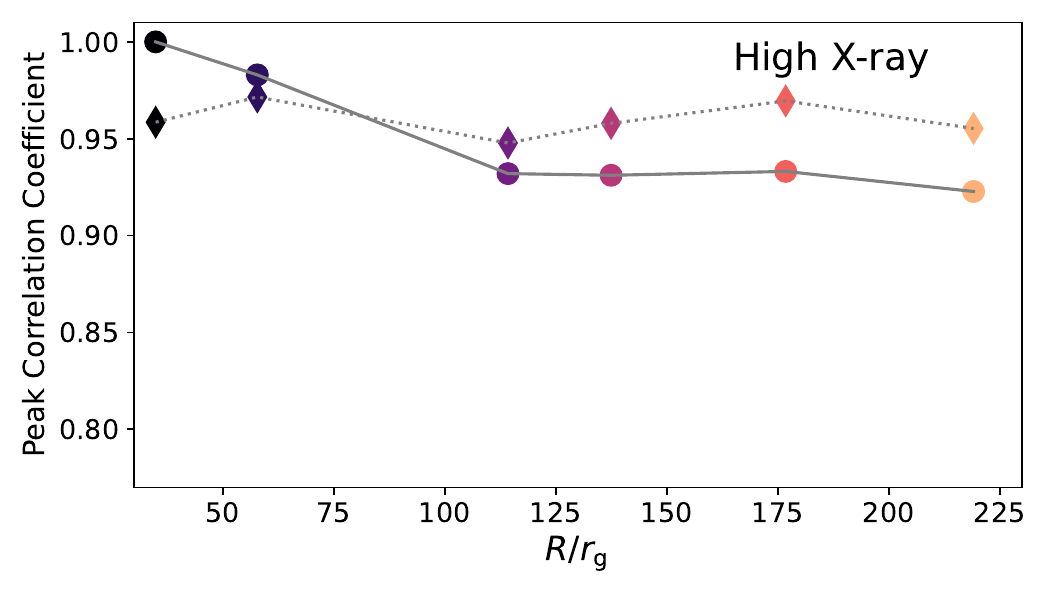} \\
    \includegraphics[width=0.49\linewidth]{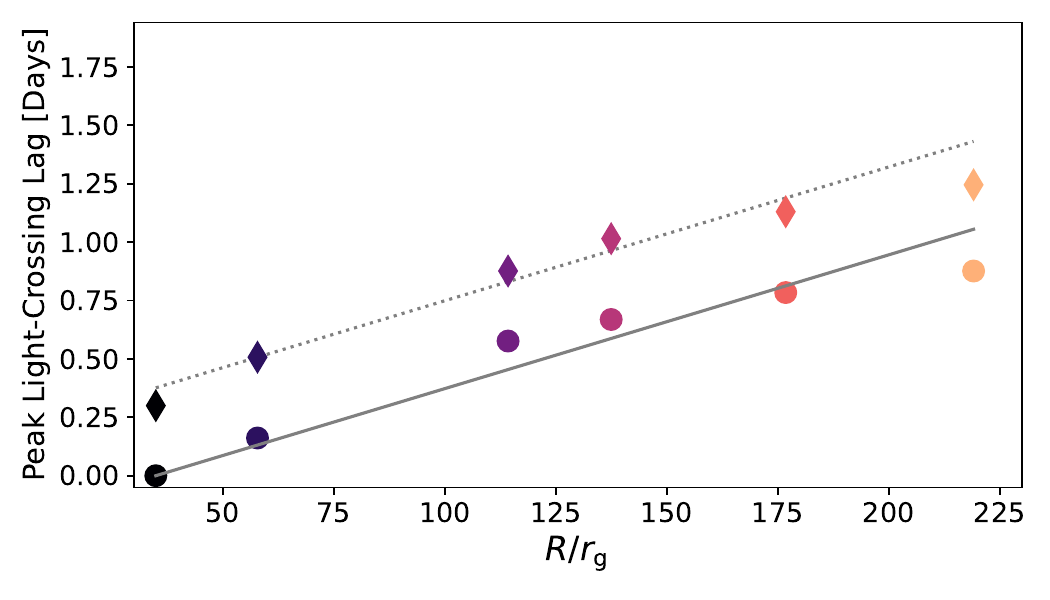}
    \includegraphics[width=0.49\linewidth]{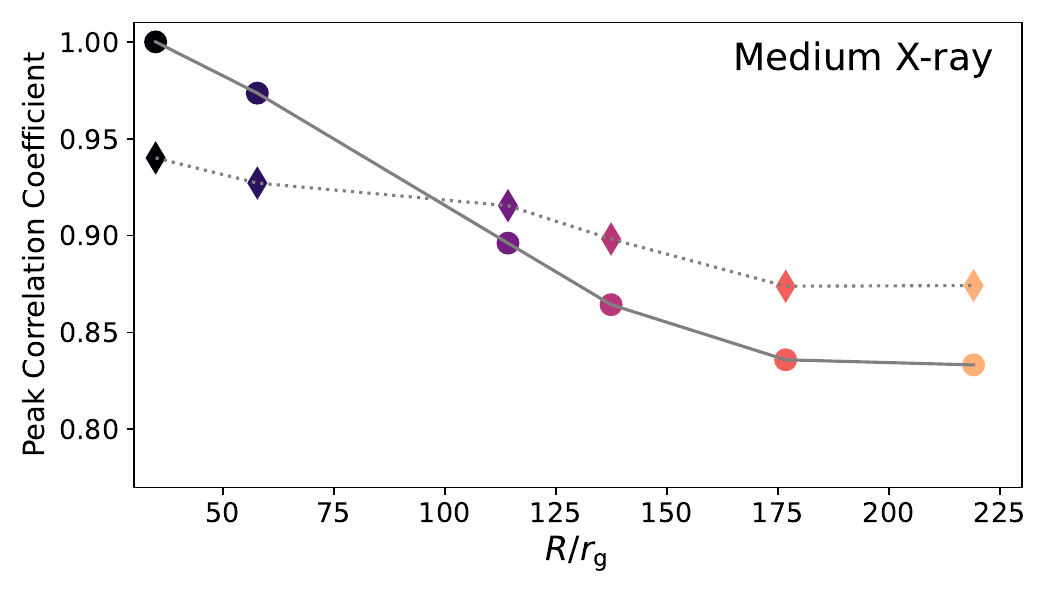} \\
    \includegraphics[width=0.49\linewidth]{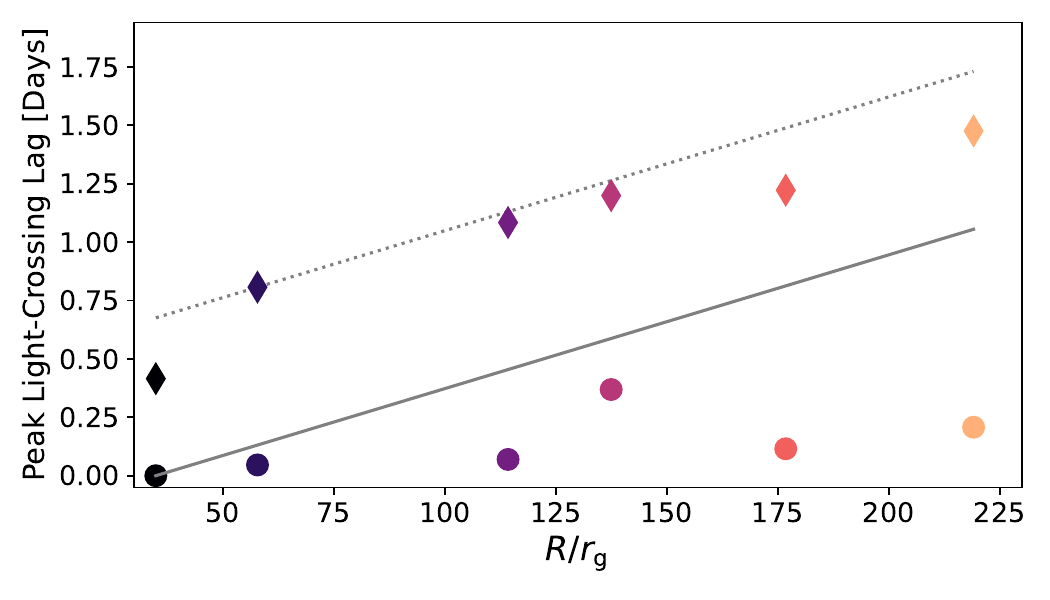}
    \includegraphics[width=0.49\linewidth]{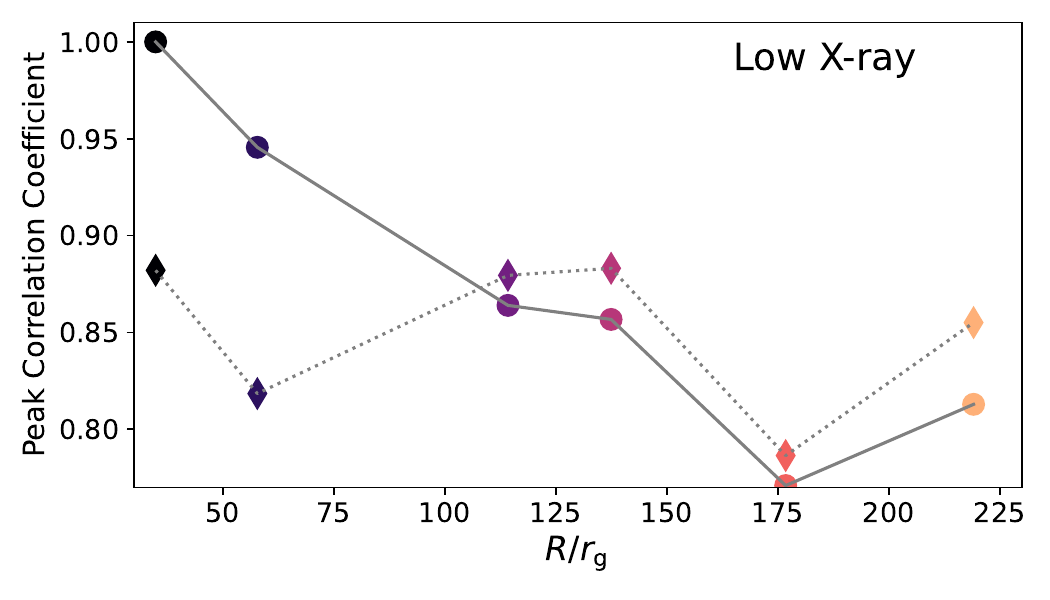}
    \caption{\textbf{Left panel:} The light-crossing lag, measured as the peak of the cross-correlation function between $\pm5$~days, as a function of radius, relative to the X-ray light curve (diamonds) and the light curve emitted at $R=36~r_g$ (circles) for light curves emitted from the top of the disk for the three runs with injected X-rays. The solid line shows the light-crossing time from $R=36~r_g$. The dotted line shows the light-crossing time for the X-ray irradiation. \textbf{Right panel:} The correlation coefficient of the cross-correlation lags in the left panel. The High and Medium X-ray runs have highly correlated lags on the light-crossing timescale. Lags are slightly less well-correlated for the Low X-ray run and measured lags are shorter than the light-crossing timescale for some radii.}
    \label{fig:topshorts}
\end{figure*}

We now present the light curves from our three runs with X-ray irradiation. In the previous section, we show that X-ray irradiation is not needed to drive variability in light curves emitted from our simulated gas disk, but it is necessary for driving correlated variability between light curves emitted from different disk radii. Traditional RM models predict that emission from different disk radii should be highly correlated because time-varying X-ray irradiation is the main driver of variability in light curves emitted by the disk. We find that for our runs with X-ray irradiation the light curves emitted from the top of our simulated disks largely agree with these traditional RM model predictions. The light curves emitted from the tops of our simulated disks are highly coherent and highly correlated with the injected X-ray light curve, and we measure lags between the injected X-ray light curve and the emitted UV-optical light curves that are consistent with the light-crossing timescale.

We show the PSDs of the light curves emitted from the tops of our simulated disks in the left panel of Figure \ref{fig:psd}. For comparison, we show the PSD of the injected X-ray light curve as the solid gray line ($P_i\propto1/f^{\alpha_{\rm in}}$) and the PSD of the light curve from our Gray run without X-ray irradiation summed over all radii out to $R=250~r_g$ as the dashed gray line. The legend provides a summary of the best-fitting parameters and p-null values for a bending power-law fit to each PSD.


X-ray irradiation has a huge impact on the light curves emitted from the top of the disk, especially in runs with higher X-ray luminosities. Unlike in the Gray run, the PSDs in our runs with X-ray irradiation are no longer distinct from each other and the reprocessed PSDs have several orders of magnitude more variability than the Gray run PSDs at most frequencies. For the High X-ray run with the highest X-ray luminosity, $\alpha\approx\alpha_{\rm in}=2$ up to high frequencies and the PSDs mostly resemble the X-ray PSD at $f\lesssim1$~days$^{-1}$. In fact, in the left panel of Figure \ref{fig:thermal} the bending timescales for the High X-ray run are significantly shorter than the MRI thermal timescale at most radii, suggesting that the X-ray variability drives most of the variability in these light curves near the surface of the disk. We can also see the close resemblance of the X-ray and emitted UV-optical light curves in the left panel of Figure \ref{fig:mid_lcs}, which compares the light curves emitted from the top of the disk to the injected X-ray light curve for the Medium X-ray run. More quantitatively, we show in the left panel of Figure \ref{fig:coherence} that these light curves and the injected X-ray light curves have very high coherence at the longest timescales.

However, a bending power-law with $\alpha_l<\alpha_{\rm in}$ and $\alpha_h>\alpha_{\rm in}$ is a better fit to the reprocessed light curves than the single power-law we use to model the X-ray PSD. Because we use a single power-law to model our X-ray PSD it is clear that this break is purely due to the physics of the gas disk. As in the Gray run, MRI driven intrinsic variability produces some additional variability at frequencies $10^{-2}<f<1$~days$^{-1}$, leading to $\alpha_l<\alpha_{\rm in}$. In addition, the gas disk smooths the X-ray variability at the highest frequencies leading to $\alpha_h>\alpha_{\rm in}$. We see in the left panel of Figure \ref{fig:thermal} that the bending timescales for runs with X-ray irradiation are typically shorter than in the Gray run, but still similar to the MRI thermal timescale, except for the High X-ray run. Therefore, MRI turbulence generated within the disk still has an impact on these light curves.

In general, runs with lower X-ray luminosity have shallower low frequency slopes and lower bending frequencies leading to less variability at the lowest and highest frequencies. For this reason, the coherence in these runs decreases more steeply with frequency than the coherence for the High X-ray run for most radii. This result is consistent with the results in \cite{Secunda:2025}, who found that decreasing the X-ray luminosity decreases the correlation between the X-ray and UV-optical light curves.

Next, we can test traditional X-ray-driven RM models by measuring lags between the variability in light curves emitted from different radii using cross-correlation. The circles in Figure \ref{fig:topshorts} show the peak lags and correlation coefficients between different radii emitted light curves and our reference light curve at $R=36~r_g$. We can also now measure the lags between our injected X-ray light curve and the emitted light curves from different radii, which we show as diamonds in Figure \ref{fig:topshorts}. The dotted and solid lines show the light travel time for the X-ray\footnote{We estimate the X-ray light travel time starting from the lag measured at $R=57~r_g$.} and $R=36~r_g$ irradiation, respectively.

The measured X-ray lags are in fairly good agreement with the light travel time for all radii for all three simulations. The lags measured relative to the innermost radius light curve are reasonably consistent with the light travel time for the High and Medium X-ray runs, but inconsistent at most radii for the Low X-ray run, similar to how they are inconsistent in the run without X-ray irradiation. Inconsistent lags are more frequently underestimated than overestimated in all runs. Despite some inconsistencies with the light-crossing timescale, the light curves are well-correlated with each other and the X-ray light curve, with peak correlation coefficients, $\rm{CC_{coeff}}>0.75$ at all radii. Therefore, light curves emitted from the top of our simulated disks with X-ray irradiation are all roughly consistent with traditional RM models, especially in simulations with higher X-ray luminosities.

\subsection{Departures from Traditional RM Models: The Bottom of the Disk, Shorter Time Intervals, and Flares}
\label{sec:departures}

\begin{figure*}
    \centering
    \includegraphics[width=0.49\linewidth]{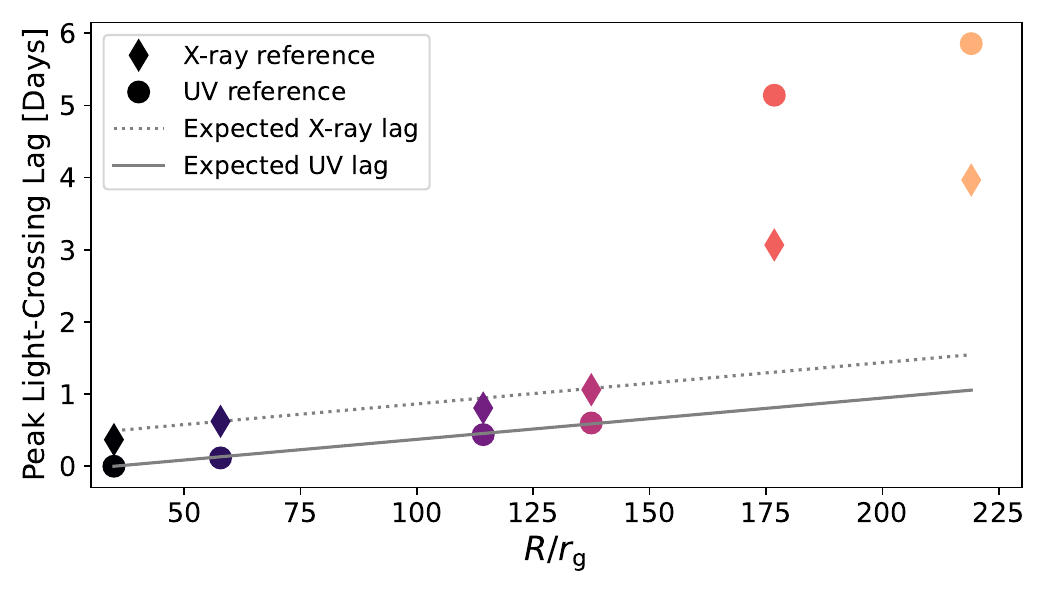}
    \includegraphics[width=0.49\linewidth]{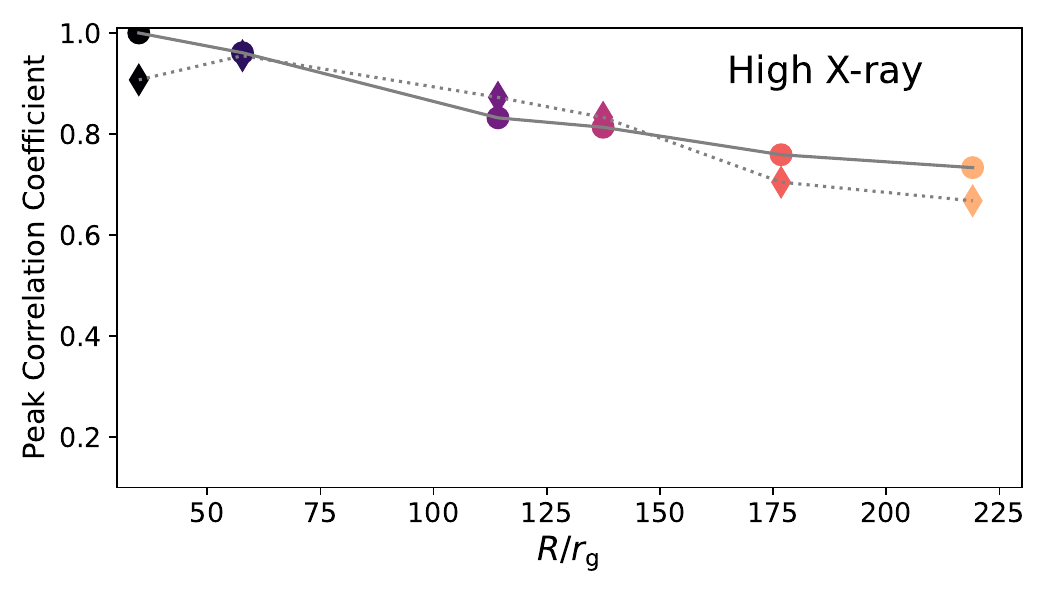} \\
    \includegraphics[width=0.49\linewidth]{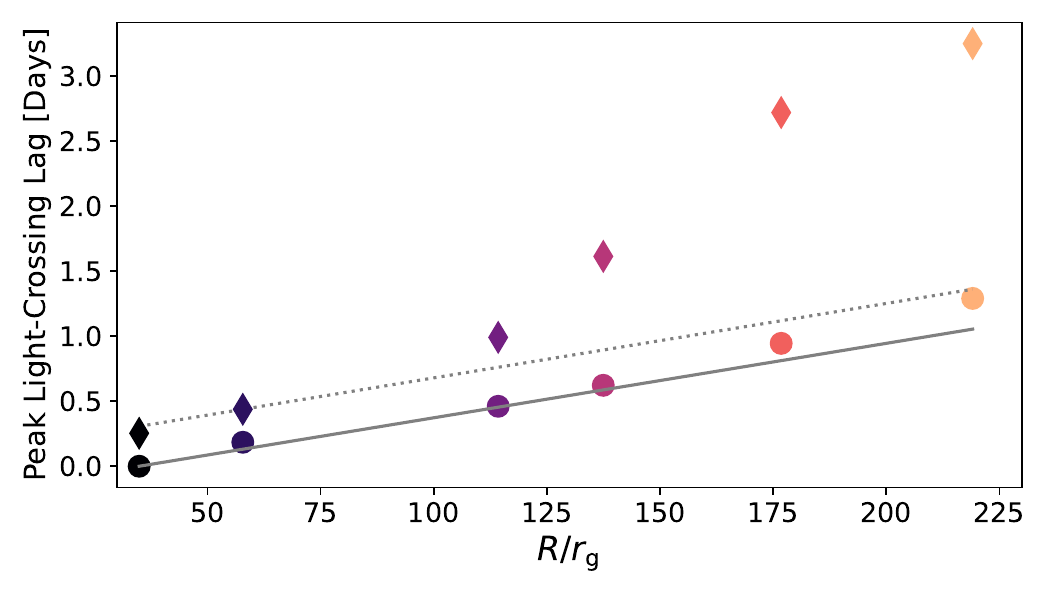}
    \includegraphics[width=0.49\linewidth]{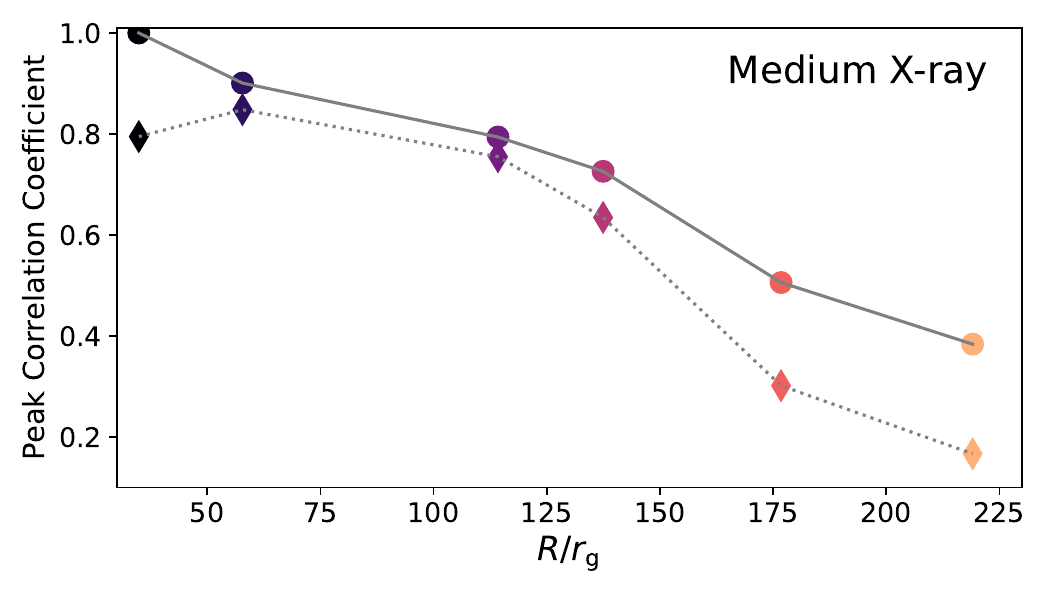} \\
    \includegraphics[width=0.49\linewidth]{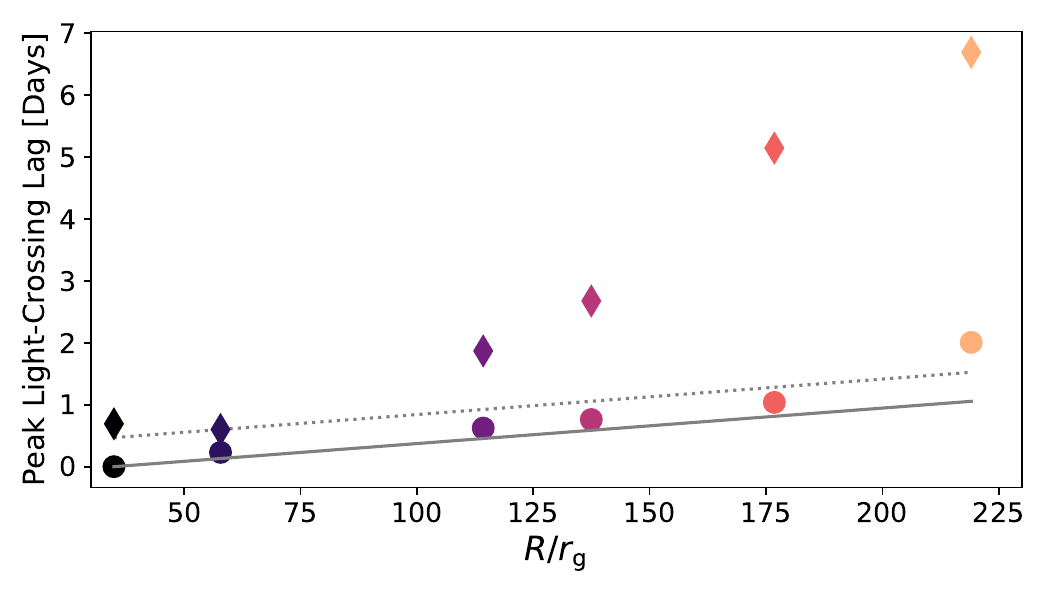}
    \includegraphics[width=0.49\linewidth]{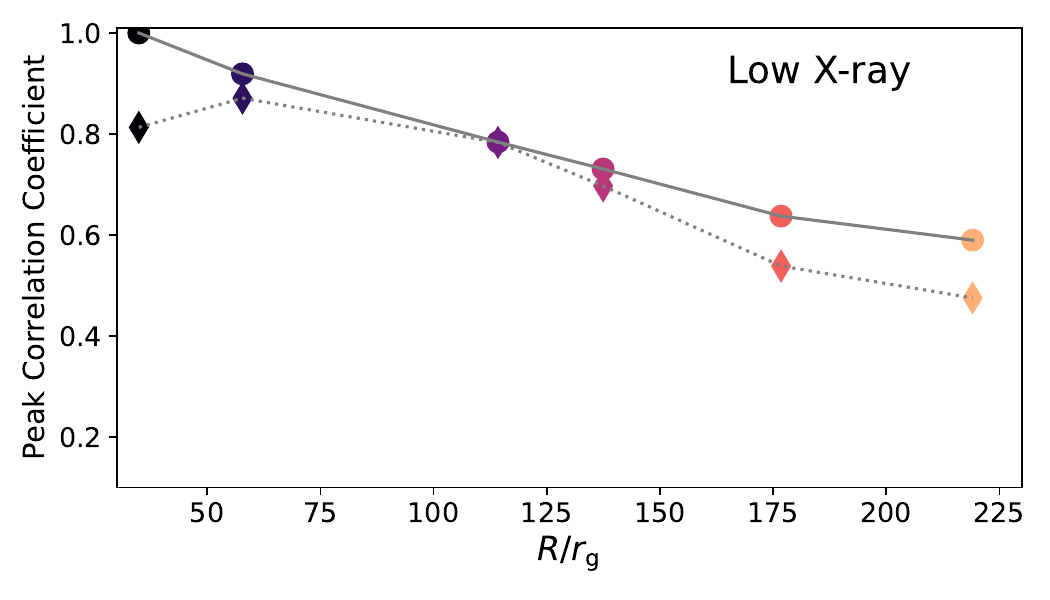}
    \caption{\textbf{Left panel:} The light-crossing lag, measured as the peak of the cross-correlation function between $\pm5$~days as a function of radius, relative to the X-ray light curve (diamonds) and the light curve emitted at $R=36~r_g$ (circles) for light curves emitted from the bottom of the disk for our three runs with injected X-rays. The solid line shows the light-crossing time from $R=36~r_g$. The dotted line shows the light-crossing time for the X-ray irradiation. \textbf{Right panel:} The correlation coefficient of the cross-correlation lags in the left panel. The correlations between light curves are weaker for the bottom of the disk and tend to decrease with radii. The lag durations are also longer than the light travel time at larger radii, especially for the X-ray reference lags.}
    \label{fig:botshorts}
\end{figure*}

\begin{figure*}
    \centering
    \includegraphics[width=0.49\linewidth]{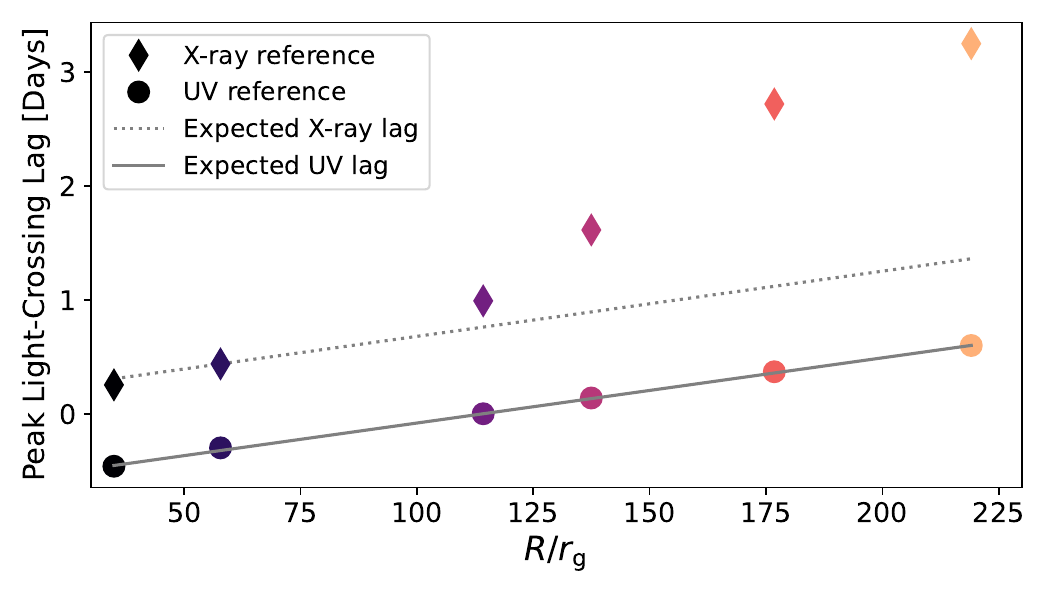}
    \includegraphics[width=0.49\linewidth]{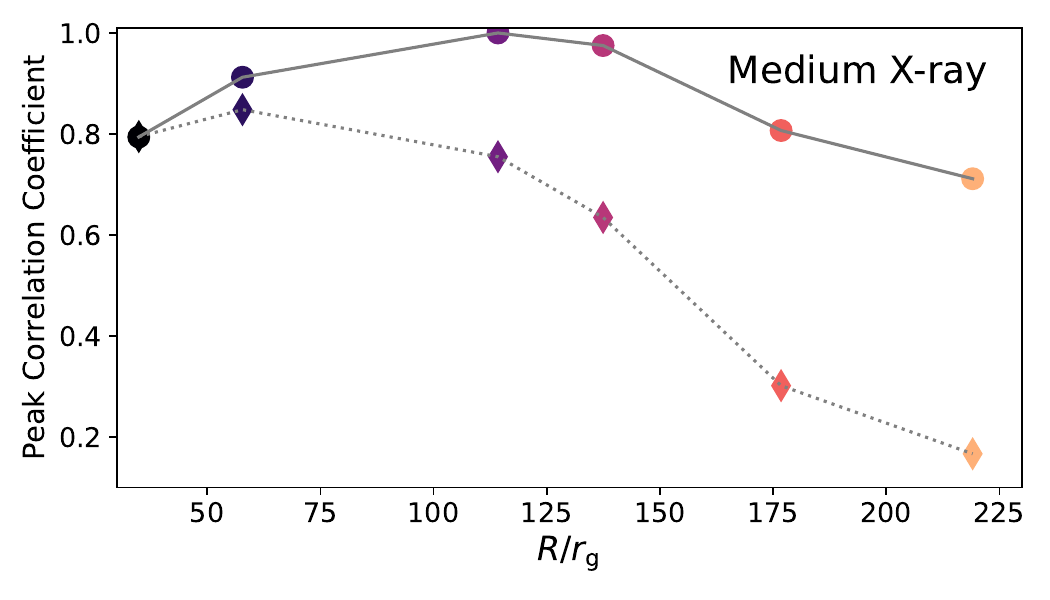} 
    \caption{Same as Figure \ref{fig:botshorts} but using the light curve emitted from $R=114~r_g$ as the reference light curve for the points shown as circles. As in the Gray run, the lag between the emitted light curves are more consistent with the light travel time (solid line) using the $R=114~r_g$ light curve as a reference than when we use the innermost radius light curve as a reference. The correlations between the light curves are also higher.}
    \label{fig:botshorts2}
\end{figure*}

\begin{figure*}
    \centering
    \includegraphics[width=0.49\linewidth]{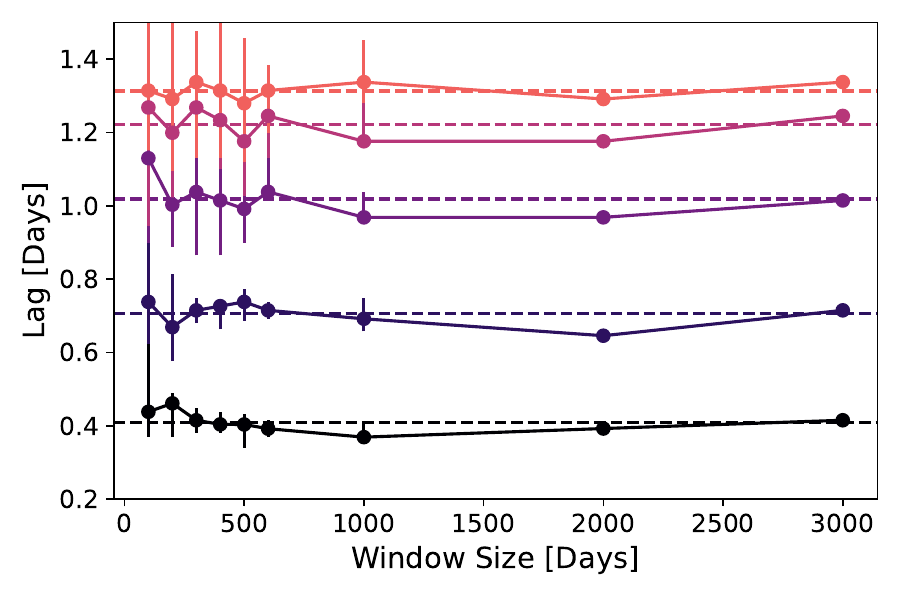}
    \includegraphics[width=0.49\linewidth]{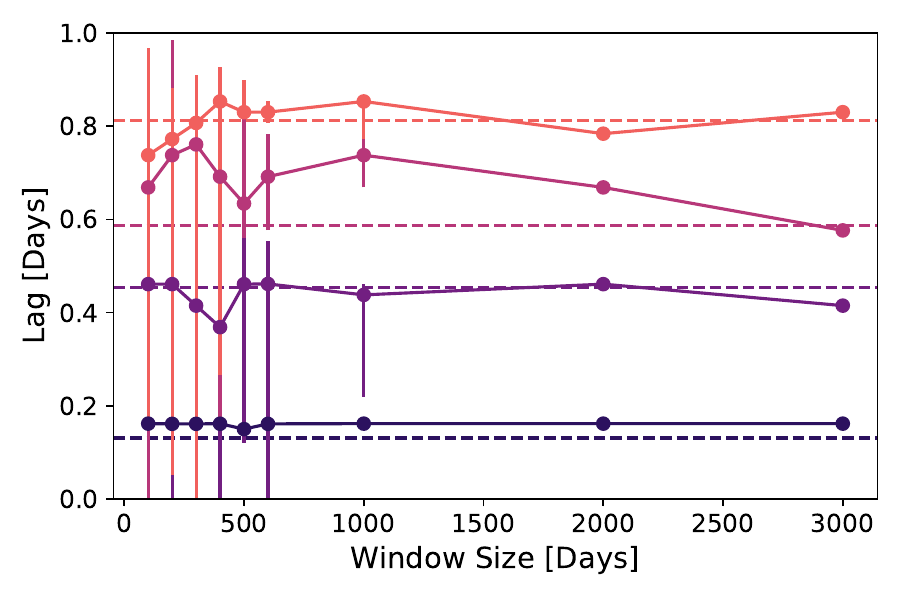} \\
    \includegraphics[width=0.49\linewidth]{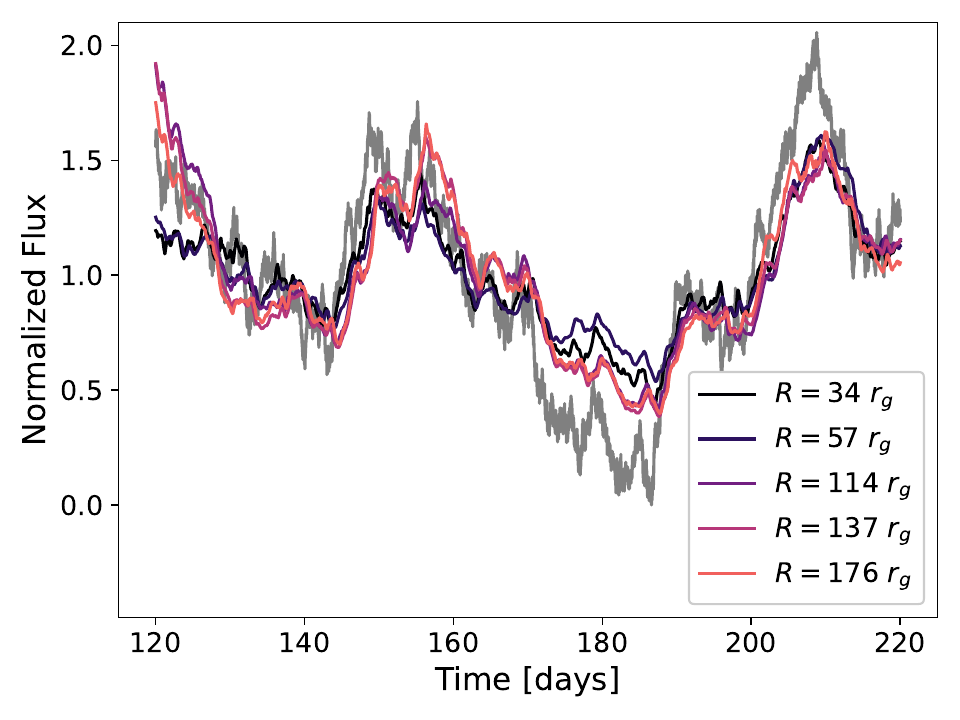}
    \includegraphics[width=0.49\linewidth]{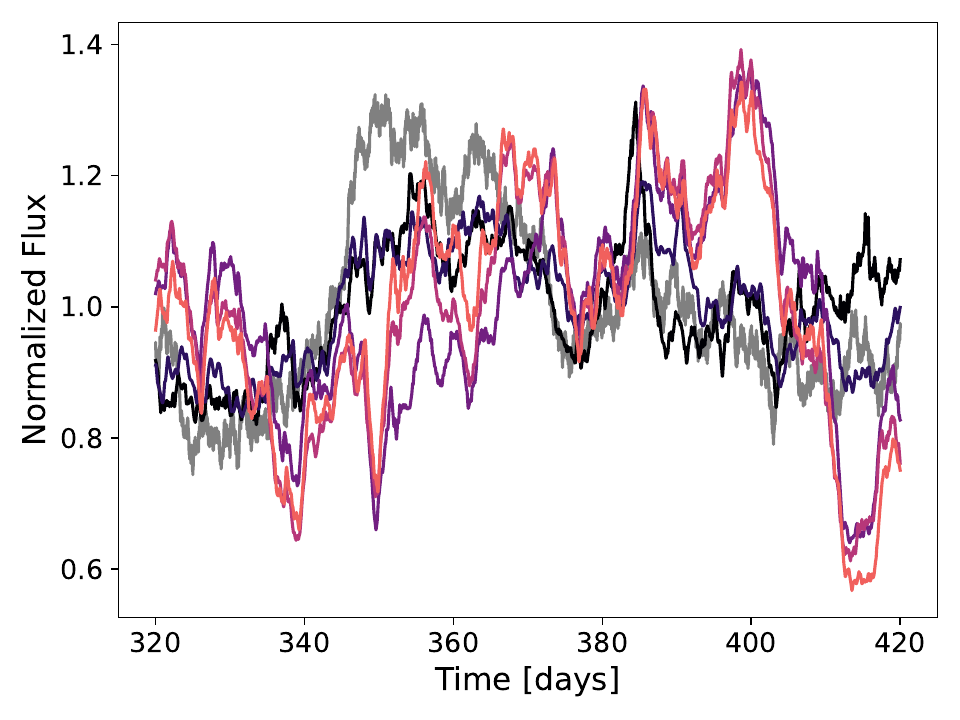}
    \caption{\textbf{Top panel:} The median (points) and interquartile range (error-bars) of the light-crossing lag, measured as the peak of the cross-correlation function between $\pm5$~days, relative to the X-ray light curve (left panel) and the light curve emitted at $R=36~r_g$ (right panel), as a function of the window size for the High X-ray run. For each window size the light curve is split into as many individual light curves of that length as the full length light curve allows. The dashed lines show the light travel time for each radius. The median lag is consistent with the light travel time for most window sizes, but the interquartile range is very large. \textbf{Bottom panel:} Two examples of 100~day light curve intervals for light curves from the High X-ray run. Cross-correlation measures lags consistent with the light crossing timescale for the interval on the left and lags inconsistent with the light-crossing timescale for the interval on the right. Over the 100~day interval on the right, intrinsic variability dominates over X-ray-driven variability at larger radii, leading to low correlations and inaccurate lag measurements for this particular segment, but not the full $>3000$~day light curve.}
    \label{fig:intervals}
\end{figure*}

In the previous section, we show that the light curves emitted from the top of the disk in our simulations are largely consistent with traditional X-ray-driven RM models, especially for runs with higher X-ray luminosities. In this section, we show how the light curves emitted from the bottom of our simulated gas disks differ significantly from traditional models. We also show that even the light curves emitted from the top of the disk differ from traditional models over shorter time intervals and due to the presence of flares.

\subsubsection{Light Curves from the Bottom of the Disk}
\label{sec:departures:bottom}

The right panel of Figure \ref{fig:psd} shows the PSDs of the light curves emitted from the bottom of the disk for our three runs with X-ray irradiation. We give a summary of the parameters for the best-fit bending power-law for these PSDs in the legend. For comparison we show the X-ray PSD as the solid gray line and the dashed gray line gives the PSD of the light curve from our Gray run without injected X-rays summed over all radii out to $R=250~r_g$. 

The PSDs of light curves emitted from the bottom of the disk show larger departures from the X-ray PSD at $f>0.1$~days$^{-1}$ than light curves emitted from the top of the disk. As in the Gray run, $f_b$ decreases with radius, leading to PSDs that are distinct from each other and the X-ray light curve at high-frequencies. In fact, smaller radii light curves have up to 3 orders of magnitude more variability than larger radii light curves at these high-frequencies. The increase in bending time scale with radius suggests that as you move outward in the disk the main source of variability goes from high-frequency variability driven by X-ray irradiation near the surface of the disk to intermediate-frequency variability driven by MRI turbulence around the disk height where the MRI peaks to low-frequency variability generated near the midplane that has been propagated to the surface through convection, as in the Gray run (see Figure \ref{fig:thermal}).



Unsurprisingly, the light curves in the right panel of Figure \ref{fig:mid_lcs} that are emitted from the bottom of the disk in the Medium X-ray run do not appear as well-correlated with the X-ray light curve as the light curves in the left panel that are emitted from the top of the disk. The diamonds in the right panel of Figure \ref{fig:botshorts} show that the peak correlations between the injected X-rays and the emitted UV-optical light curves are lower than the peak correlations of the light curves emitted from the top of the disk and can be as low as $\rm{CC_{coeff}}\approx 0.2$, especially at larger radii. We also see in the right panel of Figure \ref{fig:coherence} that the maximum coherence between the light curves emitted from the bottom of the disk and the injected X-ray light curve is weaker than the maximum coherence between the X-ray light curve and the light curves emitted from the top of the disk. The maximum coherence, like the peak correlation coefficient, decreases with radius for light curves emitted from the bottom of the disk. 


The diamonds in the left panel of Figure \ref{fig:botshorts} show that the peak lags between the injected X-rays and the emitted UV-optical light curves are up to 3--5 times longer than the light travel time, given as the dotted line, at longer radii. In addition, when we use the $R=36~r_g$ light curve as a reference, the cross-correlation lags we measure for larger radii light curves are also longer than the light travel time (solid line). In Section \ref{sec:gray}, we find that using the light curve emitted from $R=115~r_g$ as a reference band leads to higher correlations and lag measurements closer to the light travel time for $R>115~r_g$. Figure \ref{fig:botshorts2} shows the lags and correlation coefficients we measure using this $R=115~r_g$ light curve as a reference band for the Medium X-ray run, which has peak correlations of $\rm{CC_{coeff}}<0.6$ between the outer radii light curves and the light curve emitted from $R=34~r_g$. As in the Gray run case, using the $R=115~r_g$ light curve as a reference band leads to lags more consistent with the light-crossing timescale and $\rm{CC_{coeff}}>0.75$ for all light curves. 

Together, these results and the behavior of the X-ray and UV–optical flux shown in Figure \ref{fig:rho} suggest that the X-ray irradiation is not the direct driver of the variability in the UV–optical light curves emitted at larger radii from the bottom of the disk. Instead, we see that the X-ray irradiation is absorbed at $R\lesssim100~r_g$ driving variability in the inner radii light curves. The emission from this region of the disk can then self-irradiate the outer regions of the disk driving variability at larger radii, leading to longer than anticipated lags and lower correlations between the larger radii emitted light curves and the X-ray light curve. In addition, it is clear that the thicker, convectively unstable, bottom disk produces intrinsic variability that has a larger impact on the emitted light curves. This intrinsic variability can also bias lags and lower the correlation between different light curves. These results are a significant departure from traditional X-ray-driven RM models. However, in Section \ref{sec:discuss:compare} we discuss how these inconsistencies with traditional models are similar to recent observations that also do not agree with traditional RM models.


\subsubsection{Other Departures}
\label{sec:departures:other}

The light curves emitted from the top of the disk are very strongly correlated with the injected X-ray light curve and have lags that are consistent with the light-crossing timescale, as predicted by traditional X-ray-driven RM models. However, we are measuring these lags by cross-correlating a $3000-4500$~day baseline light curve, when many RM studies use light curves with much shorter baselines. In the top panel of Figure \ref{fig:intervals} we show the median and interquartile range of lags measured over shorter time intervals for light curves emitted from the top of the disk in the High X-ray run, which has the highest correlated light curves. The median lag values are all roughly consistent with the light travel time (dashed lines), especially in the left panel, where we measure this lag relative to the X-ray light curve. However, there is a large interquartile range in the lags measured for shorter intervals, suggesting that some time windows have significantly less consistent lag measurements. 

To illustrate this range in consistency, in the bottom panel of Figure \ref{fig:intervals} we show the injected X-ray and different radii emitted light curves for a 100~day interval with a lag measurement consistent with the light travel time, on the left, and a lag measurement much longer than the light travel time, on the right. The light curves are clearly very well-correlated in the left panel but not well-correlated in the right panel, despite these two light curves coming from the top of the disk in the same simulation separated by only 100 days. These changes between correlated and uncorrelated light curves occur because the top of the disk is also turbulent. This turbulence can lead to the top of the disk becoming convectively unstable and puffed up like the bottom of the disk over shorter timescales, although it is never as thick as the bottom of the disk.

A final departure from traditional RM models is the presence of flares in the emitted light curves. Flares in our simulations are caused by strong magnetic reconnection events that occur near the X-ray photosphere of the disk. Panel D of Figure \ref{fig:rho} shows one such magnetic reconnection event that causes a strong flare at just short of 3000~days in our light curves emitted from the top of the disk at $100~r_g<R<150~r_g$ in our Medium X-ray run (see left panel of Figure \ref{fig:mid_lcs}). Although these flares can always occur at smaller radii, they only occur in larger radii light curves emitted by the top of the disk, because the top of the disk experiences stronger reconnection events lower down towards the X-ray photosphere of the disk. These flares last for roughly 50~days and make it very difficult to accurately measure time lags over the portion of the light curve when they occur (see Section \ref{sec:methods:lcs}). However, it appears that they are infrequent, occurring once a simulation for a rate of $\sim 0.1$/year.

\subsection{Lags on the Inflow Timescale}
\label{sec:results:long}

\begin{figure}
    \centering
    \includegraphics[width=\columnwidth]{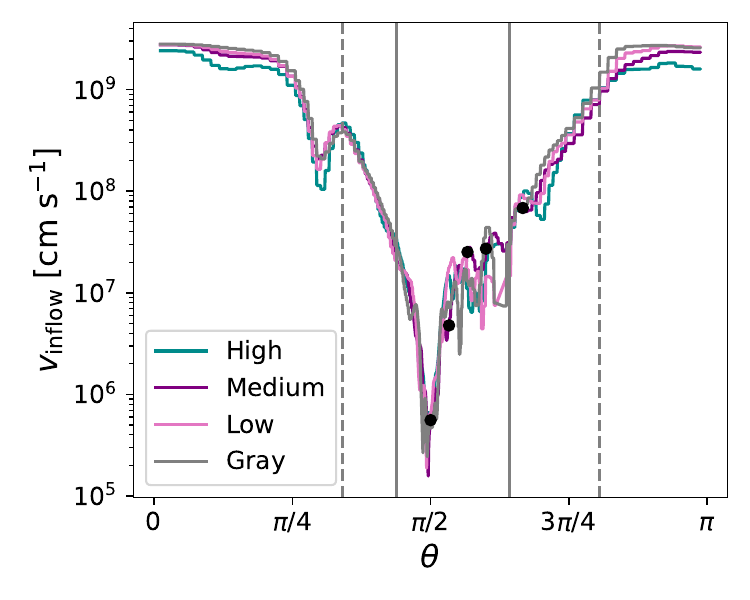}
    \caption{Horizontally- and time-averaged bulk inflow speed as a function of polar angle $\theta$ for our four different simulation runs. The solid vertical lines show the average location of the UV-optical photosphere, and the dashed lines show the average location of the X-ray photosphere. While the inflow speed is very slow at the midplane at $\theta=\pi/2$ ($\sim 10^{-5}c$), it is much faster at the photosphere of the disk ($\sim 10^{-3}c$). The black dots give the locations of the $\theta$-slices in Figures \ref{fig:gdfluct} and \ref{fig:mdfluct}. The propagation speeds in Figures \ref{fig:gdfluct} and \ref{fig:mdfluct} and the lag timescales in Figures \ref{fig:gray_bot} -- \ref{fig:low_top} are roughly between $10^7-10^8$~cm~s$^{-1}$, consistent with the inflow speed near the photosphere.}
    \label{fig:tvisc}
\end{figure}

\begin{figure*}
    \centering
    \includegraphics[width=0.9\textwidth]{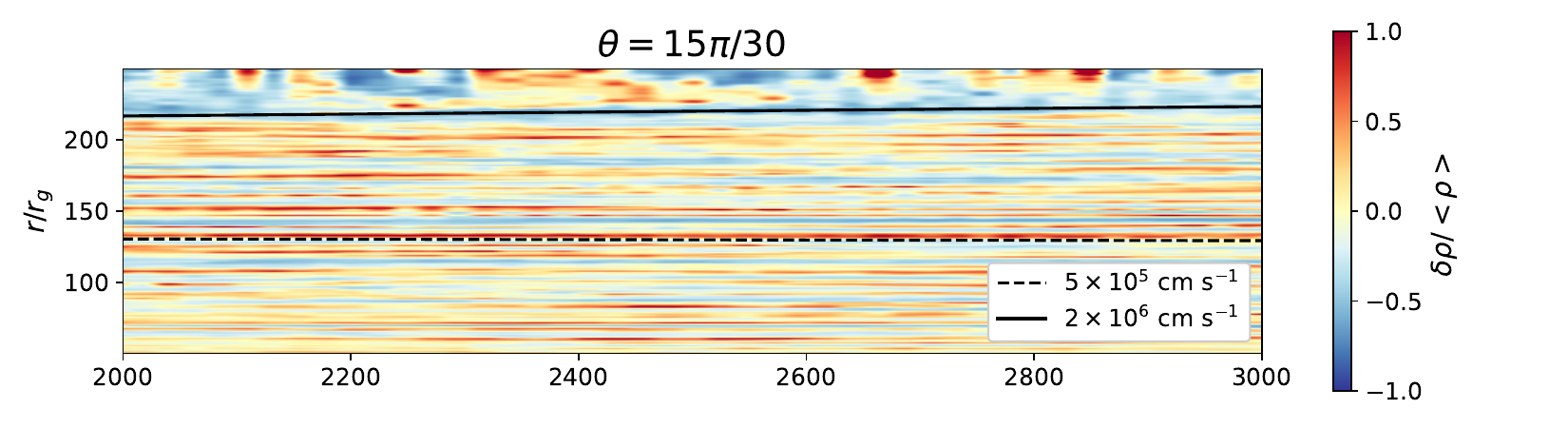} 
    \includegraphics[width=0.9\textwidth]{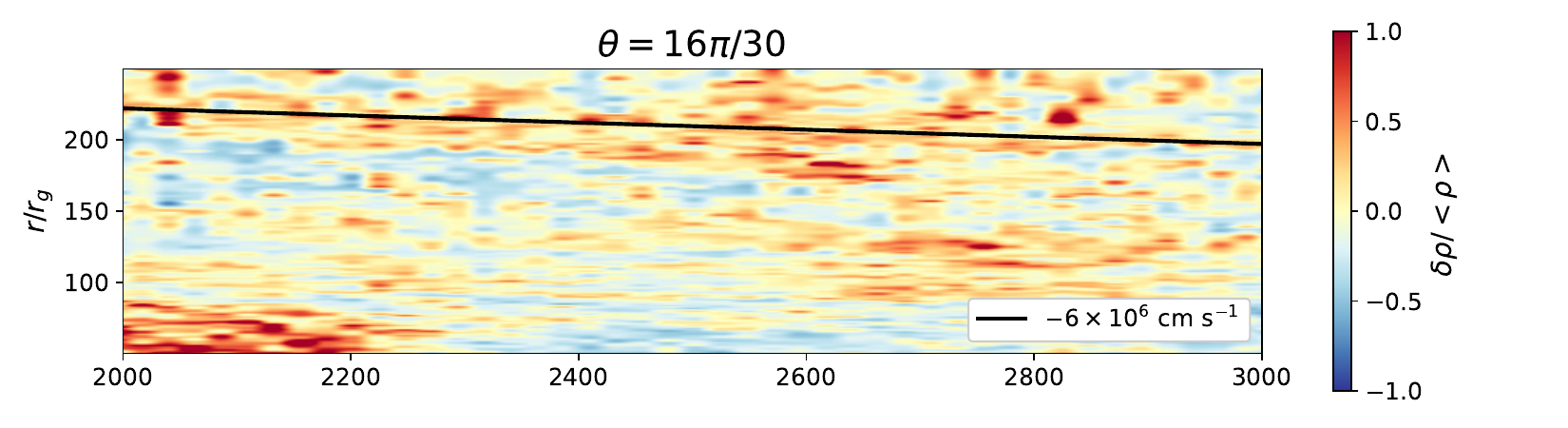}
    \includegraphics[width=0.9\textwidth]{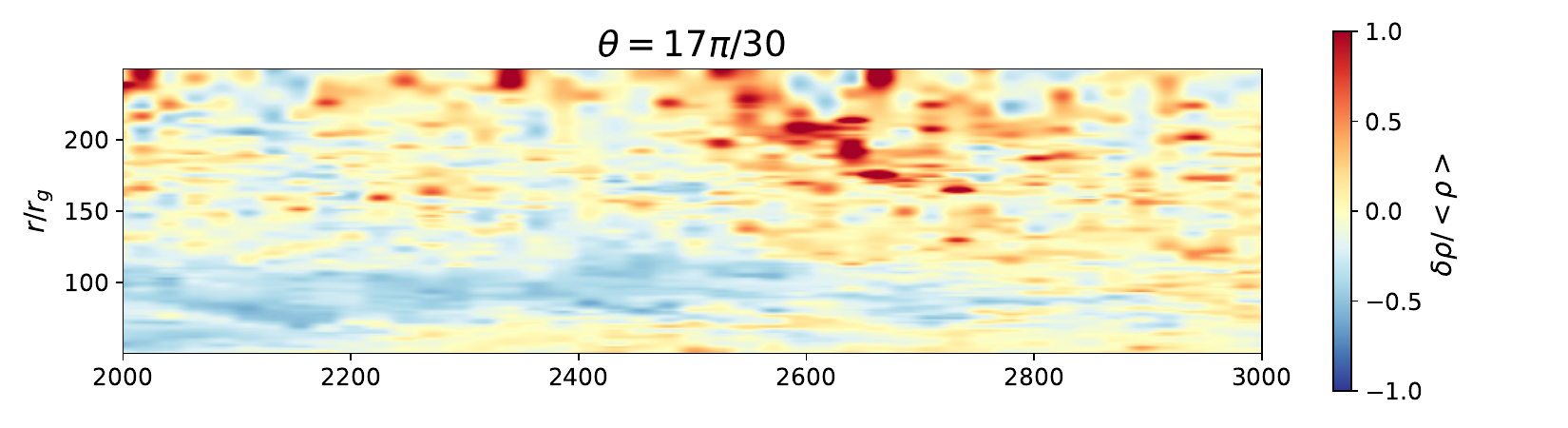}
    \includegraphics[width=0.9\textwidth]{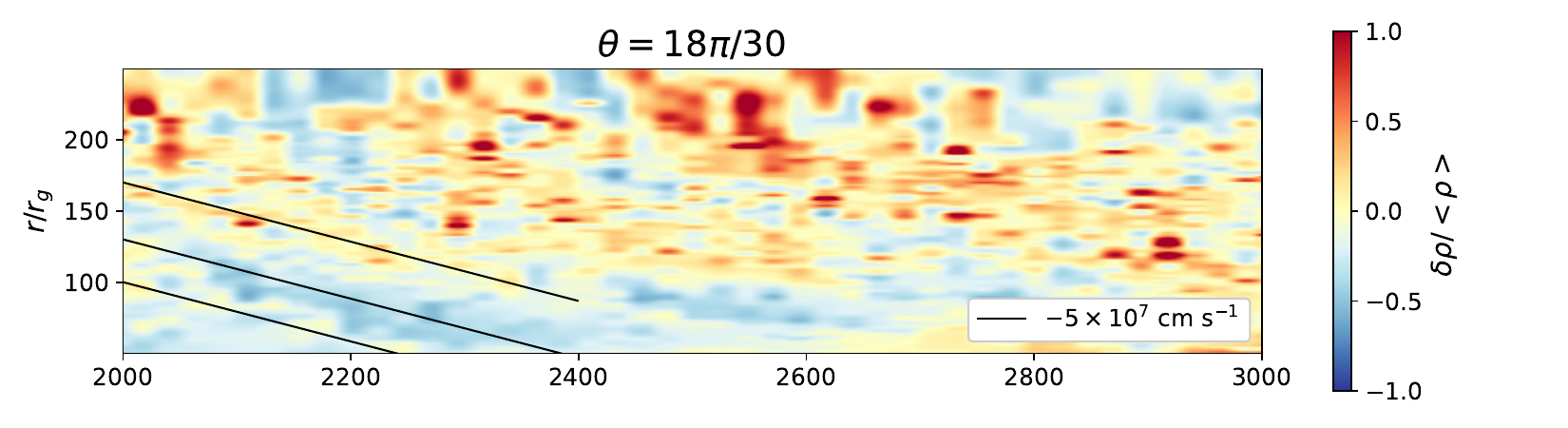}
    \includegraphics[width=0.9\textwidth]{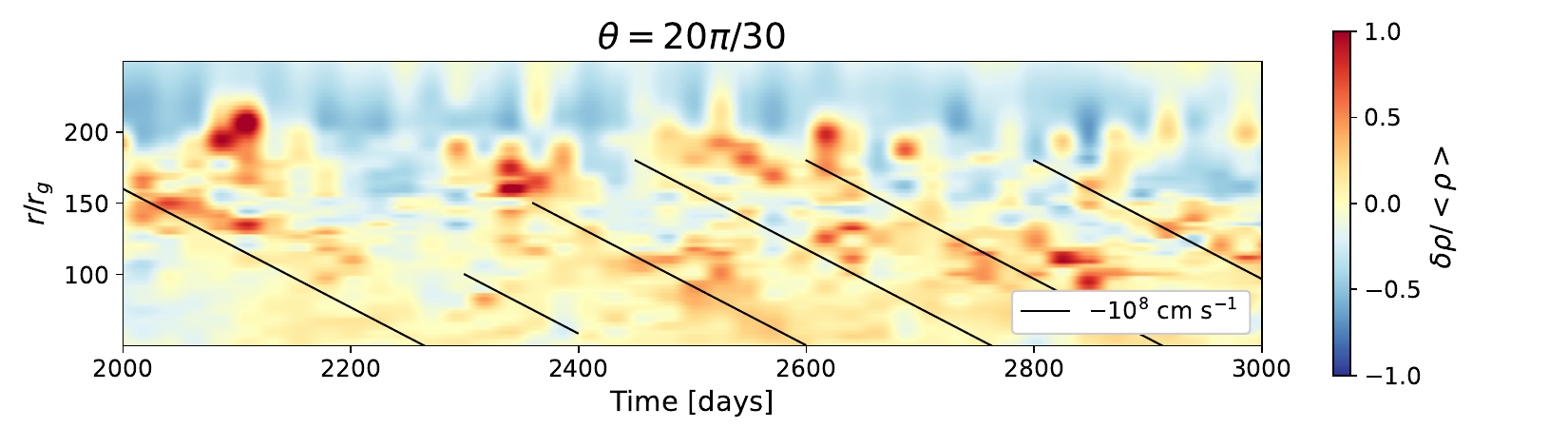}
\caption{Density fluctuations, $\delta \rho$, over mean density, $\langle\rho\rangle$, as a function of time and radius at specific polar angles, $\theta$, over 1000~days for the Gray run without injected X-rays. The solid and dotted lines guide the eye towards inward or outward propagating fluctuations of different speeds. The propagation speeds increase as function of $\theta$, correspond well to the inflow velocity in Figure \ref{fig:tvisc}, and are consistent with the long lags measured in Figures \ref{fig:gray_bot} -- \ref{fig:low_top}.}
    \label{fig:gdfluct}
\end{figure*}

\begin{figure*}
    \centering
    \includegraphics[width=0.9\textwidth]{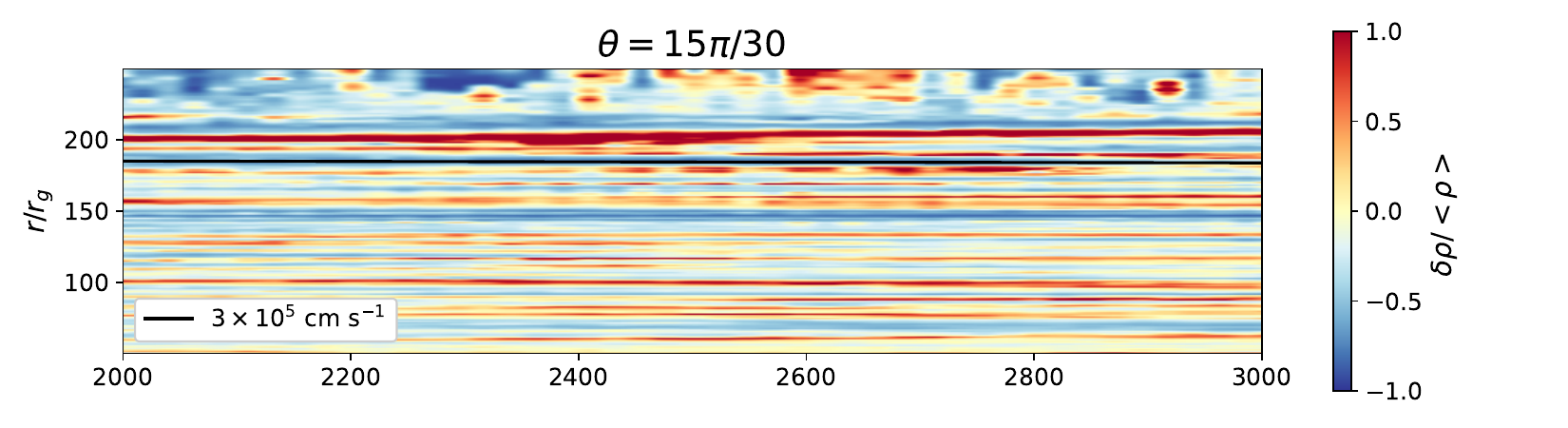} 
    \includegraphics[width=0.9\textwidth]{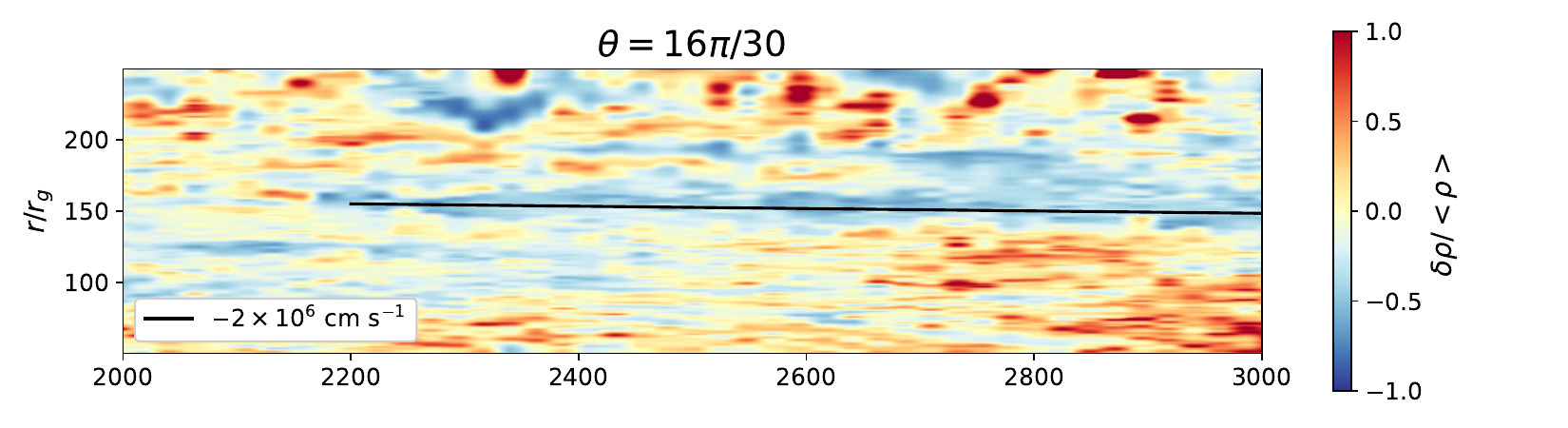}
    \includegraphics[width=0.9\textwidth]{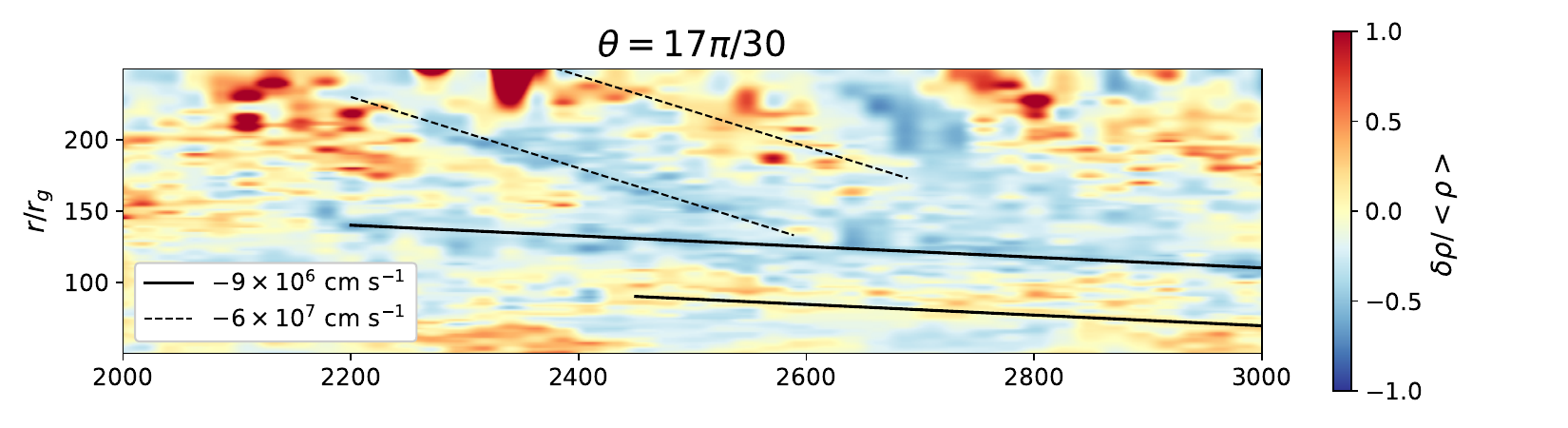}
    \includegraphics[width=0.9\textwidth]{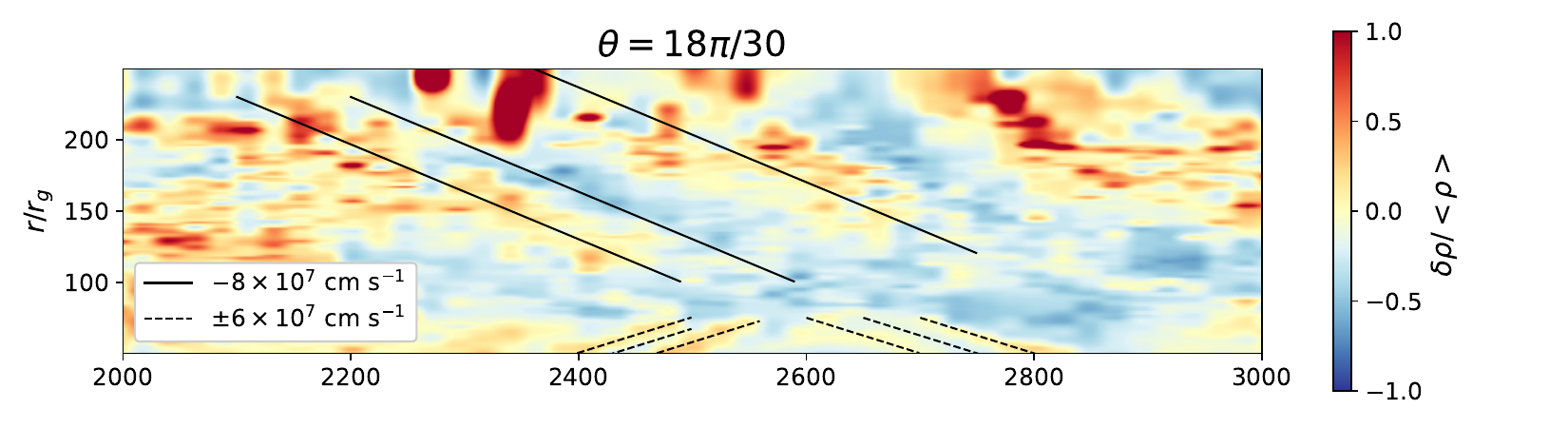}
    \includegraphics[width=0.9\textwidth]{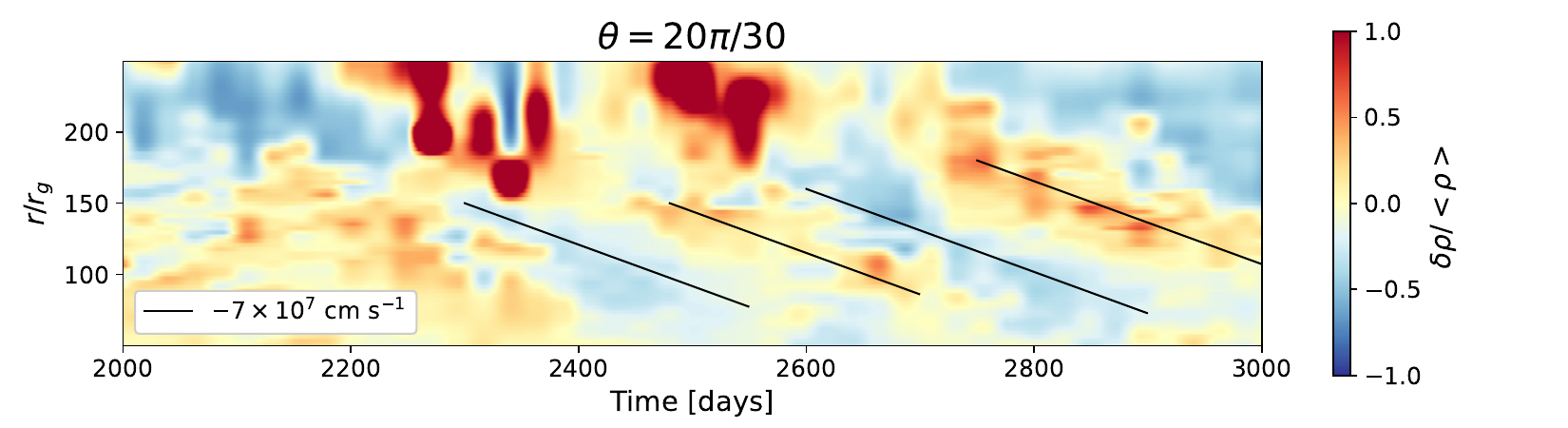}
\caption{Density fluctuations, $\delta \rho$, over mean density, $\langle\rho\rangle$, as a function of time and radius at specific polar angles, $\theta$, over 1000~days for the Medium X-ray run. The solid and dotted lines guide the eye towards inward or outward propagating fluctuations of different speeds. The propagation speeds increase as function of $\theta$, correspond well to the inflow velocity in Figure \ref{fig:tvisc}, and are consistent with the long lags measured in Figures \ref{fig:gray_bot} -- \ref{fig:low_top}.}
    \label{fig:mdfluct}
\end{figure*}

\begin{figure}
    \centering
    \includegraphics[width=0.92\linewidth]{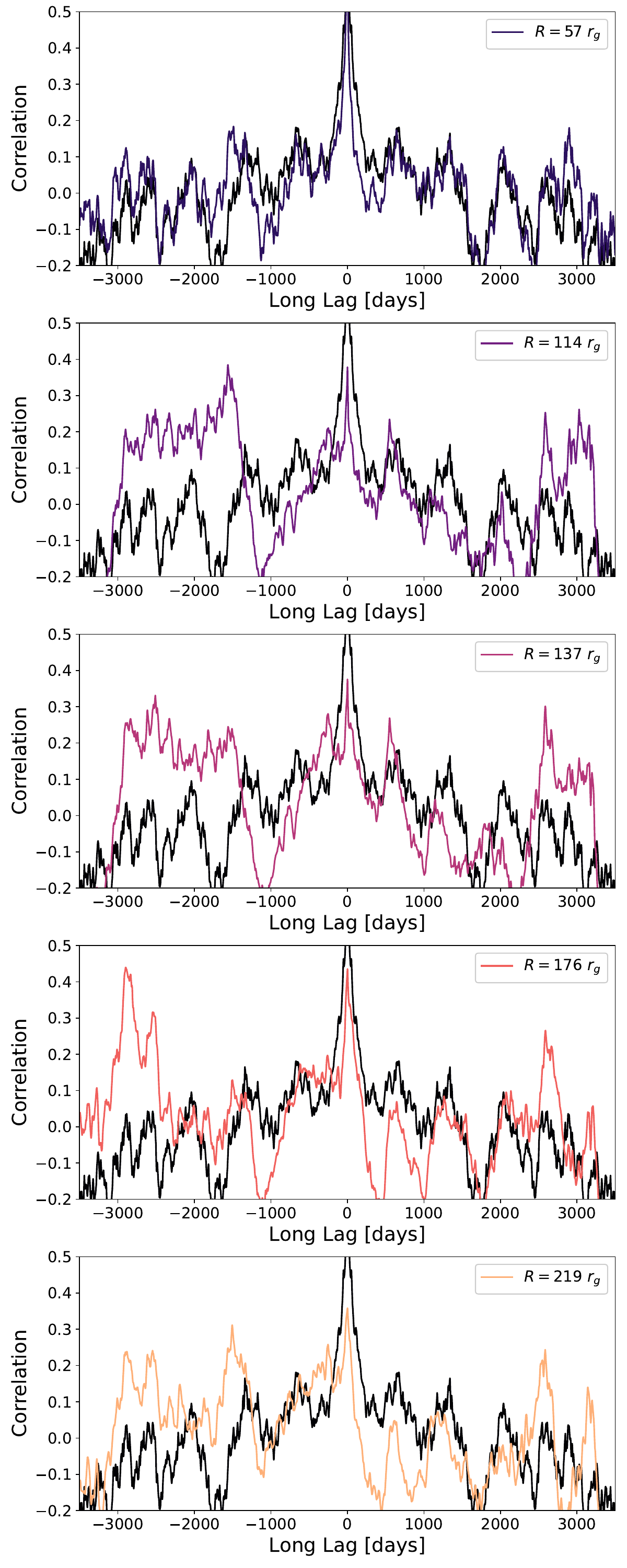}
    \caption{Cross-Correlation functions between the light curves emitted by the bottom of the disk from five different radii and the light curve emitted from $R=34~r_g$ from the Gray run. The auto-correlation function for the reference light curve is shown in each panel in black for comparison. The cross-correlation functions peak near zero for all radii and also around $-3000$ to $-1500$~days for light curves emitted from $R>57~r_g$.}
    \label{fig:gray_bot}
\end{figure}

\begin{figure}
    \centering
    \includegraphics[width=\linewidth]{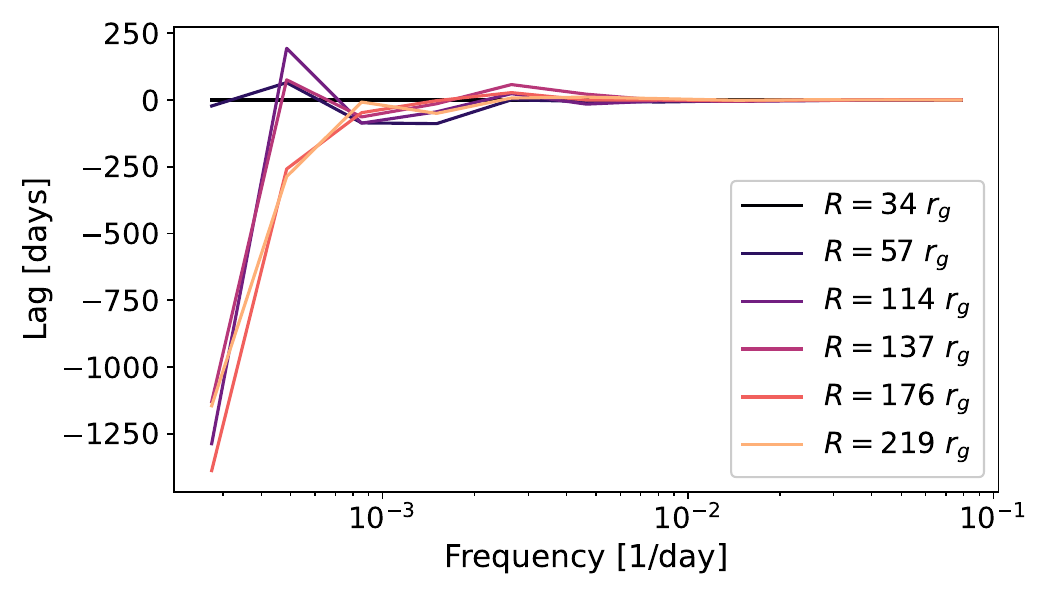}
    \caption{Lags measured using the Fourier method between light curves emitted by the bottom of the disk from six different radii and the light curve emitted from $R=34~r_g$ from the Gray run. There appears to be a $\sim -1250$~day lag between the $R=34~r_g$ and light curves at $R>57~r_g$. This lag would correspond to a propagation speed of $\sim 2\times10^7$~cm~s$^{-1}$.}
    \label{fig:gray_four_bot}
\end{figure}

\begin{figure*}[h!]
    \centering
    \includegraphics[width=0.43\textwidth]{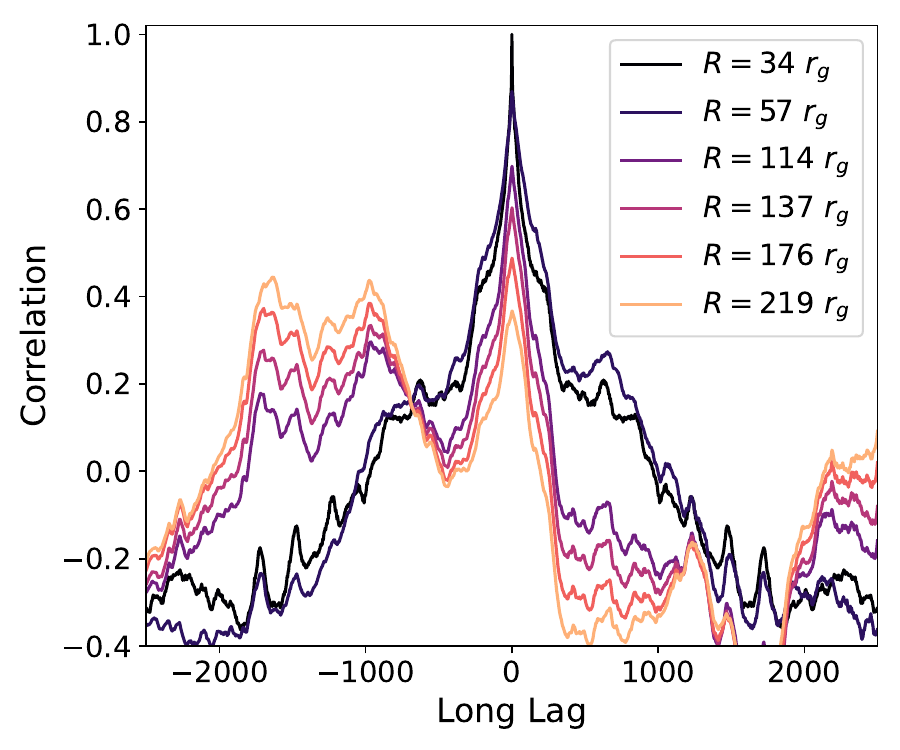}
    \includegraphics[width=0.55\textwidth]{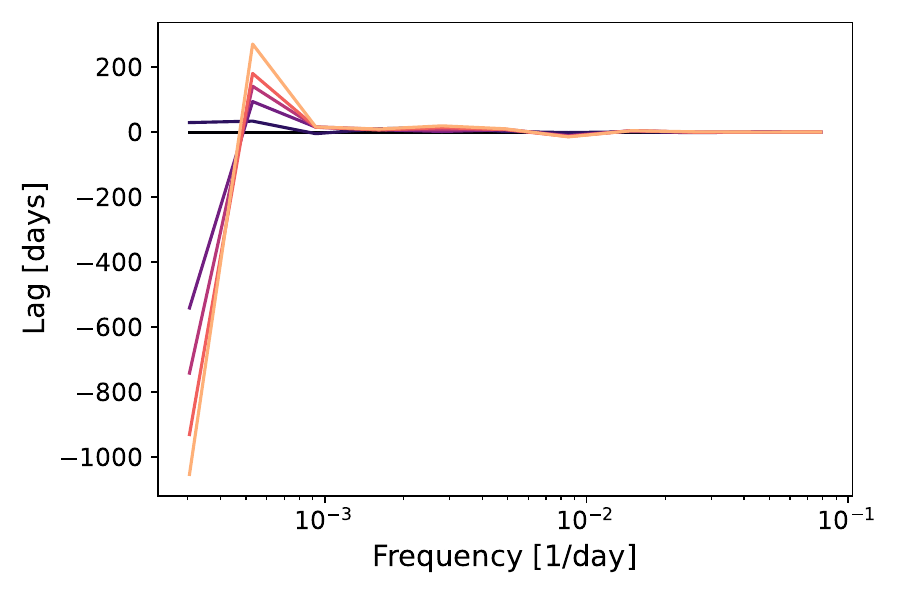}
    \caption{Long lags observed between light curves emitted by the bottom of the disk from six different radii and the light curve emitted from $R=34~r_g$ from the Medium X-ray run. The left panel shows the cross-correlation lag and the right panel shows the Fourier method lag. Both methods support the presence of a long lag of around $-500$ to $-2000$~days. The lag duration roughly increases as a function of radius and is not present in the light curve emitted from $57~r_g$. The lags give a propagation speed of roughly $10^7$~cm~s$^{-1}$.}
    \label{fig:mid_bot}
\end{figure*}

\begin{figure*}
    \centering
    \includegraphics[width=0.88\columnwidth]{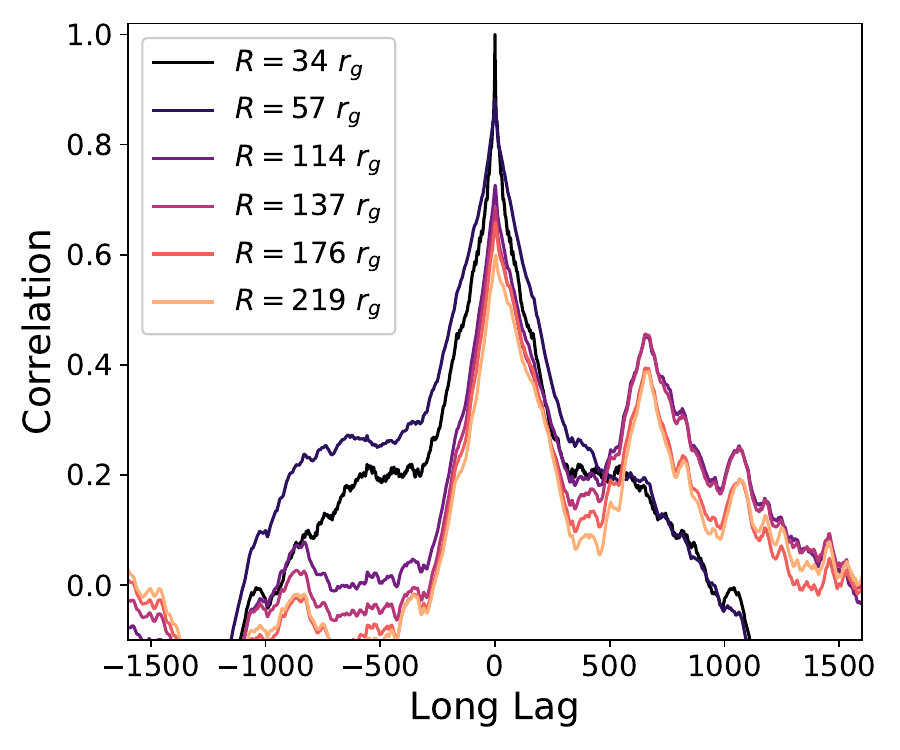}
    \includegraphics[width=1.12\columnwidth]{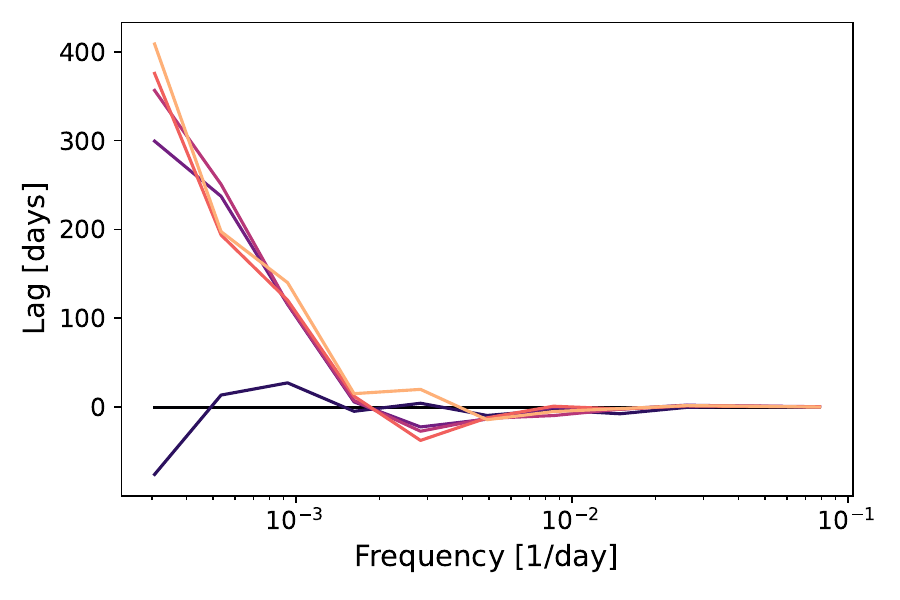}
    \caption{Long lags observed between light curves emitted by the bottom of the disk from six different radii and the light curve emitted from $R=34~r_g$ from the Low X-ray run. The left panel shows the cross-correlation lag and the right panel shows the Fourier method lag. Both methods support the presence of a long positive lag for light curves emitted at $R>57~r_g$, although the long lag predicted by the Fourier method of around 300--400~days is shorter than the peak in the cross-correlated seen around 700~days. The light curve emitted from $R=57~r_g$ has a negative lag relative to the light curve emitted from $R=34~r_g$, suggesting that the fluctuation producing this lag may originate around this radius and propagate inwards and outwards. The lags give a propagation rate of around $10^7-10^8$~cm~s$^{-1}$.}
    \label{fig:low_bot}
\end{figure*}

\begin{figure*}
    \centering
    \includegraphics[width=0.29\textwidth]{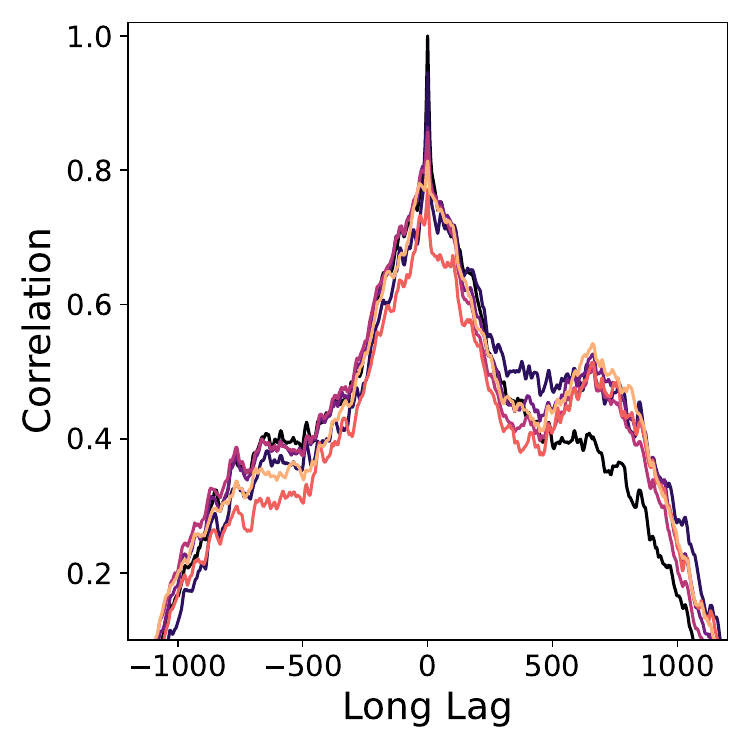}
    \includegraphics[width=0.29\textwidth]{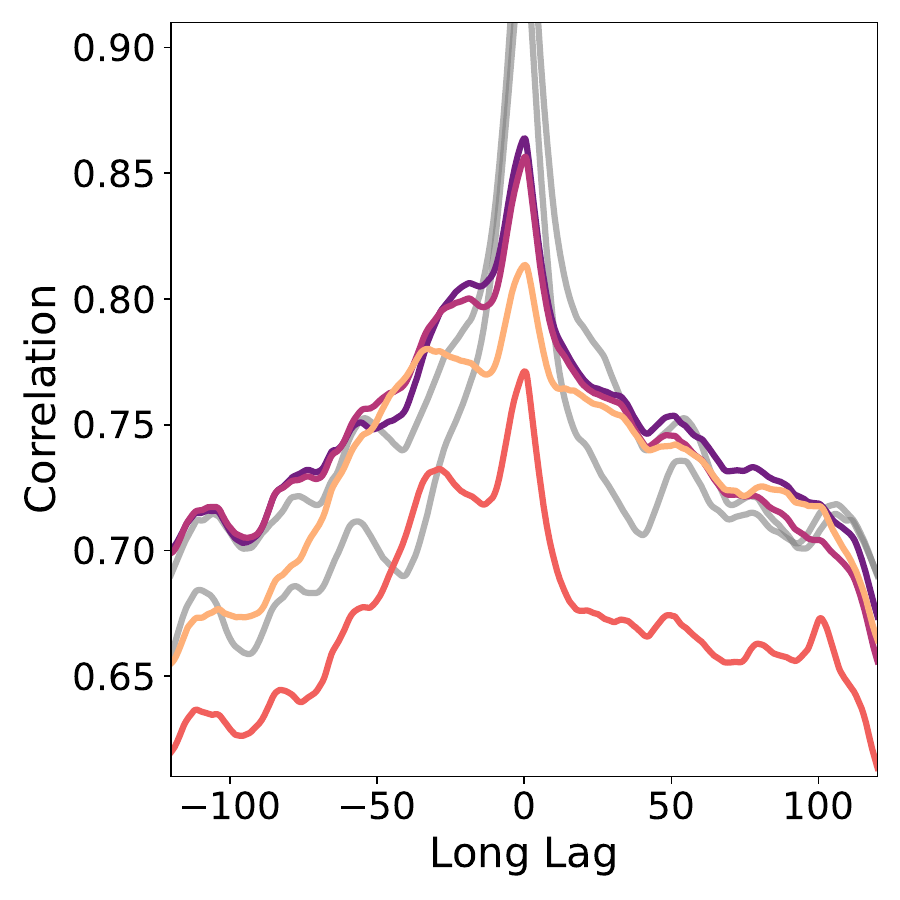}
    \includegraphics[width=0.39\textwidth]{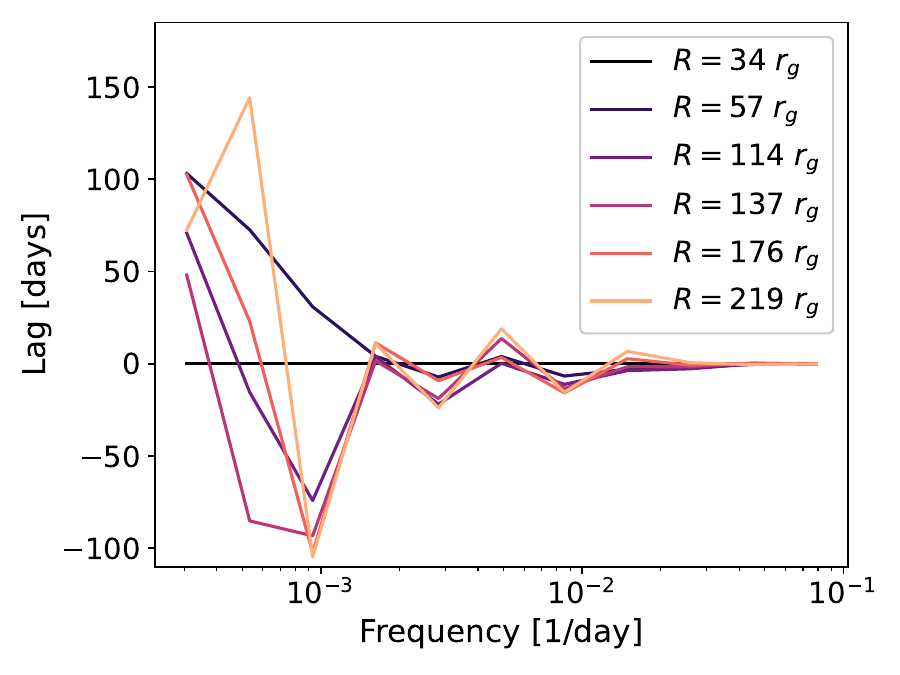}
    \caption{Long lags observed between light curves emitted by the top of the disk from six different radii and the light curve emitted from $R=34~r_g$ from the Low X-ray run. The left and center panels show the cross-correlation lag and the right panel shows the Fourier method lag. The Fourier method shows a positive lag at the lowest frequencies that appears to correspond with a longer positive lag in the cross-correlation in the leftmost panel. At around $10^{-3}$~days$^{-1}$ there also appears to be a negative lag between $-50$ and $-100$~days for light curves emitted at $R>57~r_g$. The zoomed in cross-correlation in the middle panel shows that these lags are subtle but present in the cross-correlation as well. The lags give a propagation speed of roughly $10^8$~cm~s$^{-1}$.}
    \label{fig:low_top}
\end{figure*}


The results of the previous sections are consistent with the shearing box simulations of the UV-emitting regions of AGN disks in \cite{Secunda:2024} and \cite{Secunda:2025} that show that fluctuations in the AGN disk produce intrinsic variability in AGN light curves emitted from the disk. Moving beyond radially local shearing box simulations allows us to examine how these fluctuations propagate inward or outward throughout the AGN disk and see if we can measure this propagation as a long lag on the inflow timescale between different radii light curves. In this section, we piece together a complete picture of this long lag on the inflow timescale. We first measure the bulk inflow speed for our simulated disk (Figure \ref{fig:tvisc}). Next, we explicitly track the propagation of density fluctuations to different radial regions of the disk (Figures \ref{fig:gdfluct} and \ref{fig:mdfluct}). Finally, we look for long lags between different radii light curves to see if we can measure the inflow speed of these propagating fluctuations using only AGN light curves (Figures \ref{fig:gray_bot} -- \ref{fig:low_top}).



Figure \ref{fig:tvisc} shows the horizontally- and time-averaged inflow speeds as a function of polar angle, $\theta$, for our four runs. The solid (dashed) vertical lines show the average location of the UV-optical (X-ray) photosphere. The inflow speed in the midplane is $\sim10^{-5}c$, giving an inflow timescale of nearly 1000~years, similar to predictions from the thin disk model. However, the inflow speed increases by two orders of magnitude to nearly 0.1\% the speed of light at the UV-optical photosphere and continues to increase at the highest polar angles. 

We can compare this average inflow speed to the propagation speed of turbulent density fluctuations in the simulated disk. We show these density fluctuations, $\delta \rho/\langle\rho\rangle$, as a function of radius and time over 1000 days for slices at the various polar angles marked with black dots in Figure \ref{fig:tvisc} for our Gray run without injected X-rays in Figure \ref{fig:gdfluct}. Near the midplane there are small scale density fluctuations coming from the MRI. Around the photosphere, convection becomes important (see Figure \ref{fig:entropy}), leading to larger more coherent density fluctuations, contrary to the results in \cite{Kaul:2025}.

The dashed and solid black lines in Figure \ref{fig:gdfluct} highlight some of the visible inward and outward propagating density fluctuations in these simulations. Although it is typically assumed that most of these fluctuations should propagate inward in the direction of accretion \citep{Hagen:2024,Hagen:2025,Secunda:2023,Lyubarskii1997},  \cite{Neustadt:2022}, \cite{Neustadt:2023}, and \cite{Stone:2023} all observed both inward and outward propagating temperature fluctuations. The propagation speeds we estimate in Figure \ref{fig:gdfluct} are roughly consistent with the averaged inflow speeds at the corresponding heights marked by the black dots in Figure \ref{fig:tvisc}.

Figure \ref{fig:mdfluct} shows the propagation of turbulent density fluctuations in our Medium X-ray run. Adding X-ray irradiation seems to increase the amount of larger scale density fluctuations. Similarly to our Gray run without injected X-rays, the estimated propagation speeds are roughly consistent with the average inflow speeds in Figure \ref{fig:tvisc} and the density fluctuations become more coherent towards the photosphere of the disk. The increase in the propagation speed and the coherence of the fluctuations with $\theta$, should make them easier to observe in light curves emitted at the photosphere.

The density fluctuations and propagation speeds for our High and Low X-ray runs are very similar to the propagating fluctuations shown in Figure \ref{fig:mdfluct} from our Medium X-ray run. We are also able to observe propagating fluctuations at polar angles above the disk midplane ($\theta<\pi/2$). However, these fluctuations are less coherent, perhaps because convection plays less of a role in driving fluctuations above the midplane, making them less obvious to visually identify.

The next step is to attempt to observe these propagating fluctuations as long lags in the light curves emitted from our simulated disks. Long lags are commonly assumed to be \textit{negative} and result from the \textit{inward} propagation along with the accretion flow of fluctuations from large to small disk radii (or long to short wavelength light curves in actual observations). However, just as we observe \textit{outward} propagating density fluctuations in Figures \ref{fig:gdfluct} and \ref{fig:mdfluct}, we also observe \textit{positive} long lags in our simulated light curves. We are able to measure long lags between different radii light curves and the light curve emitted from $R=34~r_g$ for both sides of the disk in all simulation runs using both cross-correlation and a Fourier method described in Section \ref{sec:methods:lcs}.

We show the cross-correlation functions between the light curve emitted from $R=34~r_g$ and five different radii light curves emitted from the bottom of the disk in our Gray run without X-ray irradiation in Figure \ref{fig:gray_bot}. For comparison we show the auto-correlation for the $R=34~r_g$ light curve. These cross-correlations are very noisy, because without X-ray irradiation these light curves are less coherent with each other. However, all light curves show a peak around 0~days, and there appears to be a prominent peak that is larger than the auto-correlation at $-3000$ to $-1500$~days for light curves emitted from $R=114~r_g$, $R=137~r_g$ and $R=176~r_g$. This lag corresponds to a propagation speed of $1-2\times10^7$~cm~s$^{-1}$, which falls somewhere in between the propagation timescale we measure at $\theta=16\pi/30$ and $\theta=18\pi/30$ in Figure \ref{fig:gdfluct}. For the light curves emitted from $R=114~r_g$ and $R=176~r_g$ this long lag peak is slightly higher than the peak around 0~days. These cross-correlations show that even though there is intrinsic variability in these light curves coming from the fluctuations themselves on timescales much shorter than the inflow timescale (see Section \ref{sec:gray}), propagating fluctuations still play a large role in driving the variability in the disk.

Unfortunately, the noise in the cross-correlation functions makes it hard to make strong statements about the presence of a long lag. Validating these lags by observing them with an additional method, such as the Fourier method, can increase our confidence in these lag detections. Figure \ref{fig:gray_four_bot} shows the lags measured with the Fourier method for light curves emitted from the bottom of the disk for the Gray run. The Fourier method measures a lag of around $-550$ to $-1250$~days between the light curve emitted from $R=34~r_g$ and all light curves at $R>57~r_g$, which is shorter than the lag measured using cross-correlation. \cite{Secunda:2023} showed that long negative lags measured with the Fourier method are often underestimated when there are other signals in the AGN light curves. Therefore, even without any signal from X-ray reprocessing and noiseless even cadence light curves, the weak overall coherence of the light curves and the fact that there is still a confounding signal around 0~days from disk self-irradiation make it difficult to measure a robust and accurate long negative lag.

The cross-correlation functions from our runs with X-ray irradiation are much smoother, which actually makes it easier to detect long negative lags in these runs than in the Gray run. We show clean examples of long negative lags and long positive lags measured between the light curve emitted at $R=34~r_g$ and all other light curves emitted at $R>60~r_g$ in Figures \ref{fig:mid_bot} and \ref{fig:low_bot}, respectively. The lags measured by cross-correlation and the lags measured at the lowest frequencies by the Fourier method agree reasonably well with each other, although again the lags measured by cross-correlation tend to be slightly longer. Interestingly, the long lag in the light curve emitted from $R=57~r_g$ in Figure \ref{fig:low_bot} is in the opposite, negative direction, suggesting that the fluctuation originates at $R\approx57~r_g$ and propagates inward to $R=34~r_g$ and outward to $R>60~r_g$.

For a few runs, such as the example from the Low X-ray run we show in Figure \ref{fig:low_top}, we see evidence of multiple long lag signals. In the right panel, we see a positive lag at the lowest frequencies, which may correspond to the long positive lag we see in the cross-correlation in the left panel. At $f\approx10^{-3}$~days$^{-1}$, we see a negative lag in all light curves at $R>60~r_g$ of around $-50$ to $-100$~days. This second long lag is similar to a peak we can see in the zoomed-in cross-correlation we show in the middle panel. 

The lags that we measure in the light curves from all four of our simulations suggest an inflow speed on the order of $10^7-10^8$~cm~s$^{-1}$. This speed is consistent with the propagation and inflow speeds we measure in Figures \ref{fig:tvisc}, \ref{fig:gdfluct}, and \ref{fig:mdfluct} for $16\pi/30\leq\theta\leq20\pi/30$. Our simulations therefore reveal a unified picture of density fluctuations driven by the MRI and convection propagating on the inflow speed producing identifiable lag signals as a function of radius in the emitted light curves. Measuring the inflow timescale of the disk through the observation of long lags would allow us to probe the vertical structure of the disk and test the standard thin disk model, because the inflow timescale depends on the disk aspect ratio (see Equation \ref{eq:inf}). We find that the inflow speed we measure from the long lags in our different radii light curves increases as a function of radius, which is consistent with the increase in aspect ratio as a function of radius in our simulations.

\section{Comparison to Observations}
\label{sec:discuss:compare}

In traditional models for RM, time-varying X-rays emitted by the corona are reprocessed by different radial regions of the gas disk and re-emitted at different wavelengths based on the local blackbody temperature of that radial region of the disk. In this model the X-ray variability is the main driver of variability in the UV-optical light curves emitted by the disk, leading to highly correlated X-ray, UV, and optical light curves and inter-waveband lags that are on the light-crossing timescale. The light curves emitted from the top of our simulated AGN disks from the High and Medium X-ray runs are consistent with this traditional model. For these light curves, the PSDs resemble the X-ray PSD up to high frequencies, we measure cross-correlation lags on the light-crossing timescale, and the correlation coefficients between the different emitted light curves and the X-ray light curve are $\rm{CC_{coeff}}>0.88$ for all radii. 

Conversely, light curves emitted from the thicker bottom of the disk are poorly correlated with the X-ray light curve at larger radii and have longer than anticipated continuum lags, which is not consistent with traditional RM models. However, there are many examples of AGN with variability that cannot be explained by X-ray reprocessing alone \citep[e.g.,][]{Arevalo:2009,Burke:2023,Beard:2025}. In addition, recent RM campaigns have observed moderately \citep[e.g.,][]{Edelson:2019,Kara:2023,Edelson:2024} or weakly \citep{Cackett:2023,Buisson:2018,Morales:2019} correlated X-ray and UV light curves, as well as lags $\gtrsim3\times$ longer than the predicted light travel time \citep[e.g.,][]{Steyn:2026,Guo:2022,Wang:2025,Lewin:2025}. 

In this section, we compare the results from our simulations to observed AGN light curves to provide a physically-motivated model for observations. We also discuss how we can use this model to predict the structure of AGN disks.



\subsection{Light Curve PSDs and Intrinsic Variability}

There is observational evidence for intrinsic variability in many AGN light curves \citep[e.g.,][]{Arevalo:2009,F92020,Hagen:2025,Edelson:2024}. For example, \cite{Beard:2025} studied the X-ray and optical PSDs of NGC 4395 over several decades of frequency. They found that the low-frequency slope of the optical PSD was $\alpha_l=1$, which is significantly steeper than the low-frequency slope of the DRW ($\alpha_l=0$). They also found that this additional low-frequency variability could not be explained by reprocessed X-ray variability due to the lack of low-frequency variability in the X-ray PSD. \cite{Petrecca:2026} studied thousands of AGN PSDs and also find low frequencies slopes of $\alpha_l\approx-1$. Our PSDs have $\alpha_l\approx0.5-1.5$, even in our Gray run without X-ray irradiation, which is consistent with observational results and points to the importance of intrinsic variability in the low frequency portion of UV-optical light curve PSDs.

While X-ray AGN light curves also have breaks that scale with SMBH mass \citep[e.g.,][]{Uttley:2002,Papadakis:2002,Markowitz:2003,McHardy:2004,McHardy:2006,McHardy2007}, \cite{Burke2021} and others have proposed that the longer bending timescales of UV-optical AGN light curves may depend on the disk thermal timescale instead of the X-ray PSD \citep[see also][]{Kelly:2009,Kozlowski2010,MacLeod:2010,Burke:2023,Arevalo:2024A&A...684A.133A}. The dependence of this bending timescale on the thermal timescale suggests that disk fluctuations on the thermal timescale have a direct impact on the variability of the emitted light curves, which is a departure from traditional X-ray-driven RM models.


We find that the bending timescales we fit to the light curves from our High X-ray run, with the highest X-ray luminosity, are typically shorter than the thermal timescales at most radii, suggesting that the main driver of variability at these radii is the X-ray variability reprocessed in the photosphere of the disk. However, in our other simulation runs the light curves emitted from the top of the disk tend to have bending timescales similar to the thermal timescale calculated from where the MRI turbulence peaks, indicating that MRI turbulence plays a large role in driving the variability in those light curves. Light curves emitted from the bottom of the disk have bending timescales that increase with radius similar to the thermal timescale, although only the Gray run has bending timescales that long, except at the largest radii. This trend may point to the important role that convection plays in propagating variability generated by fluctuations deeper down in the disk to the photosphere in these light curves. The difference in the PSDs for the top and bottom light curves offers the tantalizing possibility of using the bending timescale of AGN light curves to study MRI and convection in AGN disks.


\subsection{Longer than Anticipated Lags with No Broad Line Region}

Various models involving gas obscuration \citep{Kara:2021,Lewin:2025} or a changing corona height \citep{Panagiotou:2022B} have been proposed to explain weakly correlated X-ray and UV-optical light curves, and contamination from diffuse continuum in the BLR is often invoked to explain observed lags that are longer than anticipated lags \citep[e.g.][]{Steyn:2026,Lewin:2023,Cackett2021,Wang:2025}. However, our simulations have a static corona location, do not include a BLR, and still produce weakly correlated light curves and longer than anticipated lags.

Comparing the differences between the physical structure of the top and bottom of our simulated disk provides a physically-motivated model that can explain these departures from traditional X-ray-driven RM models. Because the top of our simulated disk is thin, X-ray irradiation is directly reprocessed at the X-ray photosphere into UV-optical emission, allowing the X-rays to be the main driver of variability in the emitted light curves. As a result, we generally find the light curves emitted from the top of the disk are well-correlated with the X-ray light curve with lags between the variability consistent with the light-crossing timescale. For the bottom of the disk, we find that gas obscuration is at least part of the story because the puffed-out structure of the bottom of the disk leads to X-ray irradiation only reaching the innermost disk radii. These inner disk radii, $R<100~r_g$, remain well-correlated with the X-ray light curve ($\rm{CC_{coeff}}\gtrsim 0.8$) and have lags that are roughly consistent with the light travel time. This region of the disk then emits time-varying far-UV radiation that is reprocessed at outer radii.

This additional reprocessing step is responsible for the lower correlations between the X-ray and outer UV-optical light curves emitted from the bottom of the disk for two reasons. First, when the inner disk reprocesses the X-ray variability, it will add the intrinsic variability of that region to the re-emitted light curve that will go on to drive variability in outer radii light curves. This behavior explains why, compared to the X-ray light curve, the different emitted UV-optical light curves are more correlated with the $R=36~r_g$ light curve and most correlated with the light curves emitted from nearby radii (i.e., Figure \ref{fig:botshorts2}). Second, this repeated reprocessing weakens the X-ray driving signal. When the X-ray driving signal is weaker, the intrinsic variability signal will be stronger in each emitted light curve. This second element is clear from the PSDs of these light curves in the right panel of Figure \ref{fig:psd}, which are each distinct from each other at higher frequencies because they depend more strongly on the intrinsic variability, which varies with radius.

It is potentially this additional reprocessing step, from X-ray to far-UV, to UV-optical, that leads to the longer than expected light-crossing lags in our simulations. In particular, while the X-ray irradiation takes a very short time to be reprocessed, the higher UV-optical scattering opacity could delay the far-UV reaching the outer disk. At $50~r_g<R<100~r_g$ where this additional reprocessing step occurs, the thermal and diffusion timescales at the photosphere are on the order of 1~day, which is roughly consistent with the additional lag duration we measure. 

Additionally, because the intrinsic variability signal is stronger in these light curves, our light-crossing lag measurements may be biased by the confounding signal of the long lag due to propagating disk fluctuations, especially in runs with less X-ray irradiation. For example, for our Low X-ray run the lag measurements between different emitted light curves underestimate and over-estimate the light-crossing timescale for the top and bottom of the disk, respectively, despite these light curves having moderate peak correlations $\rm{CC_{coeff}}>0.6$ (see bottom panels of Figures \ref{fig:topshorts} and \ref{fig:botshorts}). The direction of this bias corresponds to the direction of the long lags that we measure for the top and bottom of the disk. In other words, the long negative lag we measure in Figure \ref{fig:low_top} leads to an under-estimate of the light-crossing timescale for the top of the disk and the long positive lag we measure in Figure \ref{fig:low_bot} leads to an over-estimate of the light-crossing timescale for the bottom of the disk. This result is consistent with \cite{Secunda:2023} who found that the presence of multiple lag signals in mock light curves can bias the measurement of one lag signal towards the other lag signal.

Even for observed AGN light curves that appear to align with X-ray-driven RM models, in some cases when longer baseline light curves are divided into shorter epochs, the X-ray and UV-optical light curves become poorly correlated over these shorter timescales and have lag durations longer than predicted \citep[e.g.,][]{Lewin:2024,Hagen:2025,Zhou:2025,Montano:2026,Edelson:2024}. Correlations can also improve when variability is smoothed over longer timescales \citep{Partington:2024}. These findings are consistent with Figure \ref{fig:intervals}, where we divide the light curve emitted from the top of our High X-ray run into shorter intervals. We find that even if the light curves are consistent with traditional RM models on long timescales and over some shorter epochs, they are inconsistent over other shorter epochs. 

These changes over shorter timescales occur because while over the course of our 10~year simulation run the top of the disk is always thinner than the bottom, the top of the gas disk is still turbulent and can become convectively unstable over shorter timescales leading to time intervals where regions of the disk are more puffed up. There are also strong magnetic reconnection events that occur near the upper photosphere of the disk, producing flares in the emitted light curves. The top of the disk also has greater variability compared to the X-ray light curve at intermediate frequencies due to MRI driven fluctuations (see left panel of Figure \ref{fig:psd}), suggesting that signatures of intrinsic variability will be present in all light curves.

\cite{Lewin:2025} found that the BLR alone could not account for the longer than anticipated lags observed for nine AGN. Instead, \cite{Lewin:2025} find that the line-of-sight gas obscuration inferred by fitting the X-ray spectra of these AGN is strongly correlated with longer lags. This observation fits well with our model here since the puffed-up gas disk can be the source of X-ray obscuration. However, we note that BLR contamination can still play a role, especially in causing the \emph{u}-band excess lag.


\subsection{Can we use our models to measure disk thickness?}

We see clear distinctions between the light curves emitted from the thin top of the disk and the thick bottom of the disk in our simulation. Therefore, it may be possible to estimate the thickness of the disk based on the features of the emitted light curves. A novel method for estimating the thickness of the AGN disk using AGN light curves is of particular interest because it would allow us to test the standard thin disk model and estimate the accretion rate of the disk which is thought to increase disk thickness. In fact, we find that the accretion rate for the thick bottom of the disk is nearly double the accretion rate for the thin top of the disk at the inner boundary of our simulation ($r=50~r_g$).

The first obvious difference between the two sides of our simulated disk is that the variability in light curves emitted from the bottom of the disk is less X-ray-driven, leading to weaker correlations between the X-ray and UV-optical light curves, and longer than anticipated lags. There is some observational evidence that AGN with higher accretion rates have lower variability amplitudes \citep[e.g.,][]{Zuo:2012,Simm:2016,Yu:2022}, pointing to less reprocessed X-ray variability in thicker AGN disks. This trend is particularly strong for high-accretion-rate iron low-ionization Broad Absorption Line quasars, which are radiating powerfully enough to drive a thick outflow like we see for the bottom half of our disk \citep{Leighly:2024}. In addition, \cite{Steyn:2026} found that lag durations scale with accretion rates in their sample of $\sim9,500$ AGN and see some evidence of disk outflows, such as CIV emission line blueshifts and broad absorption lines, in their sample of AGN light curves with longer than anticipated lags. 

There are also more subtle differences between the light curves emitted from the top and bottom of the disk that remain the same for both our Gray run and our runs with X-ray irradiation. For example, we see in the left panel of Figure \ref{fig:mid_lcs} that light curves emitted from the thin top of the disk are more likely to have flares. These flares last $\sim 50$~days and appear to be due to strong magnetic reconnection events near the upper photosphere, such as the one we show in Panel D of Figure \ref{fig:rho}. We do not often see flares in light curves emitted from the bottom of the disk because the thicker disk makes magnetic reconnection weaker and pushes it farther out from the photosphere. Therefore, it is possible that observing these bright but rare flares would be a sign of a thin disk. 

In addition, it is easier for a thin disk to cool and wash out entropy inversions. As a result, the top of the disk is convectively stable and has enhanced variability at intermediate frequencies, $10^{-2}-1$~days$^{-1}$, corresponding to the thermal timescale where the MRI peaks. We see in our light curves emitted from the thicker bottom of the disk that a thicker convectively unstable disk will propagate longer timescale variability out from the midplane where the thermal timescale is longer leading to less intermediate frequency variability (see Figures \ref{fig:thermal}, \ref{fig:gray_psd}, and \ref{fig:psd}). Therefore comparing the variability at intermediate frequencies could potentially help differentiate between a thinner and thicker disk.


Finally, long lags on the inflow timescale can also be used to measure the thickness of AGN disks because the inflow timescale depends on the height of the disk. A main motivation for performing simulations with a larger radial range than our previous shearing box simulations \citep{Secunda:2024,Secunda:2025} is to understand how fluctuations propagate throughout the AGN disk and produce long lags in the light curves emitted by the disk. The scale of the density fluctuations, $\delta \rho/\langle\rho\rangle$, we find in our simulations is consistent with the scale of temperature fluctuations \cite{Neustadt:2022} and \cite{Neustadt:2023} find for eight local AGN. These temperature fluctuations in \cite{Neustadt:2022} and \cite{Neustadt:2023} propagate inward and outward on timescales of around 100~days, with propagation speeds of $\sim2\times10^8$~cm~s$^{-1}$. These timescales are roughly consistent with the lag timescales, inflow speeds, and propagation speeds that we find in our simulations, suggesting that temperature fluctuations inferred from broad band photometry could really probe this inflow timescale.

\cite{Stone:2023} also measured the propagation speed of temperature fluctuations for 100 SDSS AGN, and \cite{Yao:2022} measured a long negative lag of $\sim-70$~days between the optical and UV emitting region of the disk for the AGN Fairall 9 directly. Both of these measurements give a propagation speed of around $10^9$~cm~s$^{-1}$, which is more rapid than most of the lags found in our simulated disk. These rapid long lags are therefore hard to explain under the disk parameters of our simulations.

Long lag detections are still very rare, because they require multi-year, multi-waveband light curves. Even when longer light curves are available, they are typically only cross-correlated over much shorter timescales, meaning these signals would be completely missed. However, it may be possible to observe long lags in the $>5$~year baseline multi-wavelength light curves from the Zwicky Transient Facility (ZTF). In addition, future multi-year light curves from the Vera C. Rubin Observatory could be ideal for observing long negative lags \citep{Secunda:2023}.

We also note that even with noiseless, extremely high-cadence, 10-year baseline light curves, it is still difficult to detect precise long lags. In Figures \ref{fig:gray_bot} and \ref{fig:mid_bot} -- \ref{fig:low_top}, we observe a second peak in the cross-correlation function, but this peak is typically broad and low amplitude ($\rm{CC_{coeff}} \approx 0.4$). It is difficult to assess the significance of a lower amplitude peak like this. While the Fourier method should help identify lags, the fact that it can be biased towards a shorter measured lag, means that it is hard to determine the precise lag from this method. These limitations are consistent with findings in \cite{Secunda:2023}, where they used simple DRW modeled light curves, which they reprocessed with a long negative and short positive lag using a Gaussian response function. Future work can employ machine learning methods that are better able to handle the confounding signal of light-crossing lags to measure more robust lags in AGN light curves \citep{Secunda:2026}.


\section{Caveats}
\label{sec:discuss:caveats}
Radiation MHD simulations are still a novel technique for modeling AGN light curves. We therefore want to be very transparent about the caveats of these simulations, many of which can be mitigated in future work. The light curves emitted in our simulation are made up of a combination of intrinsic and X-ray variability. The intrinsic variation comes from the self-consistent evolution of the gas and radiation physics, but we have to make several assumptions to model the X-ray irradiation. First, we use a fixed axisymmetric corona. The location and geometry of the corona are still heavily debated, and some models of AGN disks allow for a corona that evolves on month-long timescales \citep{Panagiotou:2022B}. We performed a few preliminary simulations with different corona geometry and locations and found that it did not have a large impact on the emitted light curves compared to changing the luminosity of the injected X-ray irradiation.

We also have to assume a PSD to model the variability of the X-ray light curve. \cite{Secunda:2025} tested different PSDs for the X-ray light curve and found that it had a small impact on the emitted UV-optical light curves. Here, we only use a power-law light curve with PSD$\propto f^{-2}$, which results in emitted light curve PSDs that are somewhat steeper at all frequencies than in observations. A shallower PSD or a bending power-law PSD with less variability on longer timescales might be a better fit to X-ray observations \citep{Uttley:2002,Papadakis:2002,Markowitz:2003,McHardy:2004,McHardy2007,Paolillo:2023,Beard:2025,Hagen:2025,Georgakakis:2024} and would likely produce emitted PSDs more consistent with observations. These alternative X-ray PSDs would also make it easier to identify intrinsic variability at longer timescales in the emitted light curves, perhaps also making it easier to detect more long lags. However, using a single power-law here allows us to easily isolate any changes in this power-law that are introduced during reprocessing. Therefore it is useful to slowly add layers of complexity, even at the expense of matching observations. Future work should test different X-ray PSDs to examine the impact they have on emitted light curves in non-shearing box simulations. 

We compare the continuum lags we measure to the light-crossing timescale based on the radius they are emitted at. However, actual observations of AGN disks use the wavelengths of the emitted light curves to model the radial temperature profile of the AGN disk. The simulations presented here use only two different frequency groups. Upcoming simulations will increase the number of frequency groups to examine the peak wavelength of each radius light curve and also examine the full spectrum emitted by the disk. In particular, the impact of the X-ray irradiation on the disk temperature and emitted disk spectrum is poorly understood and will be studied in these upcoming simulations.



Finally, we only use one gray simulation run to initialize the four simulations presented in this paper. As a result, we only probe AGN variability for a single SMBH mass and accretion rate. Future work should investigate AGN variability for different AGN parameters. Also, we find that at the specific time we restart this one initial gray simulation turbulent fluctuations have produced both a thin disk for the top of the disk and a puffed out disk for the bottom of the disk. This result illustrates that both disk structures are possible for these AGN parameters but arise stochastically. In fact, while this asymmetry remains stable over the $\sim10$~year timescale of our simulation runs, it is not constant over the $\sim100$~year timescale the original gray simulation in \cite{Jiang:2025} was run for. Over this longer timescale, the disk structure can change significantly. 

\section{Conclusion}
\label{sec:conclude}
We perform multi-frequency radiation MHD simulations of the UV-optical-emitting region of an AGN disk irradiated by time-varying X-rays modeled as if they are emitted by an AGN corona. These simulations allow us to study the source of variability in light curves emitted by the AGN disk. The emitted light curves in our simulations can be directly compared to observed AGN light curves to increase our understanding of the structure and internal physics of AGN disks. We find four main conclusions from our simulations.
\begin{itemize}
    \item Thin AGN disks with sufficient X-ray luminosity are broadly consistent with the traditional X-ray-driven RM models.
    \item Thicker disks and disks with lower X-ray luminosity are inconsistent with traditional RM models, but in line with recent observations of moderate to weakly correlated X-ray and UV light curves \citep[e.g.][]{Edelson:2019,Cackett:2023,Buisson:2018} and continuum lags longer than predicted by the standard thin disk model \citep[e.g.][]{Steyn:2026,Guo:2022}. The structure of the disk itself causes this divergence from traditional models, not diffuse continuum from the BLR, as is often suggested.
    \item Fluctuations in gas density driven by MRI turbulence and convection produce intrinsic variability on the disk's thermal timescales in AGN light curves. These fluctuations propagate inwards and outwards on the inflow timescale leading to observable long lags between the intrinsic variability in different radii light curves on this inflow timescale. The simulations presented in this paper provide intuition for observing these propagating fluctuations in nature. Measuring the inflow timescale of the disk through the observation of long lags would allow us to test the standard thin disk model and probe the vertical structure of the disk, because the inflow timescale depends on the disk scale height (see Equation \ref{eq:inf}).
    \item In addition to measuring long lags, we can use the differences we see in light curves emitted from the thin and thick sides of our simulated disks to estimate the thickness of observed AGN disks using RM and the PSDs of their light curves. Disk thickness may be correlated with AGN accretion rates.
\end{itemize}

Traditional disk continuum RM is based on a quiet thin disk, but we see in our simulations that AGN disks evolve stochastically over a broad range of timescales, and that this turbulent evolution produces intrinsic variability in UV-optical AGN light curves. Future RM studies can incorporate this intrinsic variability into their mock light curves using the PSDs in Figures \ref{fig:gray_psd} and \ref{fig:psd}. Improved models for AGN light curves are crucial for preparing for the Rubin Observatory, a high-cadence and long baseline survey that will observe up to millions of AGN \citep{Ivezi2019}. Millions of AGN light curves will allow us to explore the full AGN parameter space and use the results of our simulations to measure AGN disk structure.

\begin{acknowledgements}
    The Center for Computational Astrophysics at the Flatiron Institute is supported by the Simons Foundation. 
\end{acknowledgements}

\appendix

\section{Disk Structure}
\label{app:disk}

\begin{figure}
    \centering
    \includegraphics[width=0.49\columnwidth]{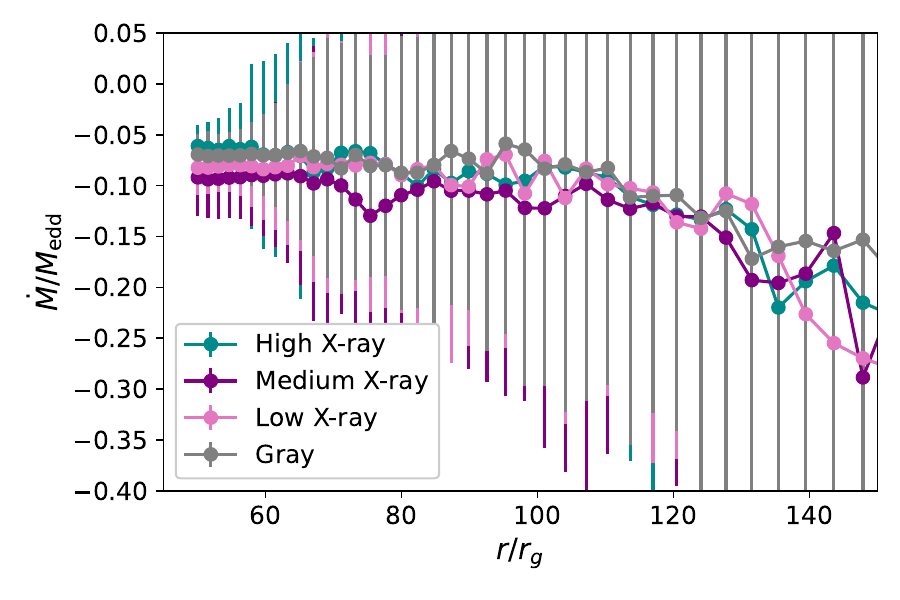} 
    \includegraphics[width=0.49\columnwidth]{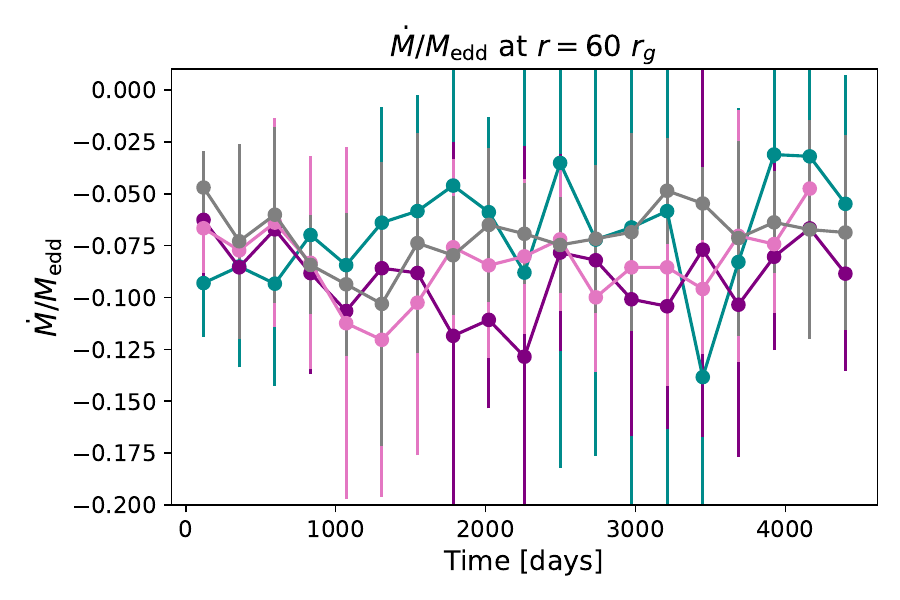} \\
    \includegraphics[width=0.49\columnwidth]{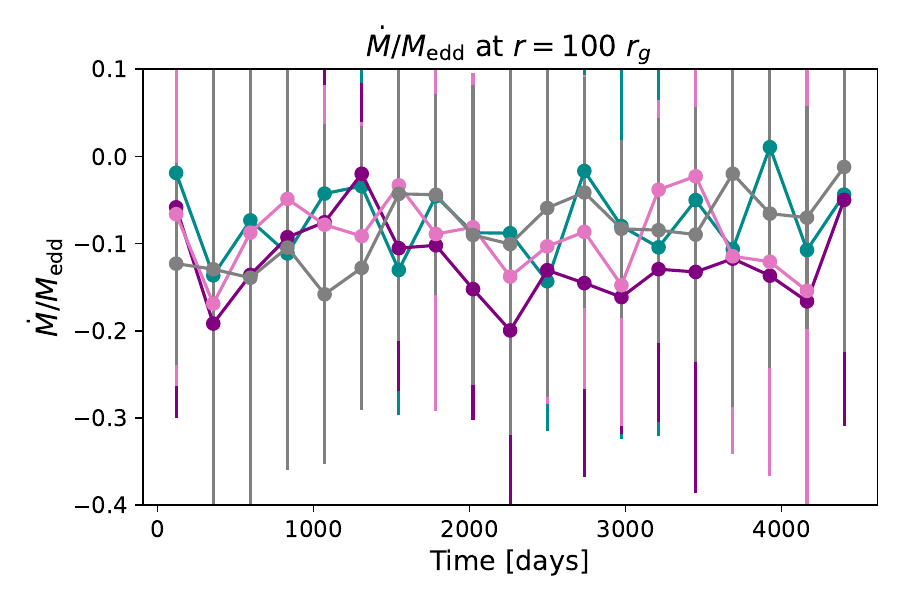}
    \includegraphics[width=0.49\columnwidth]{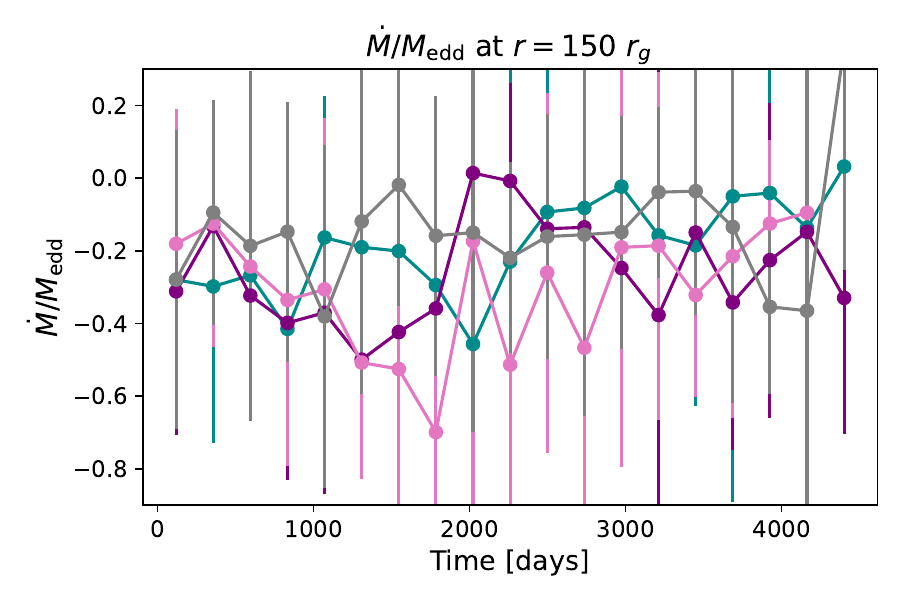}
    \caption{The Eddington accretion rate for our run without injected X-rays (in gray), our High X-ray run (in green), our Medium X-ray run (in purple) and our Low X-ray run (in pink). Top left panel shows the time-averaged accretion rate as a function of radius and the remaining panels show the accretion rate as a function of time for different radii in our simulations. $\dot{M}$ has been binned by time, with the points showing the mean and the error bars showing the standard deviation of the accretion rate in each bin. The accretion rates in the three runs are mostly consistent although the Medium X-ray run has a somewhat higher accretion rate.}
    \label{fig:mdot_t}
\end{figure}

\begin{figure*}
    \centering
    \includegraphics[width=0.49\linewidth]{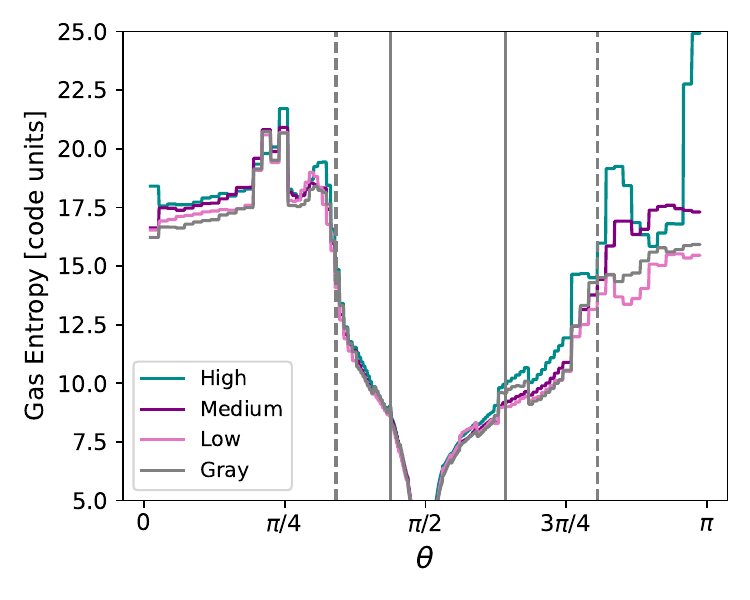}
    \includegraphics[width=0.49\linewidth]{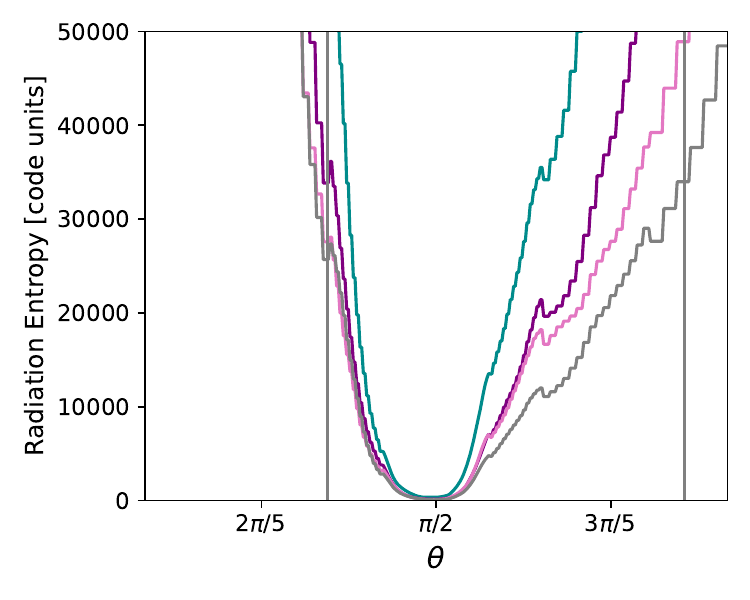} 
    \caption{Azimuthally- and time-averaged gas entropy (left panel) and radiation entropy (right panel) as a function of polar angle, $\theta$. The solid (dashed) vertical lines show the average location of the UV-optical (X-ray) photosphere. We restrict the polar angles for the radiation entropy profiles because the radiation entropy is only reliable within the photosphere. We can clearly see entropy inversions below the disk.}
    \label{fig:entropy}
\end{figure*}

\begin{figure*}
    \centering
    \includegraphics[width=0.49\linewidth]{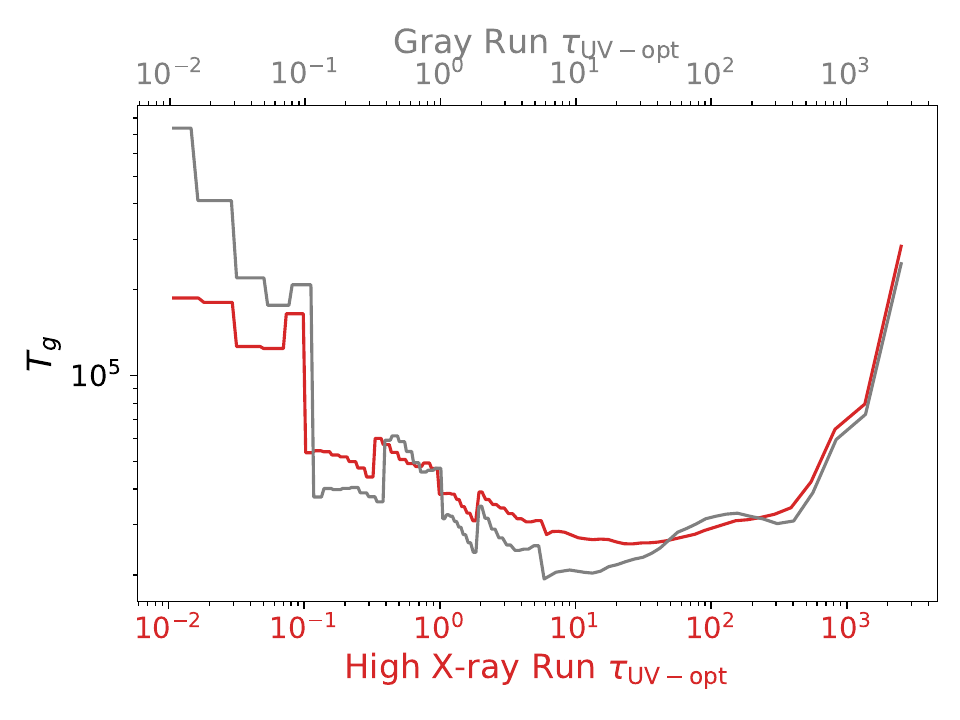}
    \includegraphics[width=0.49\linewidth]{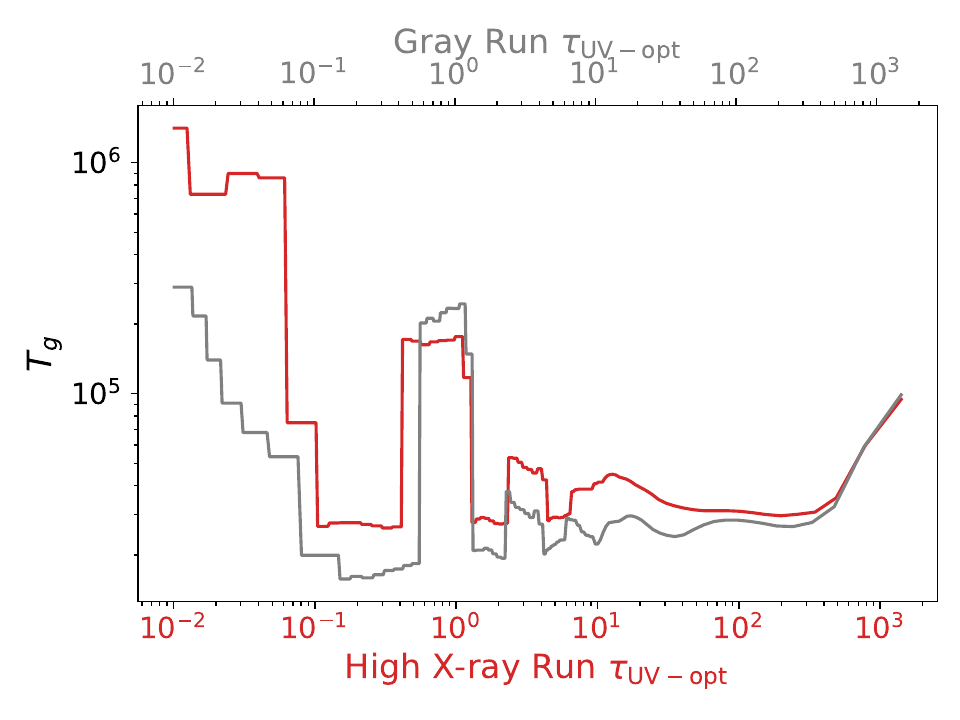} 
    \caption{Azimuthally- and time-averaged gas temperature (in Kelvin) as a function of UV-optical optical depth for our High X-ray run (top axis) and Gray run (bottom axis). Left and right panels are from the top and bottom of the disk, respectively.}
    \label{fig:t_of_tau}
\end{figure*}

\begin{figure}
    \centering
    \includegraphics[width=0.49\columnwidth]{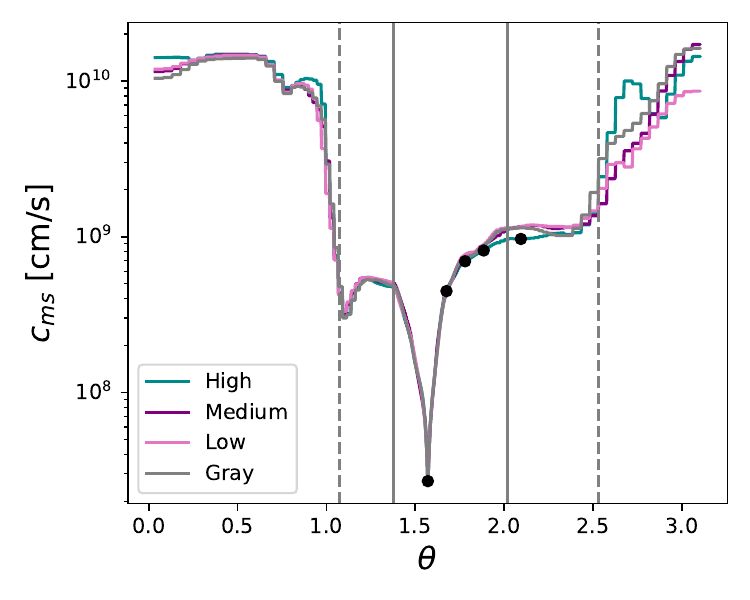}
    \caption{Horizontally- and time-averaged $\theta$-profiles of the magnetosonic speed, $c_{ms} = \sqrt{v_A^2+c_s^2}$. The solid vertical lines show the average location of the UV-optical photosphere, while the dashed lines show the average location of the X-ray photosphere. While the inflow speed is very slow at the midplane ($\sim 10^{-5}c$) it is much faster at the photosphere of the disk ($\sim 10^{-3}c$). The black dots give the $\theta$-locations in Figure \ref{fig:gdfluct} and \ref{fig:mdfluct}. The propagation speeds in Figures \ref{fig:gdfluct} and \ref{fig:mdfluct} and the lag timescales in Figures \ref{fig:mid_bot} - \ref{fig:low_top} are between $10^7$ and $4\times10^8$~cm~s$^{-1}$, consistent with the inflow speed near the photosphere and the magnetosonic speed near the midplane.}
    \label{fig:tmag}
\end{figure}

We provide additional information on the disk structure in this appendix. Figure \ref{fig:mdot_t} shows the accretion rate, $\dot{m}$, calculated as the total mass flux at the inner simulation boundary at $r=50~r_g$ for our four runs. The upper left panel shows the time- and azimuthally-averaged accretion rate as a function of radius and the remaining panels show the azimuthally-averaged accretion rate for three different radii as a function of time. We can see that the accretion rate rapidly equilibrates after we restart the Gray simulation and begin injecting X-rays. The High and Low X-ray runs and the Gray run without injected X-rays have similar accretion rates at all radii, but the Medium X-ray run does have an enhanced accretion rate at smaller radii, suggesting that X-ray irradiation can have an impact on the accretion rate of the disk.

Figure \ref{fig:entropy} shows the vertical profiles of the gas entropy ($S_g = \log(P_{\rm gas}/\rho^\gamma)/(\gamma-1)$, where $P_{\rm gas}$ is the gas pressure and $\gamma$ is the adiabatic index) and the radiation entropy ($S_r=4T_{\rm rad}^3/3\rho$, where $T_{\rm rad}$ is the radiation temperature) in the left and right panels, respectively. We see inversions in the gas and radiation entropy within the photosphere of the disk, especially for the bottom of the disk. These entropy inversions drive the convection that produces the puffed-up disk structure for the bottom of the disk.

We compare the gas temperature as a function of UV-optical optical depth for the High X-ray run and the Gray run in Figure \ref{fig:t_of_tau}. X-ray irradiation causes a slight increase in gas temperature within the photosphere. Outside the photosphere extremely low density regions produce discontinuities in the gas temperature that have a negligible impact on the simulation. Therefore, while X-ray irradiation drives temperature fluctuations near the disk photosphere, on average X-ray irradiation seems to have a small impact on the thermalized region of our simulated AGN disks. This result is consistent with the results of the shearing box simulations in \cite{Secunda:2024} and \cite{Secunda:2025}. A closer examination of the impact of X-ray irradiation on the disk structure, temperature, and spectrum emitted from the disk will be explored in future work.

\cite{Kaul:2025} found that fluctuations in their simulated AGN disks propagate inwards and outwards at the fast magnetosonic speed,
\begin{equation}
\label{eq:cms}
    c_{ms} = \sqrt{c_s^2 + v_A^2}.
\end{equation}
where $c_s$ is the sound speed, $v_A=\frac{B}{\sqrt{\rho}}$ is the Alfven speed, and $B$ is the magnitude of the magnetic field. Figure \ref{fig:tmag} shows the vertical profile for the fast magnetosonic speed.  We find that in our disk the fast magnetosonic speed at the photosphere is 1--2 orders of magnitude faster than the propagation speeds we measure at the photosphere. However, the midplane fast magnetosonic speed is consistent with the propagation speeds we measure at the photosphere. Therefore, the role this speed plays in the propagation of fluctuations is less clear than the role of the inflow speed.

\bibliography{agn_lag.bib}
\end{CJK*}
\end{document}